\documentclass[preprint2]{aastex}

\slugcomment{Submitted to ApJ, 11 June 2003}
\shorttitle{Resonant Trapping by Planet Migration}
\shortauthors{Wyatt}

\received{2003 June 11}
\begin{document}

\title{Resonant Trapping of Planetesimals by Planet Migration:
  Debris Disk Clumps and Vega's Similarity to the Solar System}

\author{M. C. Wyatt}
\affil{UK Astronomy Technology Centre, Royal Observatory,
  Edinburgh EH9 3HJ, UK}
\email{wyatt@roe.ac.uk}

\begin{abstract}
This paper describes a model which can explain the observed clumpy
structures of debris disks.
Clumps arise because after a planetary system forms its planets migrate
due to angular momentum exchange with the remaining planetesimals.
Outward migration of the outermost planet traps planetesimals outside
its orbit into its resonances and resonant forces cause azimuthal
structure in their distribution.
The model is based on numerical simulations of planets of
different masses, $M_{pl}$, migrating at different rates, $\dot{a}_{pl}$,
through a dynamically cold ($e<0.01$) planetesimal disk initially at a
semimajor axis $a$.
Trapping probabilities and the resulting azimuthal structures
are presented for a planet's 2:1, 5:3, 3:2, and 4:3 resonances.
Seven possible dynamical structures are identified from
migrations defined by $\mu=M_{pl}/M_\star$ and
$\theta=\dot{a}_{pl}\sqrt{a/M_\star}$.
Application of this model to the 850$\mu$m image of Vega's disk shows
its two clumps of unequal brightness can be explained by the migration
of a Neptune-mass planet from 40 to 65AU over 56Myr;
tight constraints are set on possible ranges of these parameters.
The clumps are caused by planetesimals in the 3:2 and 2:1
resonances;
the asymmetry arises because of the overabundance of
planetesimals in the 2:1(u) over the 2:1(l) resonance.
The similarity of this migration to that proposed for our own Neptune
hints that Vega's planetary system may be much more akin to the
solar system than previously thought.
Predictions are made which would substantiate this model, such as the
orbital motion of the clumpy pattern, the location of the planet,
and the presence of lower level clumps.
\end{abstract}


\section{Introduction}
The discovery of disks of dust around main sequence stars
showed that some grain growth must have occurred in these systems,
since this dust is short-lived and so must be continually
replenished (e.g., Backman \& Paresce 1993);
the dust is thought to originate in a collisional cascade that
starts with planetesimals a few km in size (Wyatt \& Dent 2002).
Thus the lack of warm dust in the inner $\sim 30$ AU of these
systems implies a relative paucity of planetesimals in these
regions (Laureijs et al. 2002).
This would naturally be explained if the planetesimals here grew
large enough for their gravitational perturbations to clear these
regions of any remaining debris, i.e., if they grew to $\gg 1000$
km sized planets (Kenyon \& Bromley 2002).
While the presence of such planets is still questionable, it is
widely believed that these debris disk systems are solar system
analogs, and that the dust that is observed is produced by the
destruction of analog Kuiper belts (Wyatt et al. 2003).

The most positive indication that debris disk systems
harbor planets comes from the morphology of the dust disks.
All of the disks which have been imaged exhibit clumps and
asymmetries which have been interpreted as perturbations from
a planetary system:
e.g., a brightness asymmetry in the HR4796 disk (Telesco et al.
2000) can be explained by the secular gravitational perturbations of
a planet on an eccentric orbit (Wyatt et al. 1999);
clumps observed in the $\epsilon$ Eridani, Vega and Fomalhaut
disks (Holland et al. 1998; Greaves et al. 1998; Wilner et al.
2002; Holland et al. 2003) may be indicative of dust trapped
in mean motion resonance with a planet in these systems
(Ozernoy et al. 2000; Wyatt \& Dent 2002; Quillen \& Thorndike
2002; Kuchner \& Holman 2003);
and a warp in the $\beta$ Pictoris disk (Heap et al. 2000)
may be caused by the perturbations of a planet on an inclined
orbit (Augereau et al. 2001).

This paper focuses on the possibility that clumps in debris
disks are caused by dust trapped in planetary resonances.
Simple geometrical arguments show how material that is in
such resonances is not evenly distributed in azimuth about
the star it is orbiting (e.g., Murray \& Dermott 1999;
see \S 4), and this observation has been exploited by
several authors to use debris disk clumps to infer the
presence of planets.
In the studies currently in the literature, dust is trapped
into planetary resonances when it encounters these resonances
while spiraling in towards the star due to P-R drag,
much in the same way that dust originating in the asteroid
belt becomes trapped in resonance with the Earth
(Dermott et al. 1994), and dust originating in the Kuiper
belt might become trapped in resonance with Neptune
(Moro-Mart\'{i}n \& Malhotra 2002).
However, Wyatt et al. (1999) showed that P-R drag is not
important for dust in the bright debris disks that are
currently known, since these are much denser than the zodiacal
cloud or the putative Kuiper belt dust cloud.
This means that debris disk dust is destroyed in mutual collisions
much faster than the P-R drag timescale until it is small
enough to be blown out of the system by radiation pressure.

Another reason why dust may be trapped in planetary
resonances is evident by comparison with the solar system.
About one third of known Kuiper belt objects (hereafter KBOs),
including Pluto, are trapped in 3:2 resonance with Neptune
(Jewitt 1999).
More recently KBOs have also been found in Neptune's other
resonances, including the 1:1, 4:3, 7:4, 2:1, and 5:2 resonances
Chiang et al. (2003).
These planetesimals are thought to have become trapped in these
resonances when Neptune's orbit expanded in the early
history of the solar system (Malhotra 1995).
In this scenario, the orbital expansion arises from
angular momentum exchange resulting from the scattering by Neptune
of the residual planetesimal disk left over at the end of planet
formation;
the current orbital distribution seen in the Kuiper belt
can be explained by the migration of Neptune from 23-30 AU over
a period of 50 Myr (Hahn \& Malhotra 1999). 
While this model does not completely describe the KBO
distribution, it turns out that some of its puzzling features,
such as the high inclination population of the 
classical KBOs (i.e., those not trapped in resonance with
Neptune), can now be accounted for with more detailed models
of the migration (Gomes 2003).
Besides, the success with which it explains the orbital
distribution of the resonant KBOs means that the migration of
Neptune is almost certain to have played a role in the formation
of the Kuiper belt.
Furthermore, models of the formation of such massive planets in
the absence of the substantial accretion of gaseous material
predicts such migration (Fern\'{a}ndez \& Ip 1984).
Thus if planets did form in the inner regions of debris disk
systems, some fraction of the parent planetesimals from which
the dust originates is expected to be trapped in resonance
with the outermost planet of that system.

Current studies of planet migration and the consequent resonant
trapping deal with the migration of Neptune and its effect on the
Kuiper belt (Malhotra 1995; Hahn \& Malhotra 1999; Zhou et al. 2002;
Chiang \& Jordan 2002).
This study is the first in a series which aims to determine the
effect of planet migration in extrasolar systems.
In such systems, planets of different mass may have formed at
different distances from different mass stars;
they also may have migrated at different speeds.
This paper explores the effect of planet migration on
a planetesimal disk, while the details of how the planet achieves
that migration and of the dust distribution resulting
from the destruction of these planetesimals are left for future
studies in the series.

The layout of the paper is as follows.
The numerical model of planet migration is described
in \S \ref{s:nm} and in \S \ref{s:rcp} this is used to calculate
the probability of capture of planetesimals into different
resonances for given migration scenarios.
\S \ref{s:drp} shows how the geometry of resonant planetesimal
orbits causes their distribution to be clumpy and parameters
defining this clumpiness are derived from the numerical
simulations.
In \S \ref{s:dds} I show how these results can be applied to
the structure of debris disks using the specific example of
the structure observed around Vega to set constraints on
the mass and migration history of planets in this system. 
The conclusions are given in \S \ref{s:concl}.

\section{Numerical Model}
\label{s:nm}
The dynamical evolution of a system comprised of
200 planetesimals and 1 planet was followed using
the RADAU fifteenth order integrator program
(Everhart 1985).
This integration scheme is self-starting in that
the time steps of each sequence are variable and are
chosen by the integrator based on the results of the
previous sequence;
substeps within each sequence are taken at Gauss-Radau
spacings.

In these integrations all bodies are assumed to orbit a
star of mass $M_\star$.
The planet has a mass of $M_{pl}$ while the planetesimals
are assumed to be massless, i.e., they do not affect
the evolution of the planet's orbit though they are
affected by it.
At the start of the integration, the planetesimals are
assumed to have eccentricities, $e$, chosen randomly
from the range 0 to 0.01.
Their inclinations, $I$, were chosen randomly from the
range 0 to $0.57^\circ = 0.01$ rad, and their arguments
of periastron, $\tilde{\omega}$, longitudes of ascending
node, $\Omega$, and longitudes, $\lambda$, were each
chosen randomly from the range 0 to $360^\circ$.
The planet was assumed to have all of these orbital
parameters set to zero at the start of the integration;
i.e., the planet starts on a circular orbit in the mid-plane
of the planetesimal disk.
The initial semimajor axes of the planetesimals, $a$,
and that of the planet, $a_{pl}$, were set according to
the integration being performed.

Planet migration was simulated by the addition of a
force acting in the direction of orbital motion of the
planet.
The prescription for the force used in this paper has as its
input the variable $\dot{a}_{var}$ and results
in an acceleration of:
\begin{equation}
  \dot{v} = 0.5\dot{a}_{var}\sqrt{GM_\star/a^3}. \label{eq:vdot2}
\end{equation}
Since the work done by this force results in a change in orbital
energy (see e.g., section 2.9 of Murray \& Dermott 1999), this
causes a change in the planet's semimajor axis of
$\dot{a} = 2\dot{v}\sqrt{a^3/GM_\star} = \dot{a}_{var}$;
i.e., the force defined by $\dot{v}$ results in a constant
rate of change in the planet's semimajor axis.
The planet maintains its circular orbit ($e_{pl} = 0$) in
the midplane of the disk ($I_{pl}=0$).

The integrator itself has been extensively tested, and the
addition of planet migration was tested by repeating
simulation Ia of Chiang \& Jordan (2002).
The resulting final distribution of eccentricities,
inclinations and semimajor axes of planetesimals
was qualitatively the same as that depicted in
their figure 3.

\section{Resonant Capture Probabilities}
\label{s:rcp}
Planetary resonances are locations at which planetesimals
orbit an integer $p$ number of times for every integer
$p+q$ times that the planet orbits the star.
Kepler's third law shows that resonances occur at
semimajor axes of:
\begin{equation}
  a_{p+q:p} = a_{pl}\left(\frac{p+q}{p}\right)^{2/3}. \label{eq:apq}
\end{equation}
If the planet is migrating at a rate $\dot{a}_{pl}$,
these resonances only remain in the vicinity of
planetesimals at a given semimajor axis for a short time.
However, while close to a planetary resonance, a planetesimal
receives periodic kicks to its orbit from the gravitational
perturbations of the planet.
These can impart angular momentum to the planetesimal so that
its semimajor axis increases in such a way that the planetesimal
is always orbiting at the location of the resonance;
such a planetesimal is said to be trapped in the planet's
resonance.
The resonant forces causing this trapping are discussed in
greater detail in \S \ref{s:drp}.

The first set of integrations was performed with the
aim of determining how many planetesimals initially at a
semimajor axis $a$ (in the range 30-150 AU) get trapped
in a given resonance of a planet of mass $M_{pl}$ (in the
range 1-300 $M_\oplus$) that is migrating at a constant
rate of $\dot{a}_{pl}$ (in the range 0.01-1000 AU Myr$^{-1}$),
where all bodies are orbiting a star of mass $M_\star$
(in the range 0.5-5 $M_\odot$).
Studies of the migration of the solar system's planets
showed that the most important resonances are the 3:2 and
2:1 resonances, and to a lesser extent the 4:3 and 5:3
resonances.
These are studied in individual sections below,
where a model is derived to estimate capture
probabilities using equations (\ref{eq:mu})-(\ref{eq:p2})
and Table \ref{tab:xyuv}.
Readers not interested in how this model is derived
could skip to \S \ref{ss:tpsumm} without much loss in
continuity.

The way the integrations were performed was to study
specific values of $a$, $M_{pl}$ and $M_\star$,
then determine the trapping probability, $P$, for different
migration rates chosen with the aim of obtaining
$P$ in the range 0.1-0.9.
Trapping probabilities were determined by starting the
planet at a semimajor axis d$a_1$ below the location of
the resonance and allowing the integration to continue
until the planet had migrated d$a_2$ beyond the resonance.
The parameters d$a_1$ and d$a_2$ were chosen by 
monitoring the orbital elements of the planetesimals
at 10 intervals throughout the integrations.
Since resonances have a finite libration width, d$a_1$
was chosen so that the planetesimals were started well
outside the resonance region.
At the end of all runs, d$a_2$ was chosen so that two
distinct populations of trapped and non trapped planetesimals
were easy to distinguish by the semimajor axes and 
eccentricities of their members.
Typically both d$a_1$ and d$a_2$ were set between 0.5-4 AU.

\subsection{3:2 Resonance}
\label{ss:23}

\subsubsection{Dependence on planet mass, $M_{pl}$}
\label{sss:23mpl}
\begin{figure}
  \begin{center}
    \begin{tabular}{rlc}
      \textbf{(a)} & \hspace{-0.3in}
        \epsscale{0.9} 
\plotone{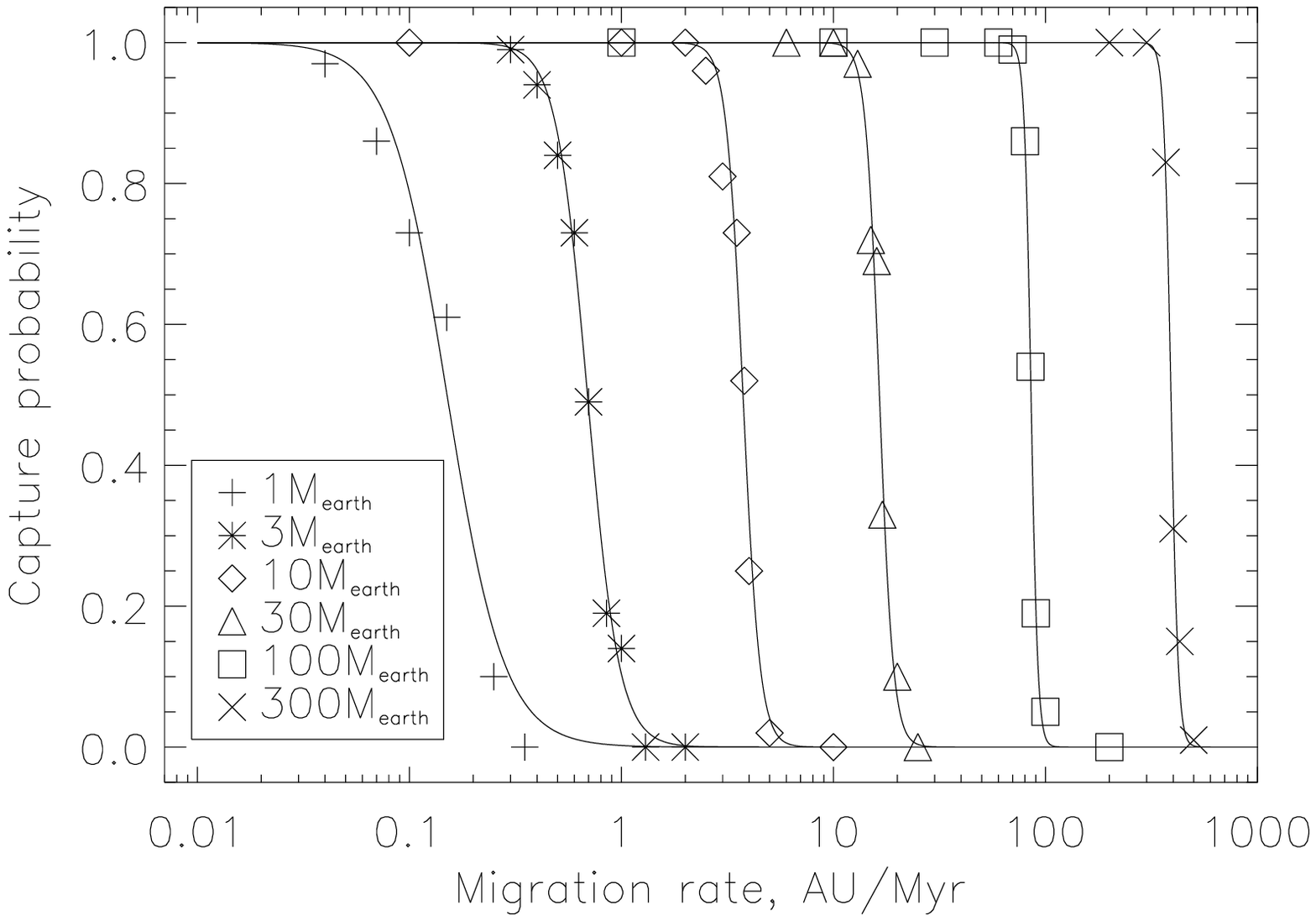} & \\[0.2in]
      \textbf{(b)} & \hspace{-0.4in} 
        \epsscale{0.92} 
\plotone{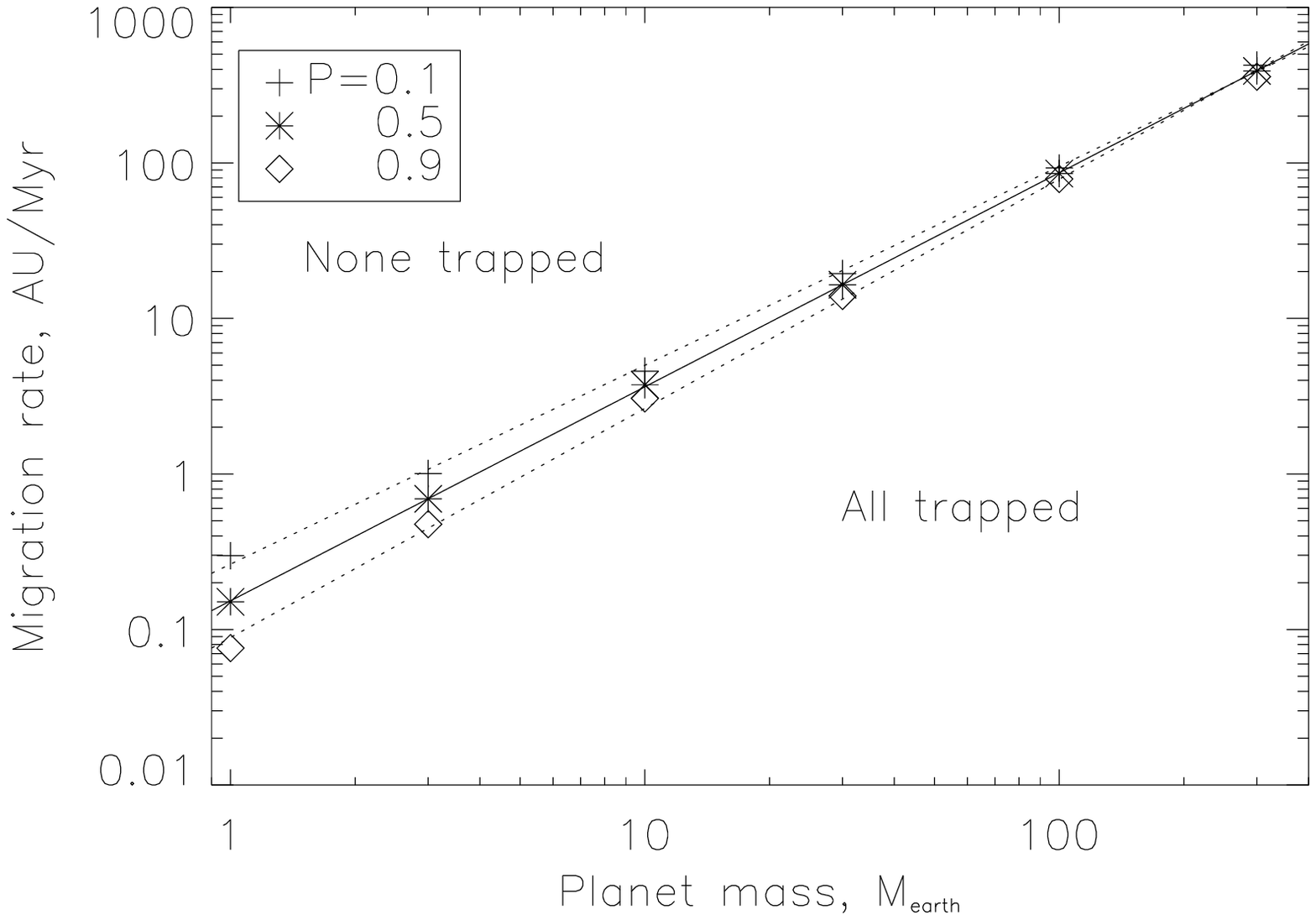} & \\[0.2in]
      \textbf{(c)} & \hspace{-0.3in}
        \epsscale{0.9} 
\plotone{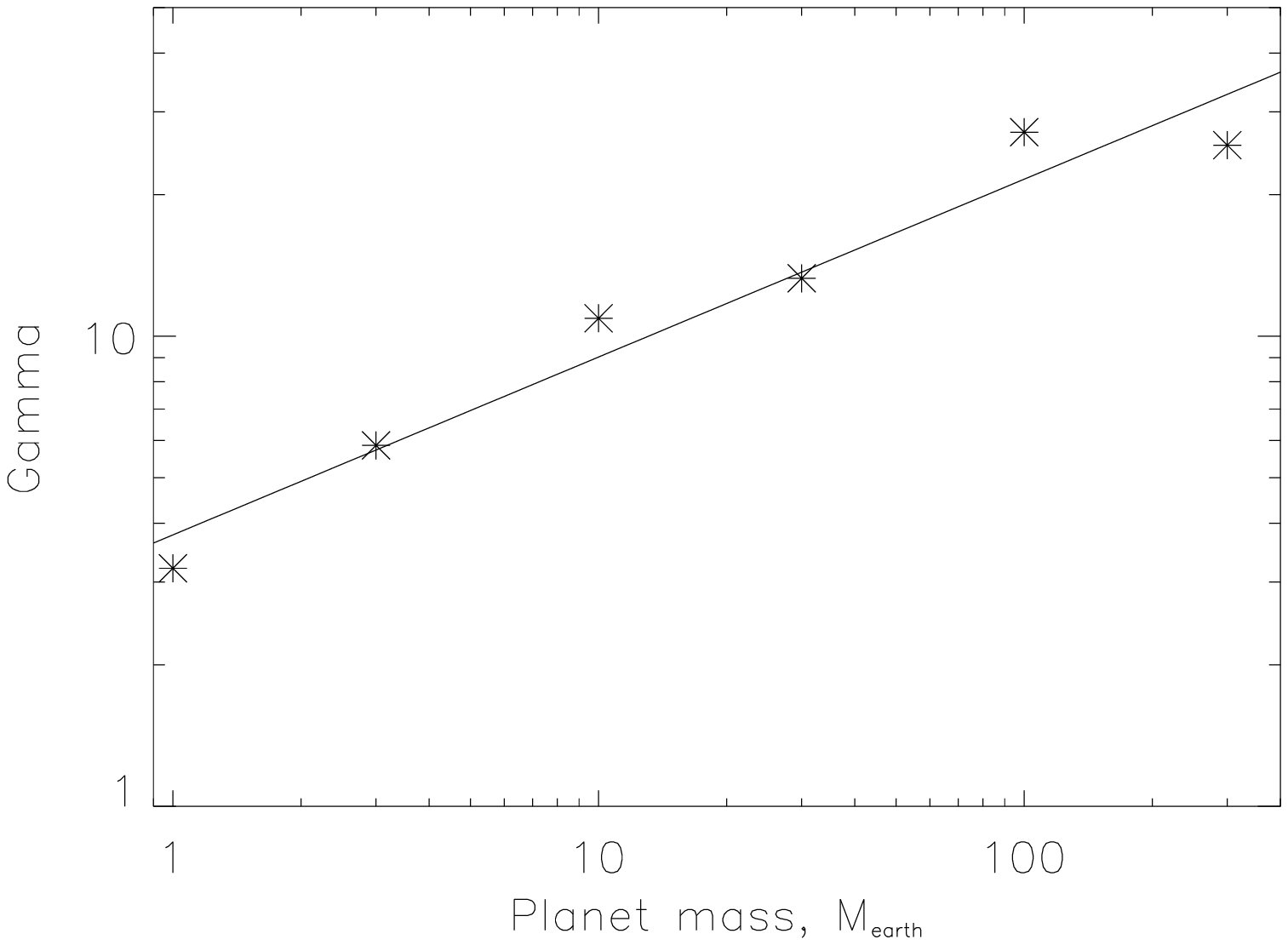} &
    \end{tabular}
  \end{center}  
  \caption{Capture probabilities for the 3:2 resonance for
  planetesimals initially orbiting at 60 AU from a $2.5M_\odot$
  star plotted for different mass planets that are migrating
  at different rates \textbf{(a)} (see \S \ref{sss:23mpl}).
  The solid lines show fits to these probabilities using
  the function given in equation (\ref{eq:p}).
  Parameters derived from these fits are shown in
  \textbf{(b)} and \textbf{(c)}.
  \textbf{(b)} shows the migration rate required for a
  planet to capture a given fraction of the planetesimals
  in its 3:2 resonance, while $\gamma$ in \textbf{(c)}
  defines how fast the capture probability drops
  with migration rate for a given planet mass.
  \label{fig:23mpl}}
\end{figure}

The first runs were undertaken with
$M_\star = 2.5 M_\odot$ and $a=60$ AU.
The nominal location of the planet for planetesimals
at 60 AU to be in its 3:2 resonance is $a_{pl} = 45.8$ AU.
Trapping probabilities were determined for planet masses
of 1, 3, 10, 30, 100, and 300$M_\oplus$, and the results are
plotted in Figure \ref{fig:23mpl}a.
For a given planet mass, capture is assured as long as
the migration rate is low enough.
The capture probability is a strong function of the
planet's migration rate, and the range of migration
rates for which the capture probability is 0.1-0.9 is
relatively small.
Such observations are consistent with analytical predictions;
e.g., a sharp dependence on the migration rate is predicted
by autoresonance theory (Friedland 2001), and
certain capture is expected in the adiabatic
limit (i.e., when the migration rate is slow, see Appendix
\ref{app:adiabatic}) provided that the planetesimals' eccentricities
are low enough (e.g., Henrard 1982; Henrard \& Lemaitre 1983).
The higher capture probabilities found with more massive planets
for any given migration rate are also expected, since their
resonant gravitational perturbations are much stronger.

A parametric fit to these trapping probabilities was
performed that has the form:
\begin{equation}
P = \left[ 1 + \left(\frac{\dot{a}_{pl}}{\dot{a}_{0.5}}
               \right)^\gamma\right]^{-1}, \label{eq:p}
\end{equation}
where $\dot{a}_{0.5}$ is the migration rate for which half
of the planetesimals are captured, and the parameter
$\gamma$ determines how fast the turnover is from a capture
probability of 0.9 to 0.1 (e.g.,
$(\dot{a}_{0.1} - \dot{a}_{0.9})/\dot{a}_{0.5} =
9^{1/\gamma} - 9^{-1/\gamma}$).
These fits are also plotted in Figure \ref{fig:23mpl}a.

The parameters derived from these fits, $\dot{a}_{0.5}$ and
$\gamma$, are plotted in Figures \ref{fig:23mpl}b and
\ref{fig:23mpl}c to show how they vary
with the mass of the migrating planet.
A linear regression fit to the logarithm of these parameters
is also shown on these plots.
Thus it is found that the capture probability for the systems
described in these runs can be approximated by equation (\ref{eq:p})
with the following parameters:
\begin{eqnarray}
  \dot{a}_{0.5} & = & 0.153 M_{pl}^{1.376 \pm 0.004}
    \label{eq:a0.5mpl} \\
  \gamma        & = & 3.8 M_{pl}^{0.38 \pm 0.04}.
    \label{eq:gammpl}
\end{eqnarray}
Equation (\ref{eq:a0.5mpl}) agrees well that the prediction
from autoresonance theory that the critical migration rate
should scale with $M_{pl}^{4/3}$ (Friedland 2001).
Equation (\ref{eq:gammpl}) shows that lower mass planets
have a larger range of migration rates resulting in
intermediate (0.1-0.9) capture probabilities.

\subsubsection{Dependence on semimajor axis, $a$}
\label{sss:23a}
\begin{figure}
  \begin{center}
    \begin{tabular}{rlc}
      \textbf{(a)} & \hspace{-0.3in}
        \epsscale{0.9} 
\plotone{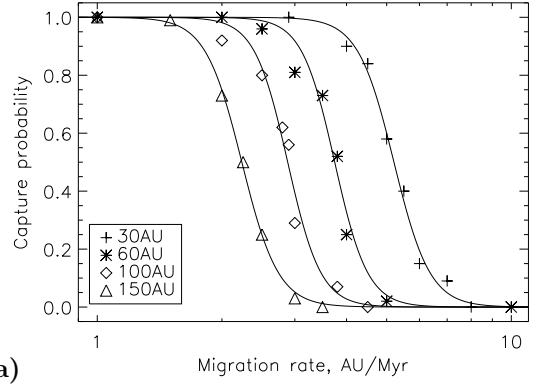} & \\[0.2in]
      \textbf{(b)} & \hspace{-0.3in} 
        \epsscale{0.9} 
\plotone{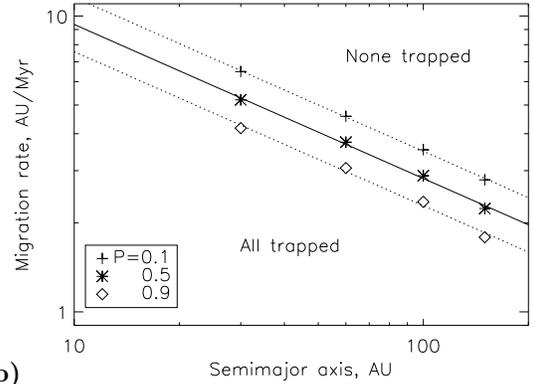} & \\[0.2in]
      \textbf{(c)} & \hspace{-0.27in}
        \epsscale{0.9} 
\plotone{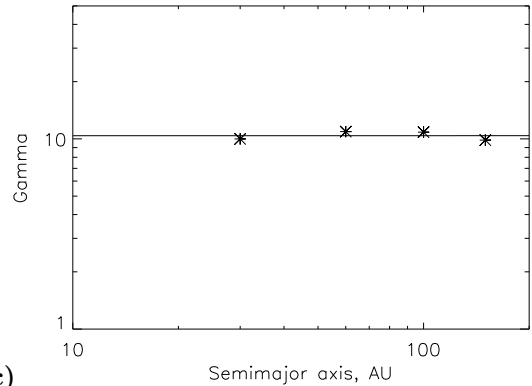} &
    \end{tabular}
  \end{center}  
  \caption{Capture probabilities for the 3:2 resonance of a
  $10M_\oplus$ planet for planetesimals initially orbiting at
  different distances from a $2.5M_\odot$ star \textbf{(a)}
  (see \S \ref{sss:23a}).
  The solid lines show fits to these probabilities using
  the function given in equation (\ref{eq:p}).
  Parameters derived from these fits are shown in
  \textbf{(b)} and \textbf{(c)}.
  \label{fig:23a}}
\end{figure}

Next the dependence of trapping probabilities on the
planetesimals' semimajor axis was tested by 
performing a set of runs with $M_\star = 2.5 M_\odot$ and
$M_{pl} = 10M_\oplus$, and placing the planetesimals
at semimajor axes of 30, (60), 100, and 150 AU.
The results are plotted in Figure \ref{fig:23a}, where I have also
plotted fits to the capture probabilities and to the
derived parameters in the same manner as in
\S \ref{sss:23mpl}.
In this way the trapping probability is found to
depend on $a$ in the sense that
$\dot{a}_{0.5} \propto a^{-0.52 \pm 0.02}$.
No dependence of $\gamma$ with $a$ was found
($\gamma \propto a^{0.00 \pm 0.05}$).

\subsubsection{Dependence on stellar mass, $M_\star$}
\label{sss:23ms}
\begin{figure}
  \begin{center}
    \begin{tabular}{rlc}
      \textbf{(a)} & \hspace{-0.3in}
        \epsscale{0.9} 
\plotone{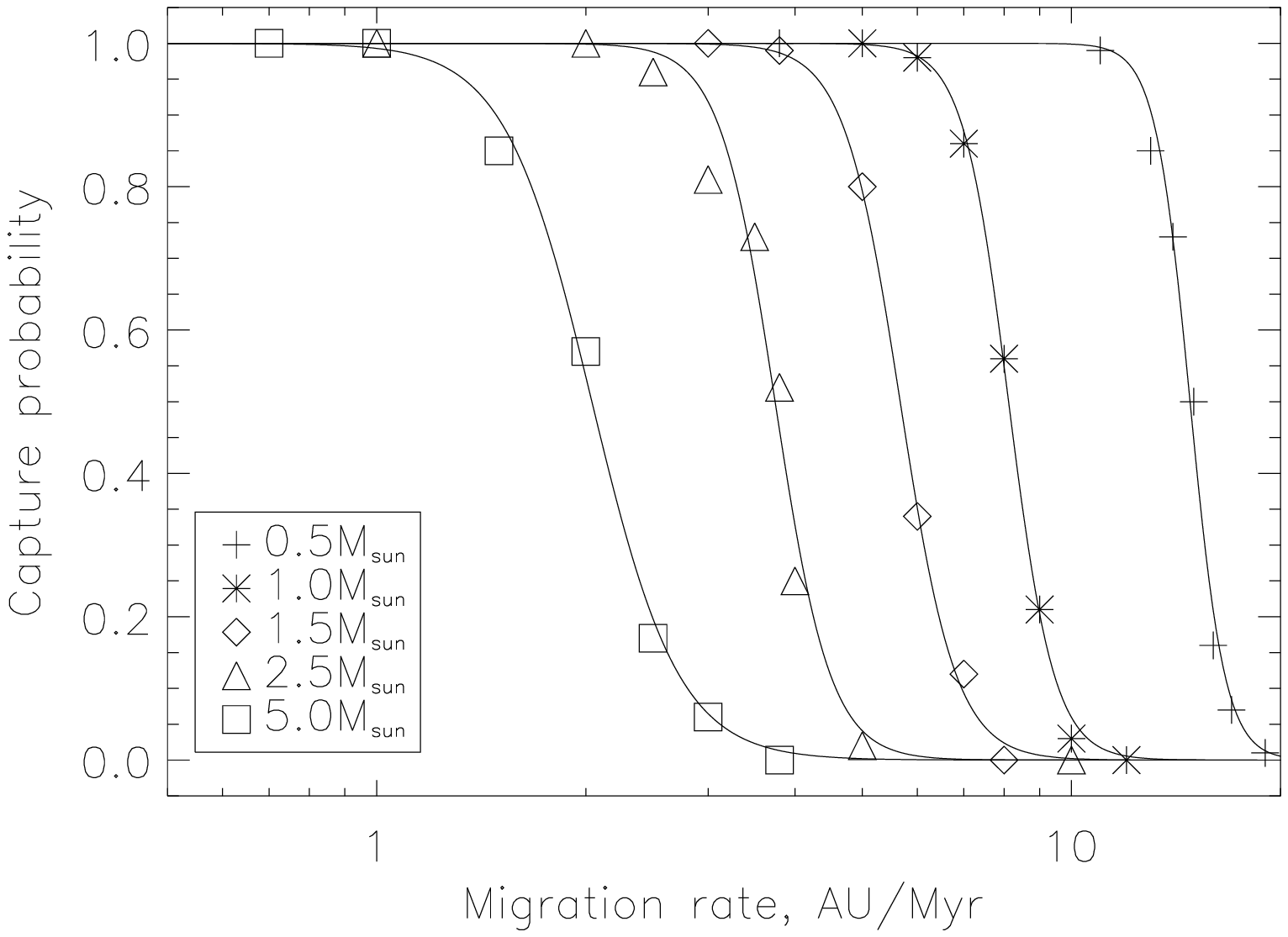} & \\[0.2in]
      \textbf{(b)} & \hspace{-0.3in} 
        \epsscale{0.92} 
\plotone{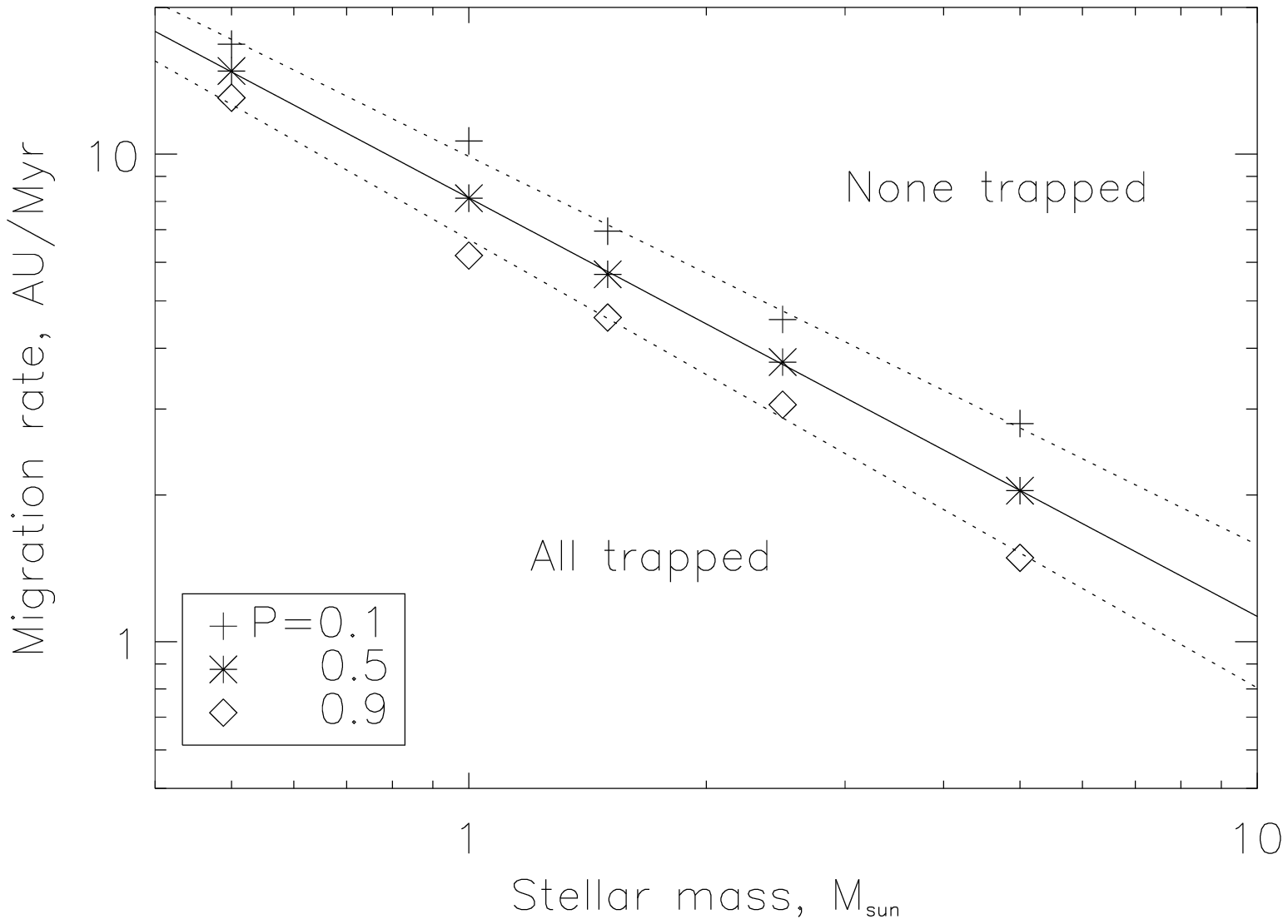} & \\[0.2in]
      \textbf{(c)} & \hspace{-0.3in}
        \epsscale{0.92} 
\plotone{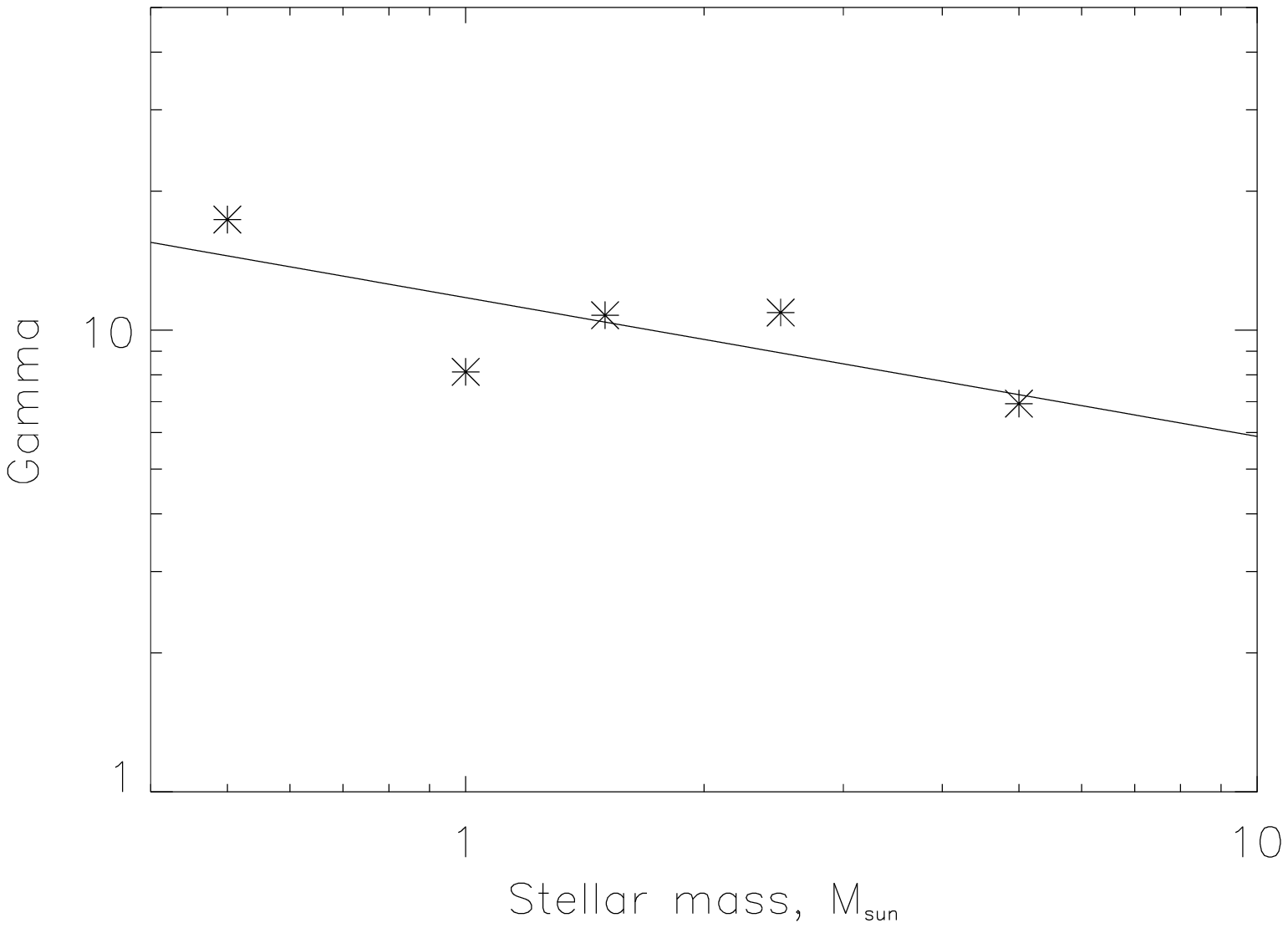} &
    \end{tabular}
  \end{center}  
  \caption{Capture probabilities for the 3:2 resonance of a
  $10M_\oplus$ planet for planetesimals initially orbiting at
  60 AU from stars of different mass \textbf{(a)}
  (see \S \ref{sss:23ms}).
  The solid lines show fits to these probabilities using
  the function given in equation (\ref{eq:p}).
  Parameters derived from these fits are shown in
  \textbf{(b)} and \textbf{(c)}.
  \label{fig:23ms}}
\end{figure}

Finally, the dependence of trapping probabilities on the
stellar mass was tested.
This was achieved with a set of runs with
$M_{pl} = 10M_\oplus$ and $a = 60$ AU, and trying
different stellar masses of 0.5, 1.0, 1.5, (2.5),
and 5.0$M_\odot$.
The results are plotted in Figure \ref{fig:23ms}, where fits to
the capture probabilities and to the derived parameters
are also plotted (see e.g., \S\S \ref{sss:23mpl} and
\ref{sss:23a}).
In this way the trapping probability is found to
depend on $M_\star$ in the sense that
$\dot{a}_{0.5} \propto M_\star^{-0.86 \pm 0.01}$.
A weak dependence of $\gamma \propto M_\star^{-0.3 \pm 0.15}$
was also found.

\subsubsection{Summary}
From the previous sections, it is evident that
the trapping probability depends only on the dimensionless
parameters:
\begin{eqnarray}
  \mu    & = & M_{pl}/M_\star, \label{eq:mu} \\
  \theta & = & \dot{a}_{pl}\sqrt{a/M_\star}, \label{eq:theta}
\end{eqnarray}
where $\mu$ determines the gravitational strength of the
planet's resonance,
and $\theta$ is the ratio of the planet's migration rate to
the planetesimal's orbital velocity, which determines
the angle at which the resonance is encountered.
\footnote{To allow you put the angle $\theta$ in perspective,
a migration rate of 1AU Myr$^{-1}$ for a $1M_\odot$ star corresponds
to meeting the orbital velocity at 100 AU at an incident
angle of 0\farcs3.}
Thus the probability that planetesimals orbiting a star of
mass $M_\star$ (in $M_\odot$) at a semimajor axis
$a$ (in AU) get captured into the 3:2 resonance of a
planet of mass $M_{pl}$ (in $M_\oplus$) that is migrating
at a rate $\dot{a}_{pl}$ (in AU Myr$^{-1}$) at a
semimajor axis of $(2/3)^{2/3}a$ can be
approximated by:
\begin{equation}
  P = \left[1+\left(X\mu^{-u}\theta
              \right)^{Y\mu^v}\right]^{-1}.
  \label{eq:p2}
\end{equation}

Armed with this knowledge, the results of all the
runs in \S\S \ref{sss:23mpl}-\ref{sss:23ms} were reanalysed
to determine $X_{3:2}$, $Y_{3:2}$, $u_{3:2}$, and $v_{3:2}$.
A least squares fit to equation (\ref{eq:p2}) was used to
obtain best fit values of:
$X_{3:2} = 0.37$, $Y_{3:2} = 5.4$, $u_{3:2} = 1.37$, and
$v_{3:2} = 0.38$.
The errors in these parameters are estimated to be
$\pm 0.02,2.0,0.02$ and 0.1 respectively.
This model for $P$ was found to be accurate to about
$\pm 0.04$ over the range of parameters tried in these
runs.
Translating back to the lexicon of equations
(\ref{eq:p})-(\ref{eq:gammpl}), which can
be compared with the parameters derived from Figs. (1-3):
\begin{eqnarray}
  \dot{a}_{0.5} & = & 2.7(M_{pl}/M_\star)^{1.37}
                      \sqrt{M_\star/a}, \label{eq:ad23}\\
  \gamma        & = & 5.4(M_{pl}/M_\star)^{0.38}. 
                      \label{eq:gam23}
\end{eqnarray}
Thus the results derived in \S \ref{sss:23mpl} 
have not changed by the inclusion of the runs in
\S\S \ref{sss:23a} and \ref{sss:23ms}.

\subsection{2:1 Resonance}
\label{ss:12}
\begin{figure}
  \begin{center}
    \begin{tabular}{rlc}
      \textbf{(a)} & \hspace{-0.3in}
        \epsscale{0.9} 
\plotone{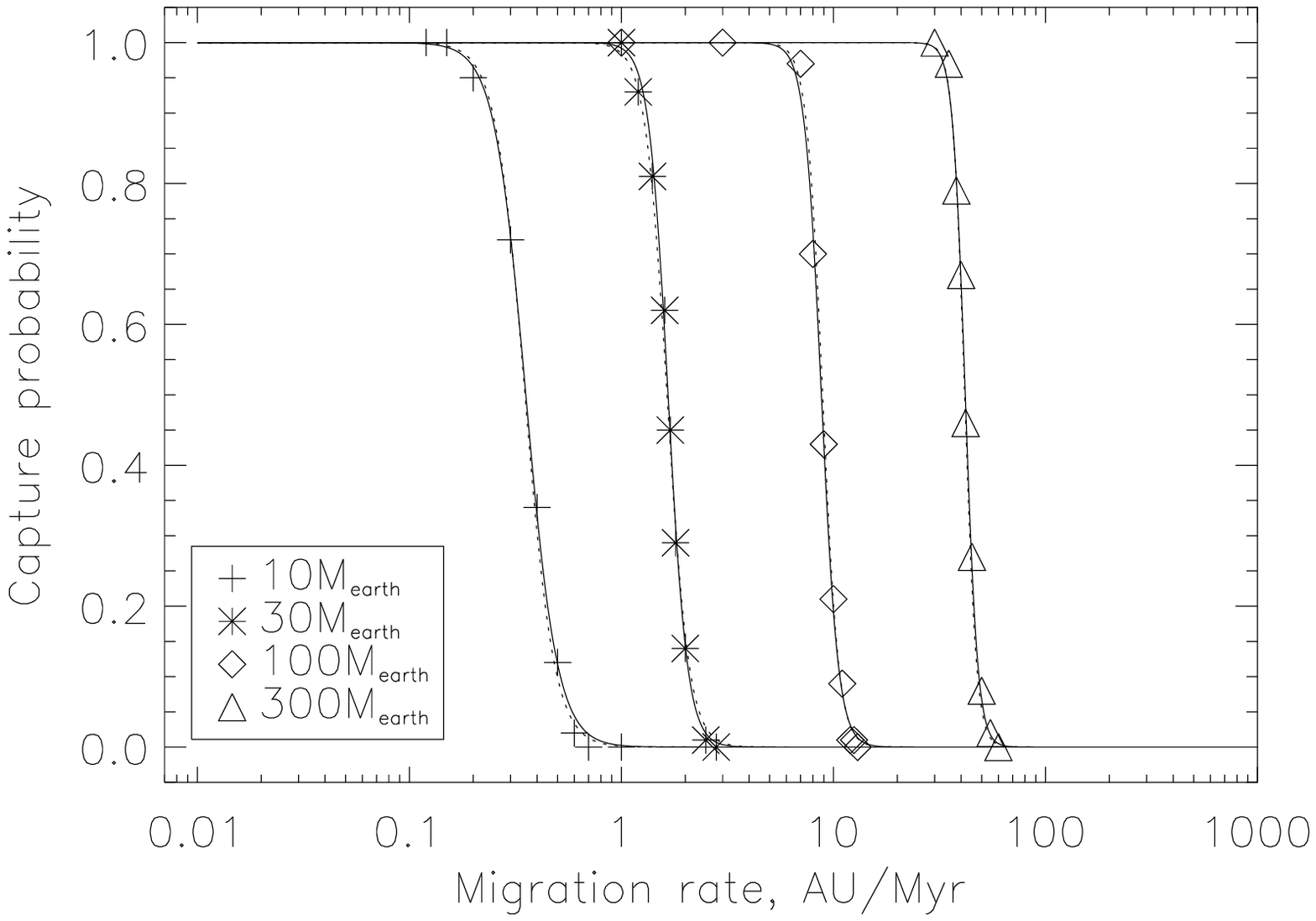} & \\[0.2in]
      \textbf{(b)} & \hspace{-0.3in} 
        \epsscale{0.92} 
\plotone{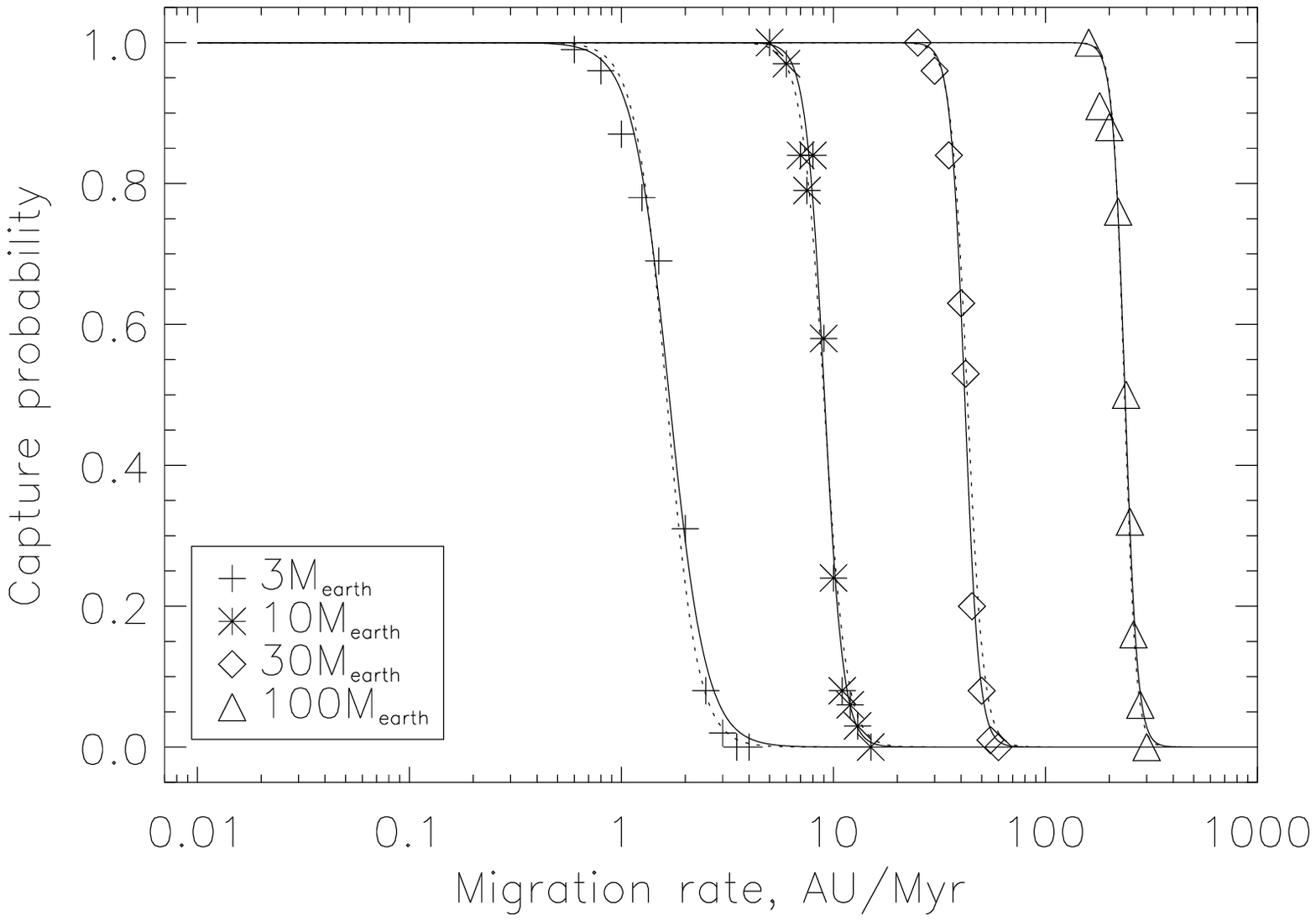} & \\[0.2in]
      \textbf{(c)} & \hspace{-0.3in}
        \epsscale{0.9} 
\plotone{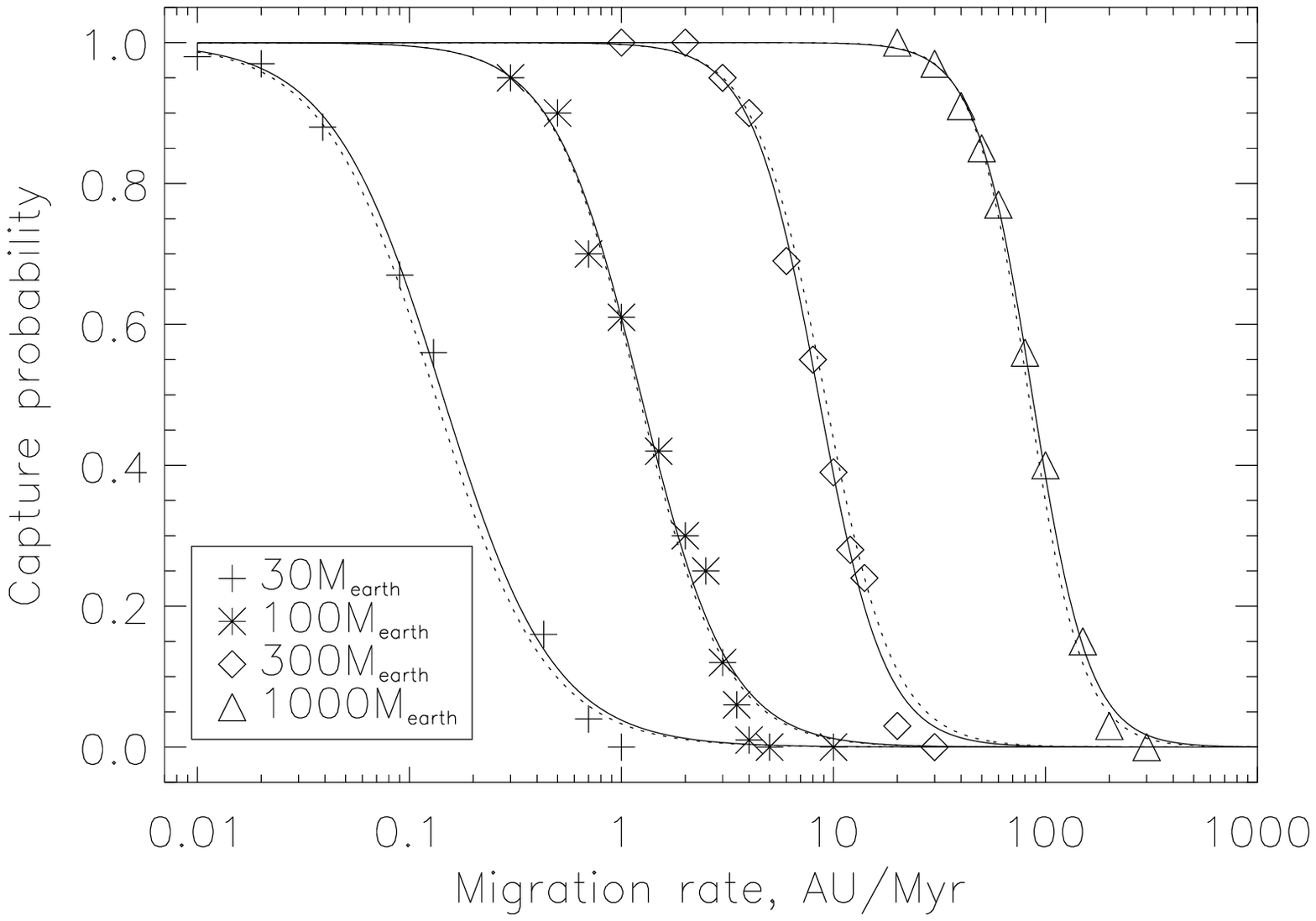} &
    \end{tabular}
  \end{center}
  \caption{Capture probabilities for planetesimals initially 30 AU
  from a $2.5M_\odot$ star for trapping into the
  \textbf{(a)} 2:1, \textbf{(b)} 4:3, and \textbf{(c)} 5:3 resonances.
  The solid lines show fits to these probabilities for each
  planet mass using equation (\ref{eq:p}).
  The dotted lines show the fit to the capture probabilities using
  equation (\ref{eq:p2}).
  \label{fig:123435mpl}}
\end{figure}

The same process was then repeated for the 2:1 resonance,
except that now that the scaling law is known, only the
equivalent runs of \S \ref{sss:23mpl} needed to be
performed.
The results of runs with $M_\star = 2.5M_\odot$ and
$a=30$ AU for different mass planets are shown in
Figure \ref{fig:123435mpl}a.
\footnote{While it might appear that integration times
could be reduced by achieving lower values of $\theta$ 
by varying $M_\star/a$ rather than reducing $\dot{a}_{pl}$,
this is not the case, since the integrator chooses a
stepsize that is $\propto \sqrt{a_{pl}^3/M_\star}$ and
integration length is $\propto \Delta a/\dot{a}_{pl}$,
where $\Delta a \propto a$ and
$\dot{a}_{pl} \propto \theta \sqrt{M_\star/a}$.}
Fits of the form given in equation (\ref{eq:p})
were performed for each planet mass and fits to the
variation of the derived parameters with $M_{pl}$
were also performed.
The resulting parameters were used to make initial
estimates of the equivalent parameters $X_{2:1}$, $Y_{2:1}$,
$u_{2:1}$, and $v_{2:1}$ in equation (\ref{eq:p2}).
A least squares fit to the capture probabilities
then found these parameters to be $X_{2:1}=5.8$,
$Y_{2:1}=4.3$, $u_{2:1}=1.40$, and $v_{2:1}=0.27$, with
respective errors estimated to be $\pm 0.2, 2.0, 0.01$, and
0.1.
This model for $P$ is shown on Figure \ref{fig:123435mpl}a
with dotted lines and is accurate to about $\pm 0.025$.

The results for different resonances are summarized in
Table \ref{tab:xyuv}.
It is immediately obvious that the functional form of the
capture probabilities for the two resonances are very similar
and differ significantly only in the parameter $X$, which
determines the strength of the resonance.
This numerical study finds that the 2:1 resonance is about
16 times weaker than the 3:2 resonance.
This is close to the factor of 12 predicted by autoresonance
theory (Friedland 2001).

\subsection{4:3 and 5:3 Resonances}
\label{ss:34}
The same analysis as for \S \ref{ss:12} was then repeated for
the 4:3 and 5:3 resonances using runs with planetesimals
orbiting initially 30 AU from a $2.5M_\odot$ star.
The results are plotted in Figures \ref{fig:123435mpl}b and
\ref{fig:123435mpl}c, and the parameters derived from these
results given in Table \ref{tab:xyuv}.
Again I find that for the 4:3 resonance, the values of $Y$, $u$,
and $v$ are similar to those of the other first order resonances
(i.e., those with $q=1$).
Also, $X_{4:3}$ is close to that anticipated from autoresonance
theory (i.e., that the 4:3 resonance is $\sim 1.6$ times stronger
than the 3:2 resonance).
The values of $Y$, $u$ and $v$ for the second order ($q=2$) 5:3
resonance are somewhat different to those of the first order resonances;
there is a steeper dependence of the strength of the resonance
on the planet mass, and the range of migration rates for intermediate
trapping probabilities is greater for a given planet mass.
In general second order resonances are expected to be weaker than
first order resonances, which is borne out by the higher value
of $X$.

\subsection{Trapping Probability Summary}
\label{ss:tpsumm}
\begin{figure}
  \begin{center}
    \begin{tabular}{c}
        \epsscale{0.9} 
\plotone{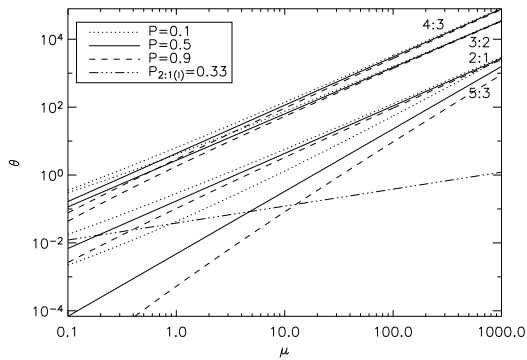}
    \end{tabular}
  \end{center}  
  \caption{Summary of capture probabilities for the 4:3, 3:2, 2:1,
  and 5:3 resonances for migrations defined by the parameters $\mu$
  and $\theta$ (eqs. [\ref{eq:mu}]-[\ref{eq:p2}] with parameters from
  Table \ref{tab:xyuv}).
  The dotted, solid, and dashed lines indicate migrations for which
  trapping probabilities are 10, 50, and 90\% respectively.
  The dash-triple dot line indicates migrations for which the 2:1(u)
  resonance has twice as many members as the 2:1(l) resonance (see \S 
  \ref{sss:tp21l}).
  \label{fig:tpsumm}}
\end{figure}

The trapping probabilities derived in the last subsections
are summarized in Figure \ref{fig:tpsumm}.
This shows, for the four resonances that were studied, the lines
of 10, 50, and 90\% trapping probability in terms of the migration
parameters $\theta$ and $\mu$.
For planetesimals at a given semimajor axis any given migration
is defined by a single point on this Figure.
Thus this Figure quickly summarizes which
resonances these planetesimals will end up in (if any).

\section{Distribution of Resonant Planetesimals}
\label{s:drp}
Planetesimals that are trapped in resonance are not
evenly distributed around star, rather they congregate at
specific longitudes relative to the perturbing planet.
This can be understood by first considering the geometry of
a generic resonant orbit (\S \ref{ss:rg}), and then considering
the action of resonant forces (\S \ref{ss:sr}).
I use these in \S \ref{ss:numdrp} to build up a model
for resonant structure caused by planet migration, the
parameters of which are determined from analysis of the
numerical integrations described in \S \ref{s:rcp}.

\subsection{Resonant Geometry}
\label{ss:rg}
\begin{figure}
  \begin{center}
    \begin{tabular}{rcccc}
     & \textbf{$e=0.1$} & \textbf{$e=0.2$} & \textbf{$e=0.3$} & \\[0.2in]
      \textbf{2:1} & \hspace{-0.3in}
        \epsscale{0.28} 
\plotone{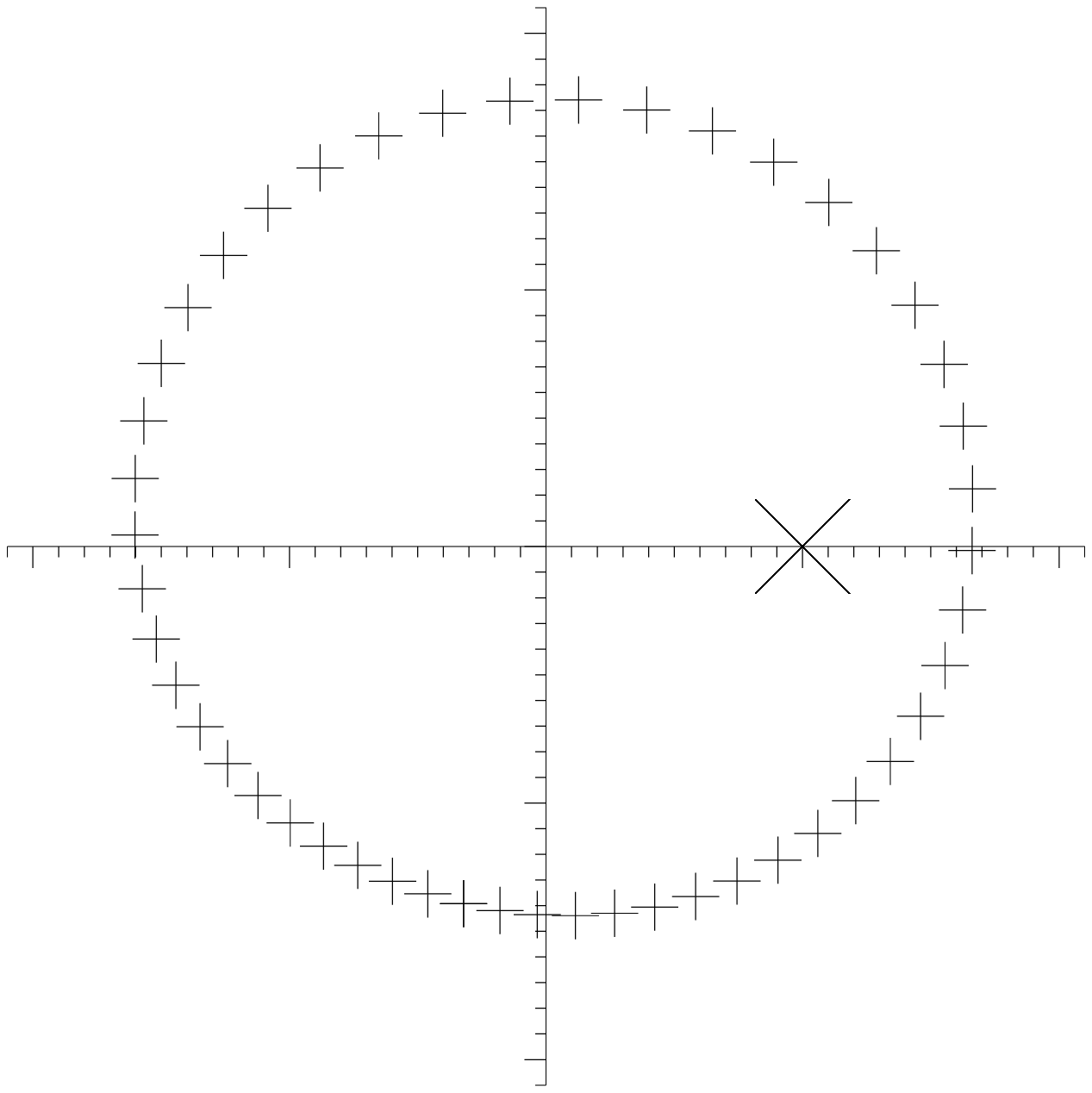} &
        \epsscale{0.28} 
\plotone{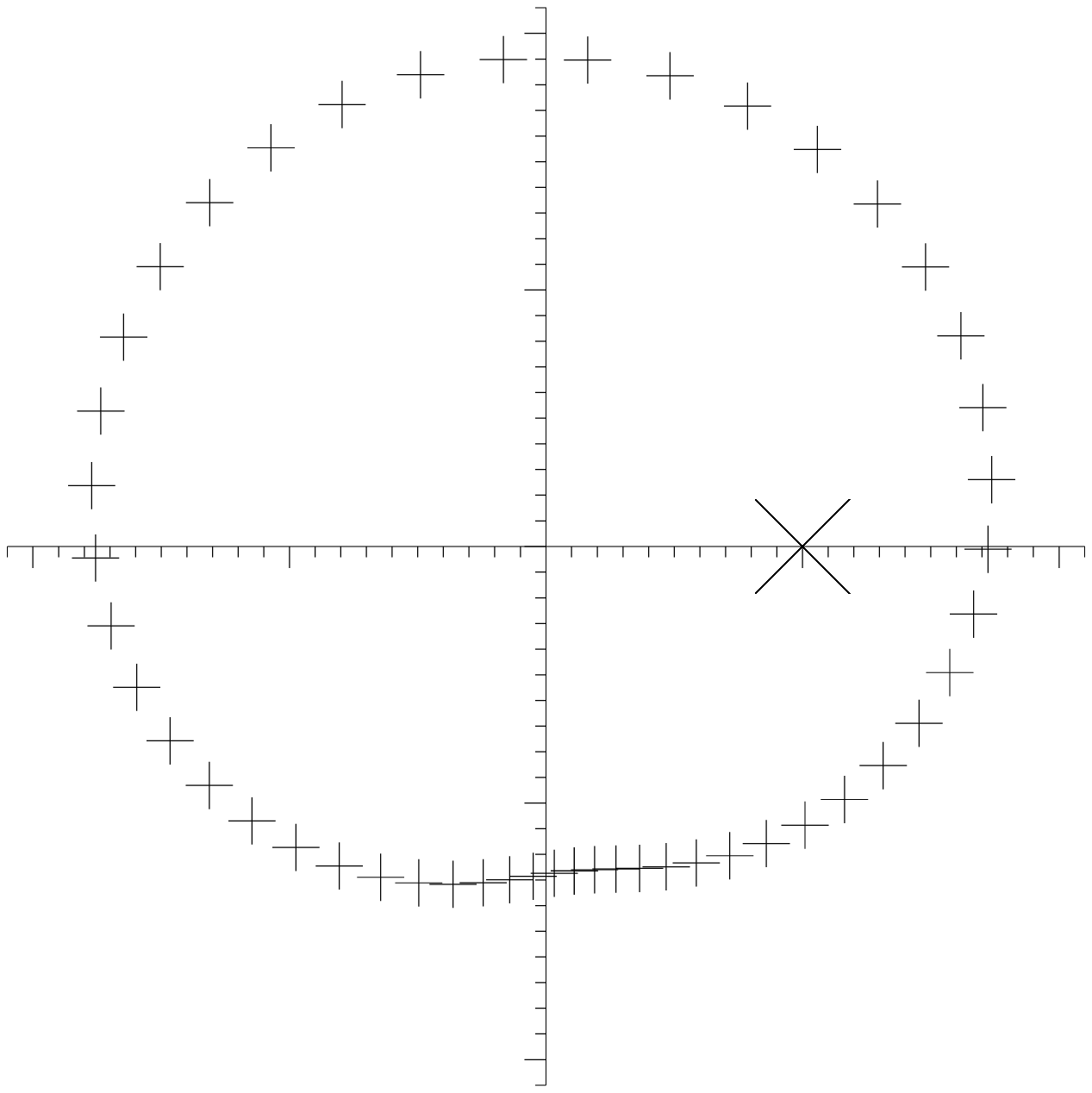} &
        \epsscale{0.28} 
\plotone{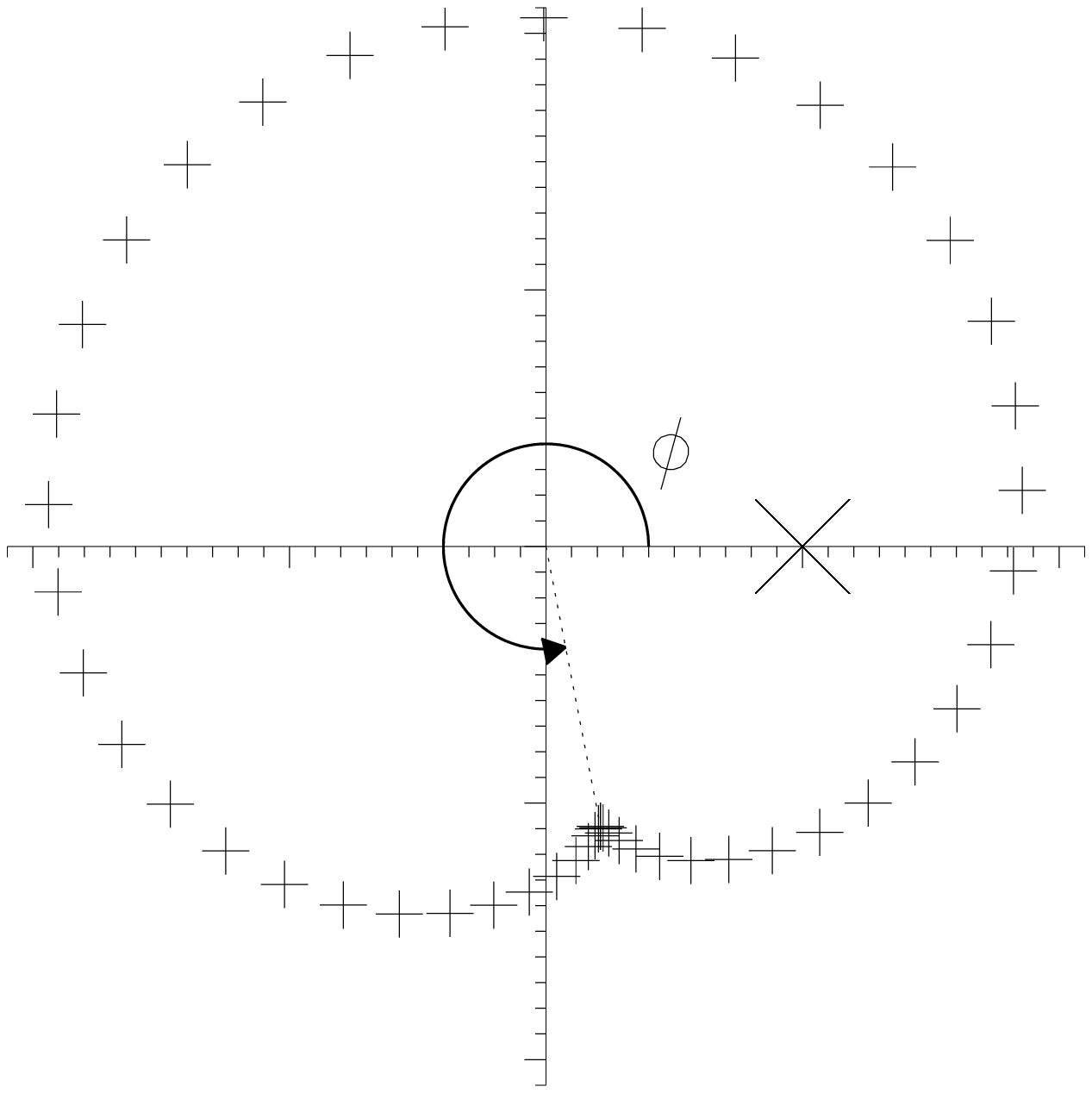} & \\[0.0in]
      \textbf{5:3} & \hspace{-0.3in}
        \epsscale{0.28} 
\plotone{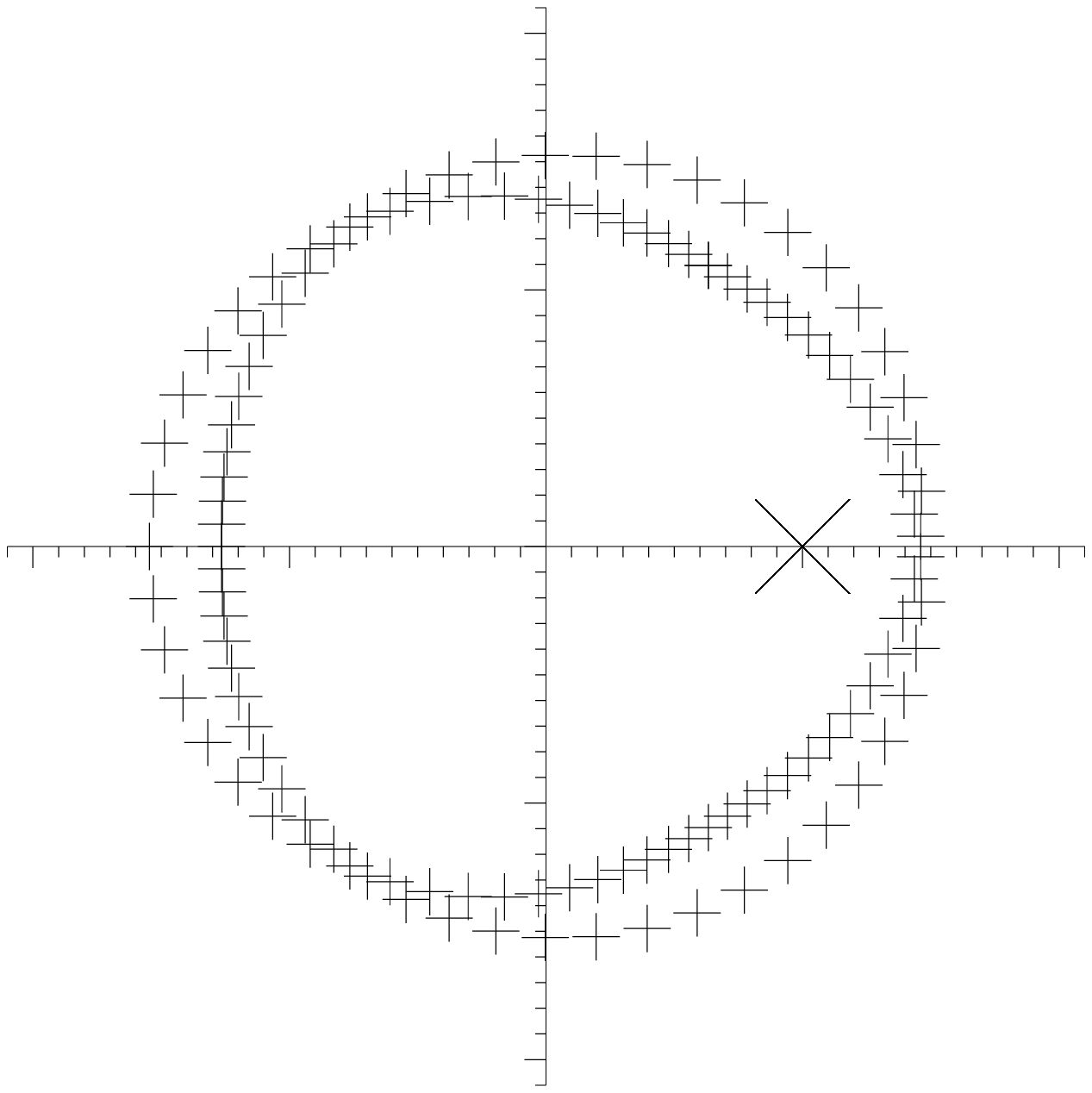} &
        \epsscale{0.28} 
\plotone{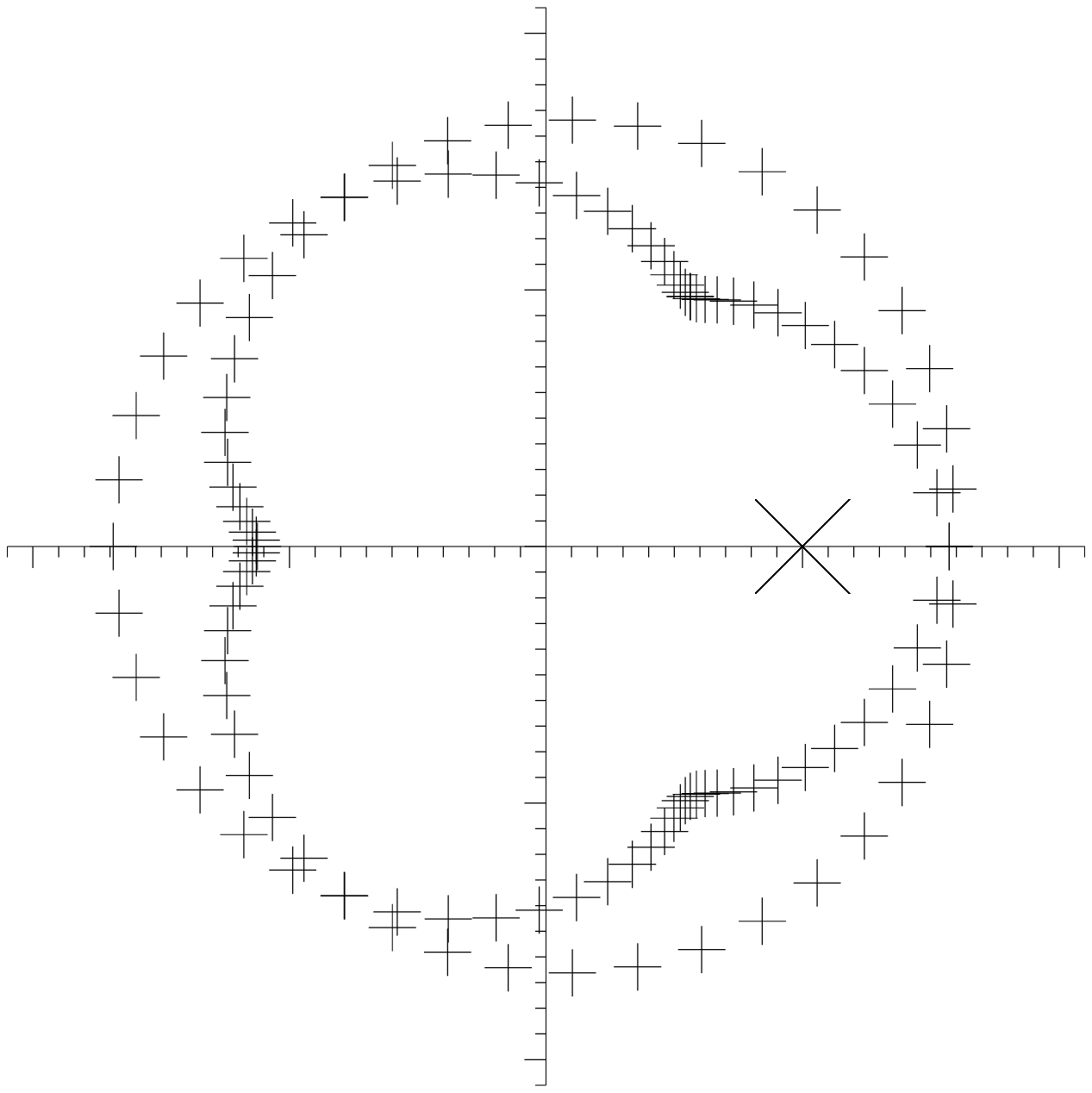} &
        \epsscale{0.28} 
\plotone{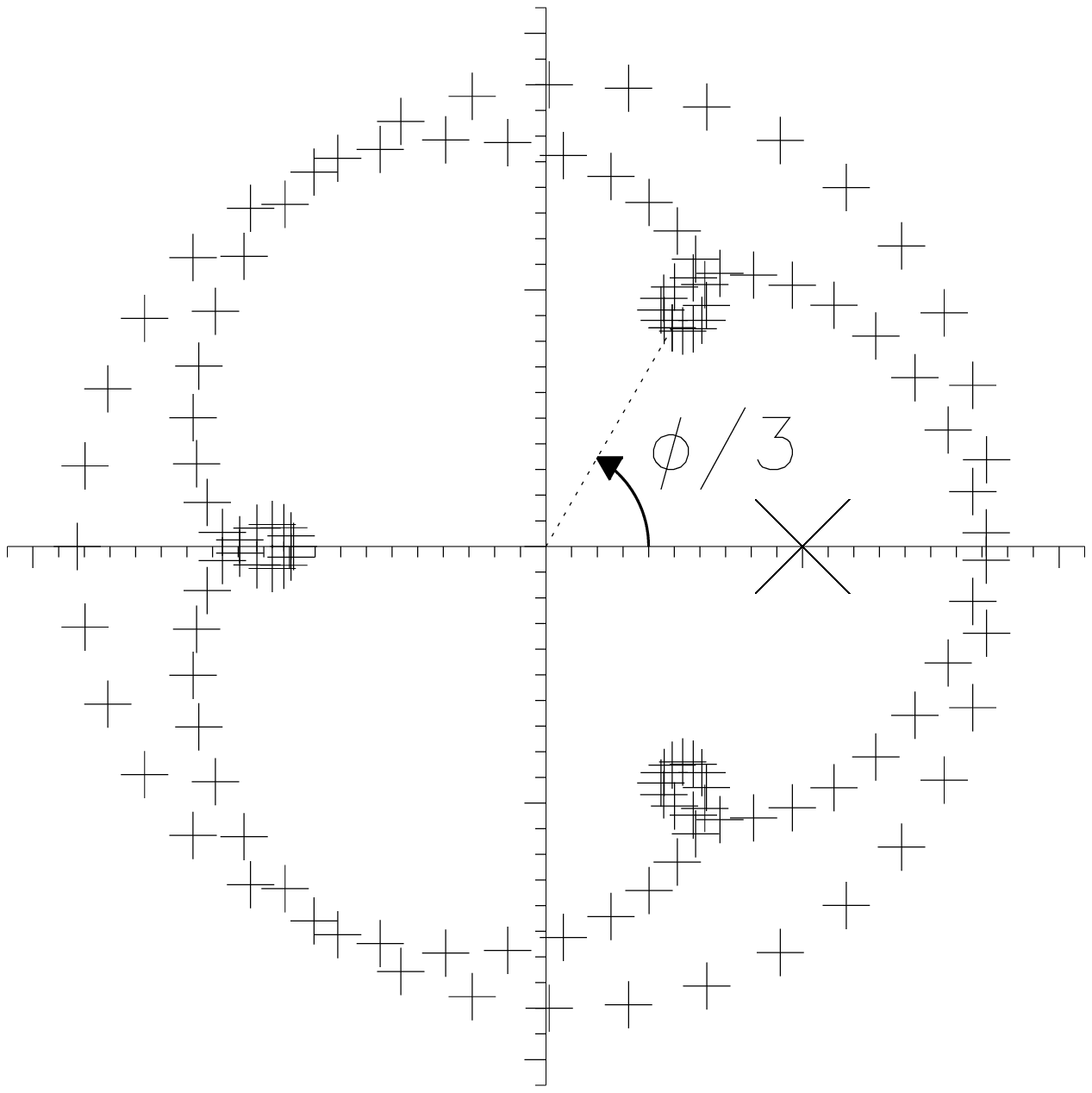} & \\[0.0in]
      \textbf{3:2} & \hspace{-0.3in}
        \epsscale{0.28} 
\plotone{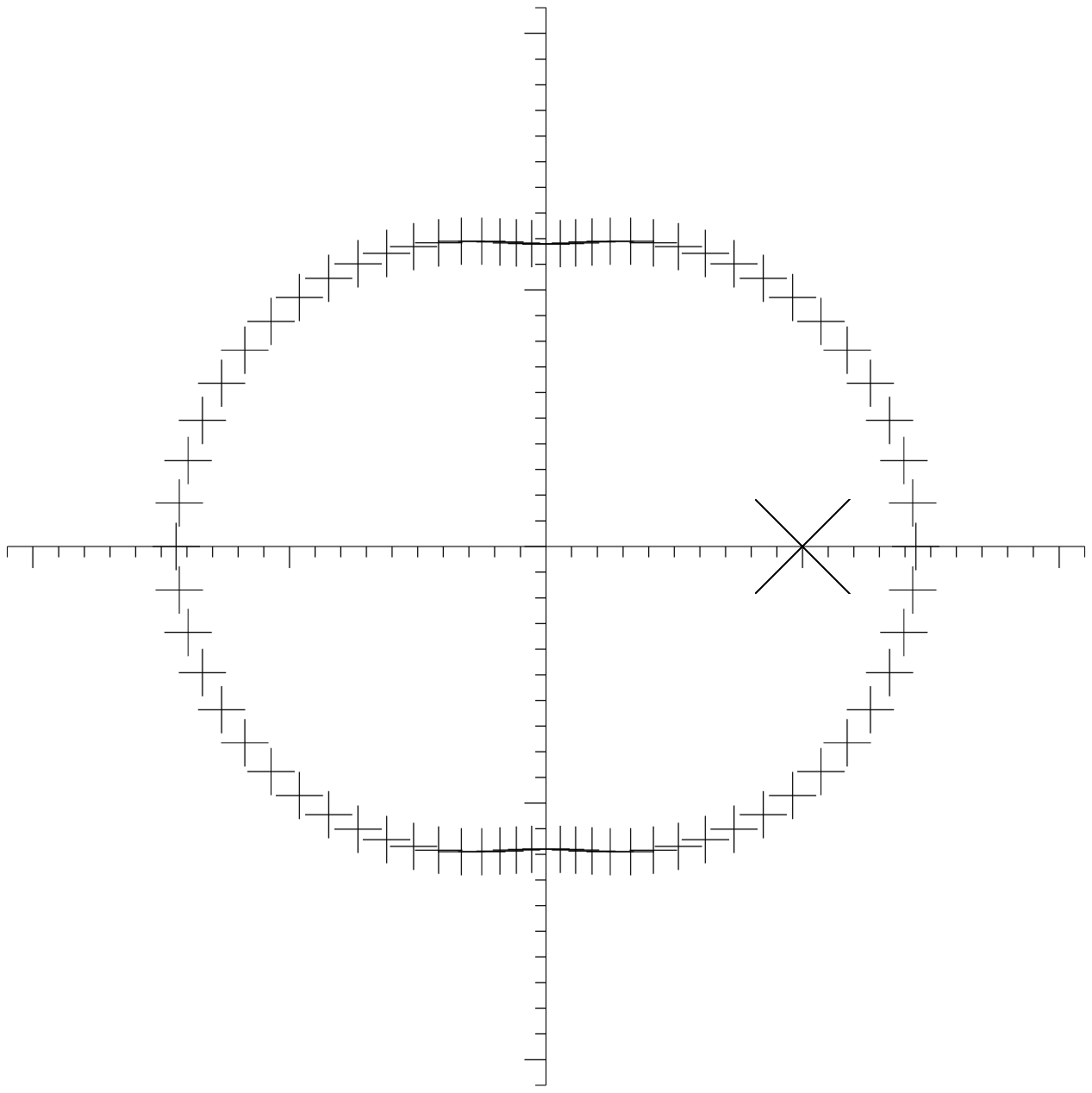} &
        \epsscale{0.28} 
\plotone{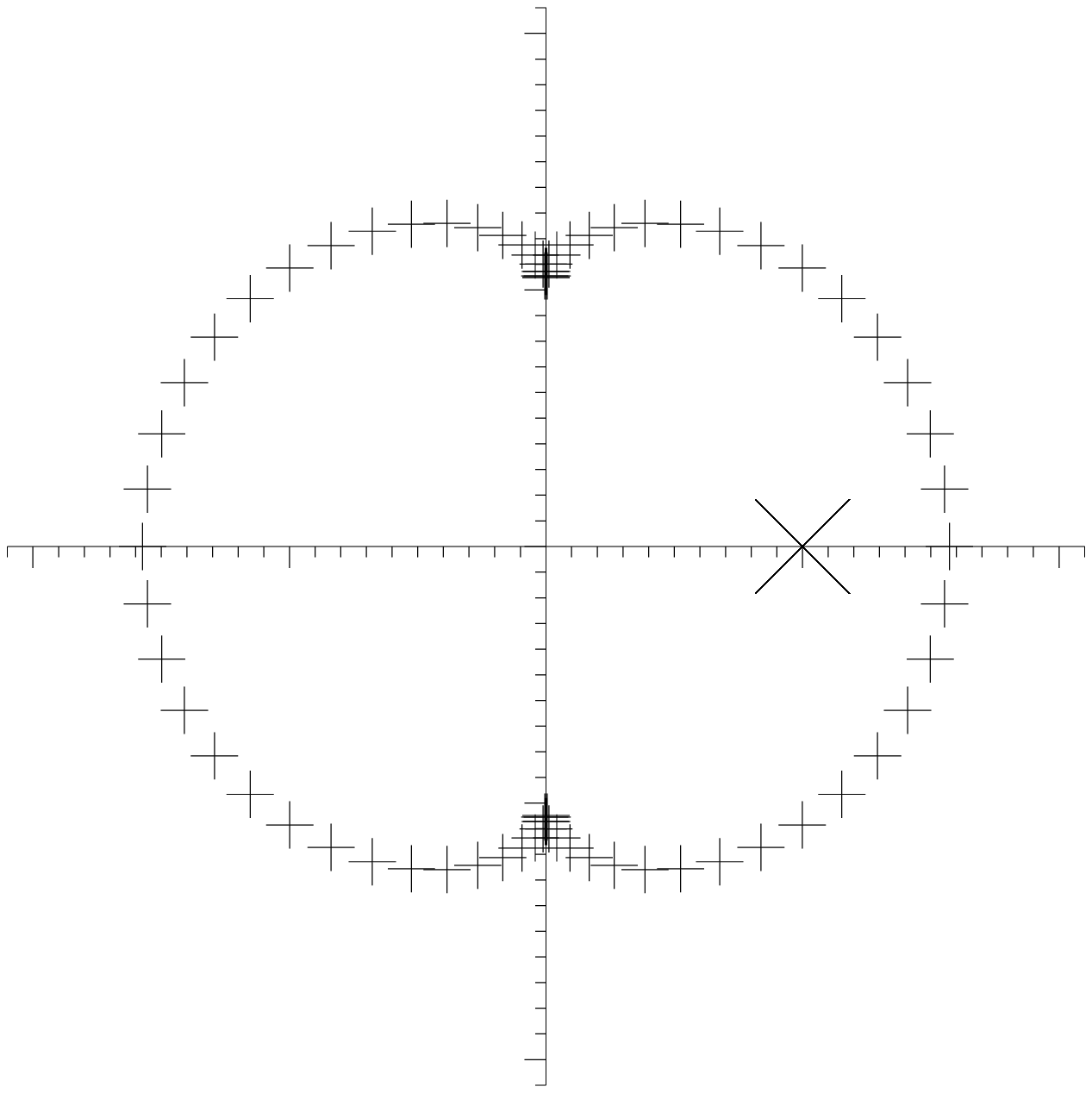} &
        \epsscale{0.28} 
\plotone{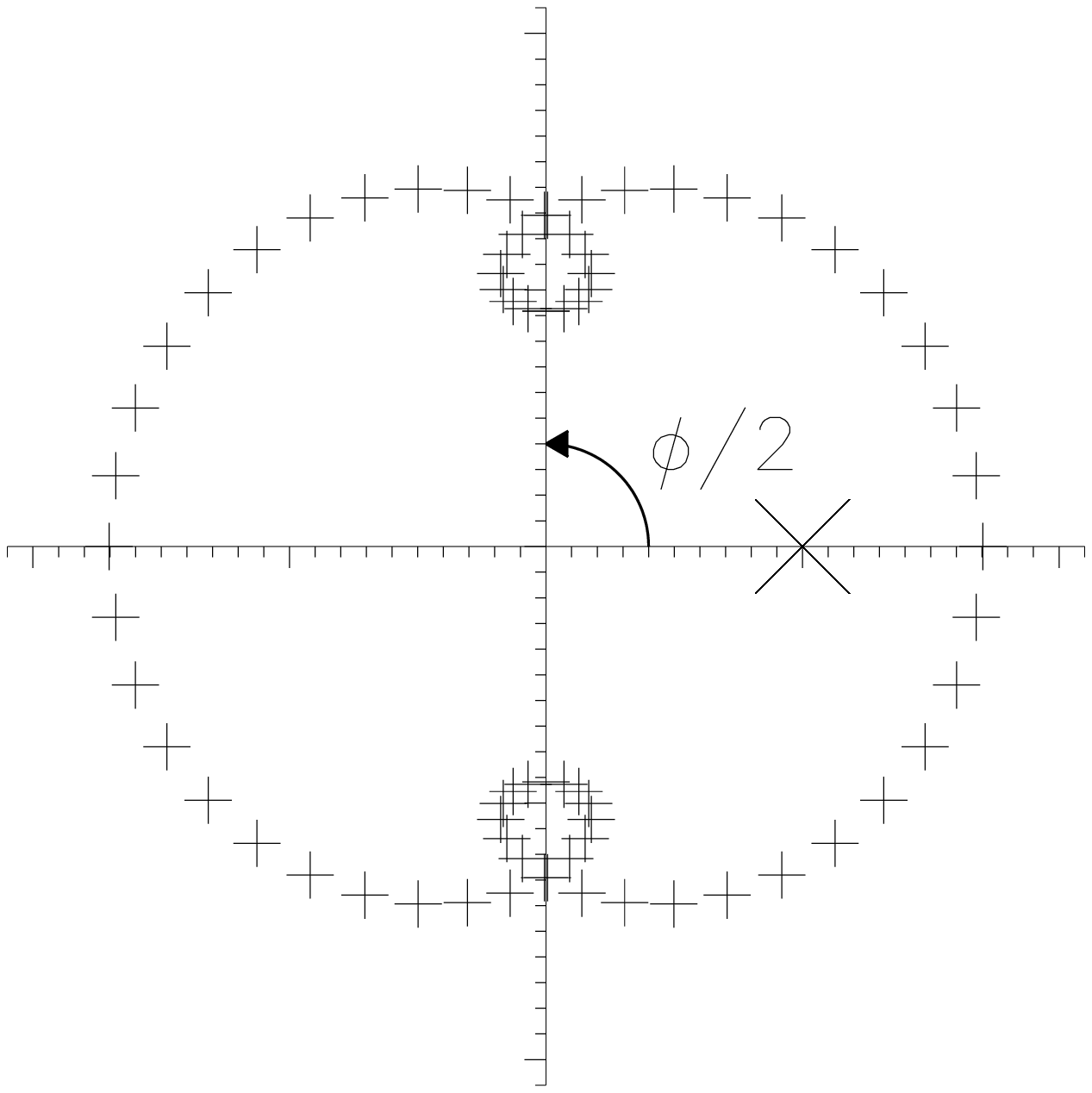} & \\[0.0in]
      \textbf{4:3} & \hspace{-0.3in}
        \epsscale{0.28} 
\plotone{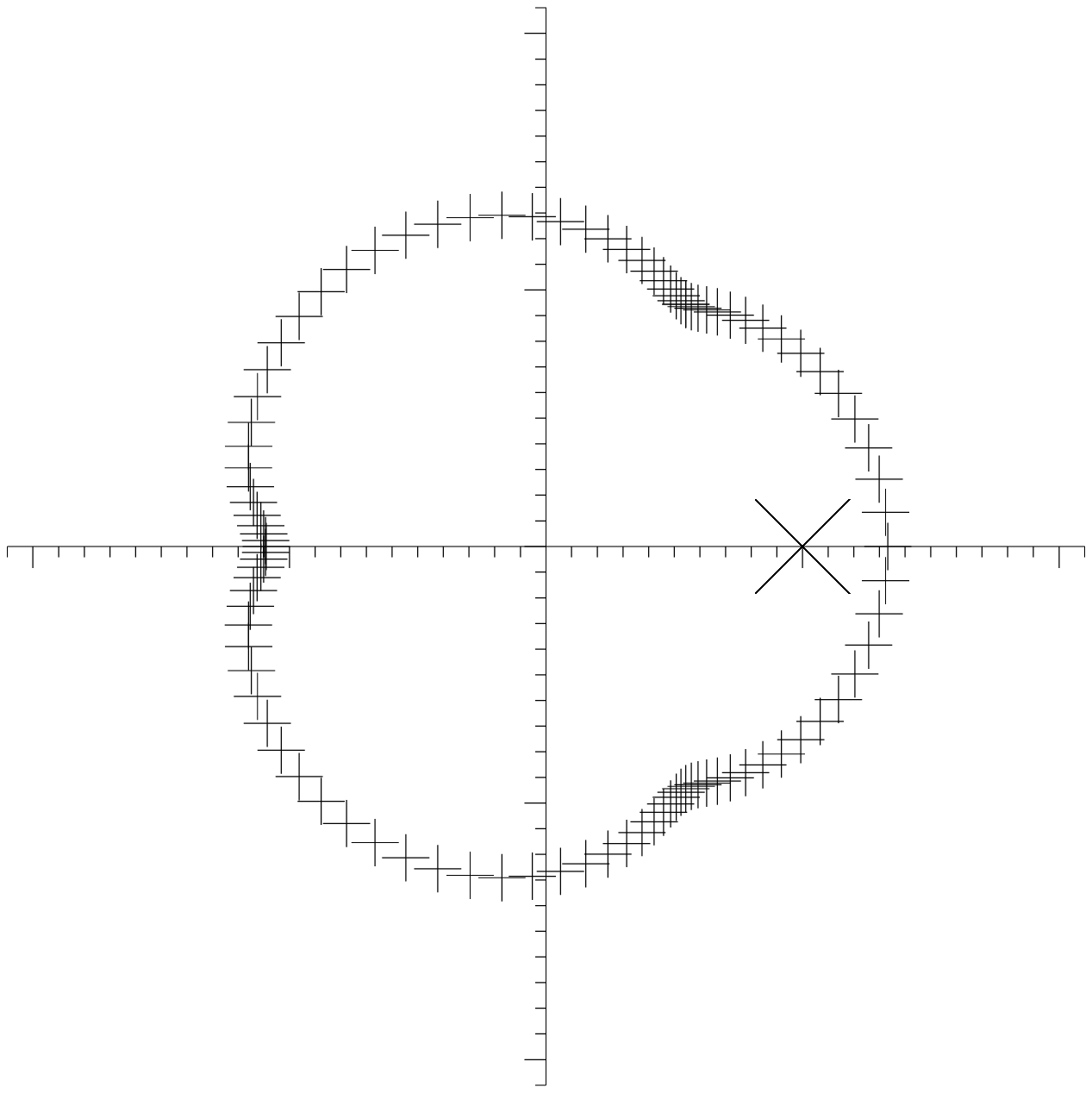} &
        \epsscale{0.28} 
\plotone{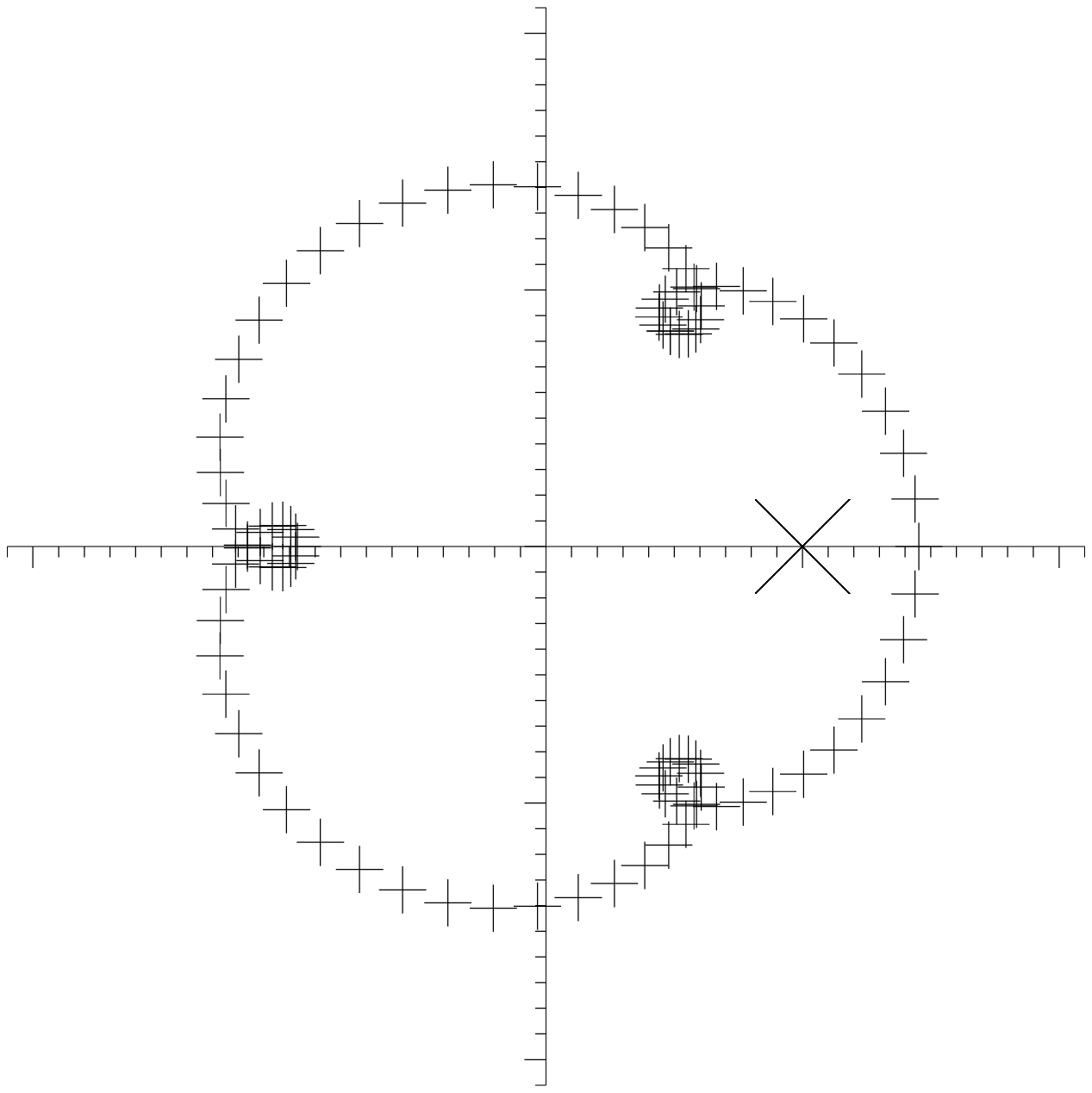} &
        \epsscale{0.28} 
\plotone{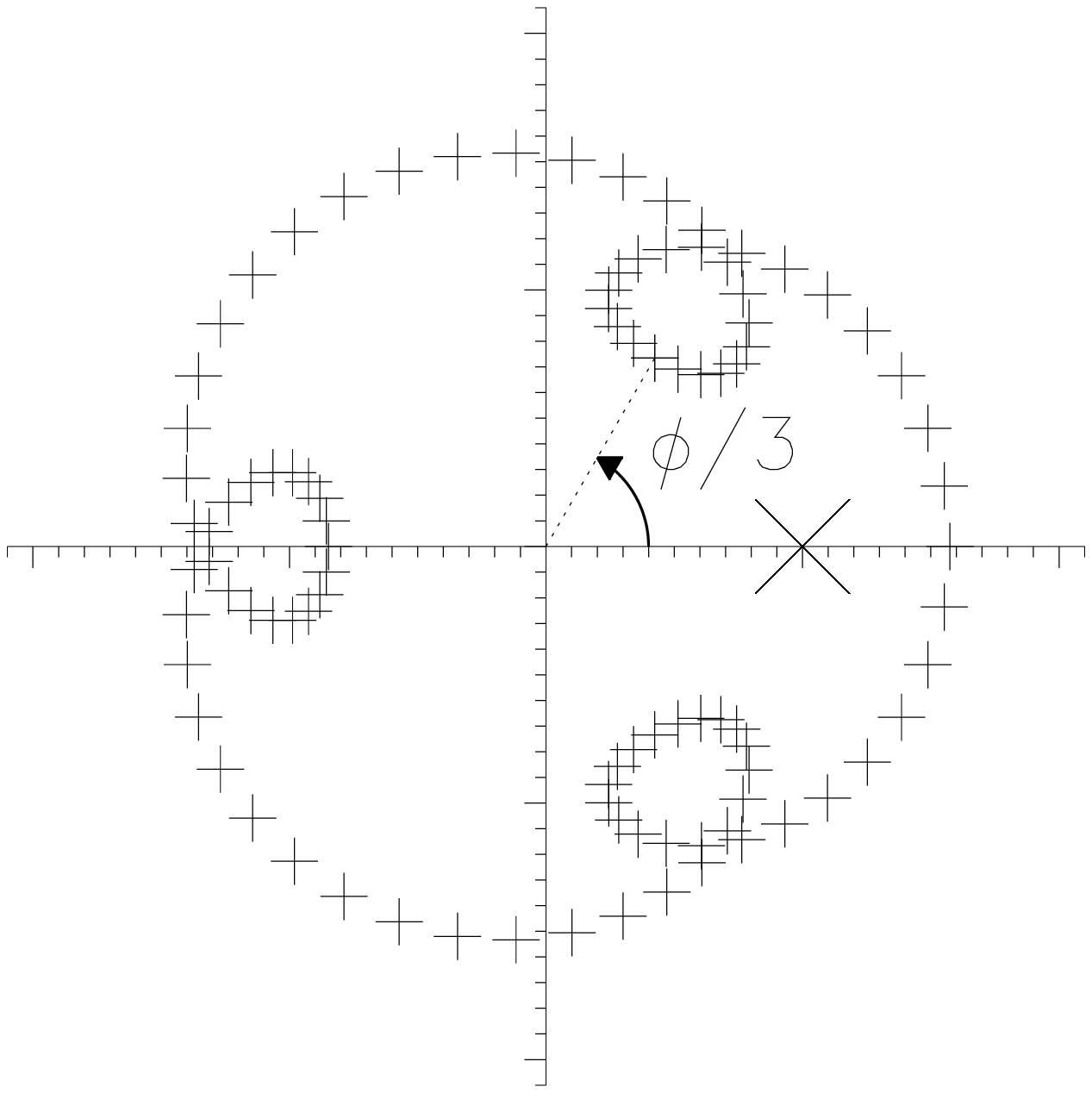} & 
    \end{tabular}
  \end{center}
\begin{minipage}{80mm}
  \caption{Paths of resonant orbits in the frame co-rotating with
  the mean motion of the planet.
  The planet is marked by a cross in these figures and it
  is stationary in this reference frame, since its orbit is circular.
  Crosses show the location of a planetesimal in the different
  resonances (2:1, 5:3, 3:2, and 4:3) at intervals of 1/24th of the
  planet's orbital period (i.e., when the planet has moved by
  $15^\circ$ in the inertial frame).
  The planet moves anticlockwise in the inertial frame, and the
  planetesimals move anticlockwise in this reference frame.
  Sufficient crosses are marked to show how this pattern repeats
  itself, which is every $p+q$ orbits of the planet.
  The three plots for each resonance show the paths for
  planetesimal eccentricities of 0.1, 0.2, and 0.3.
  Naturally, these plots show just one value of the
  planetesimal's pericenter, which always occurs at the same
  location in inertial space, but occurs $p$ times in the
  rotating frame plots (at the innermost point of the loops).
  Resonant orbits with different pericenters, which can be
  specified by the resonant angle $\phi$ (eq. [\ref{eq:phi}]),
  would show the same pattern, but rotated on this figure
  by an angle $\phi/p$.
  The three plots for the 2:1 resonance have
  $\phi=257$, 275, and $281^\circ$ (for increasing eccentricity),
  while the remainder have $\phi=180^\circ$ (see \S \ref{sss:nlc}).
  \label{fig:res}}
\end{minipage}
\end{figure}

Each of the panels in Figure \ref{fig:res} shows the path
that a planetesimal orbiting at a semimajor axis
corresponding to a planet's $p+q:p$ resonance (eq.~[\ref{eq:apq}])
takes when plotted in the frame co-rotating
with the mean motion of the planet.
While these orbits are elliptical in the inertial frame,
there are two obvious features in the rotating frame:
(i) when the planetesimal reaches pericenter
it must be at one of $p$ specific longitudes relative to
the planet;
(ii) the planetesimal spends more time at relative
longitudes close to those at which it is at pericenter,
than to those at which it is at apocenter.
The former occurs because by definition, the planetesimal's
pericenter passages occur at longitudes relative to the planet
incremented by $q/p \times 360^\circ$ from the previous
passage, a pattern which repeats itself after $p$ of the
planetesimal's orbits.
The latter occurs because when the planetesimal is close
to its pericenter, the rate of change of its longitude is
more closely matched to that of the planet.
\footnote{In fact, even though the planetesimal orbits at a larger
semimajor axis, its longitude can change faster than that
of the planet if the planetesimal's eccentricity is
high enough.
This results in backward motion in the rotating frame
and causes the loops seen in Figure \ref{fig:res} for high
eccentricity planetesimal orbits.}

While the pattern shown in the panels in Figure \ref{fig:res}
is unique, its orientation relative to the planet is not,
since this is determined by the starting longitudes of the
planet and planetesimal and the pericenter direction.
This orientation can be defined using just one variable,
the planetesimal's resonant argument:
\begin{equation}
  \phi = (p+q)\lambda_r - p\lambda_{pl} - q\bar{\omega}_r.
  \label{eq:phi}
\end{equation}
By definition, $\phi$ remains constant while a planetesimal is
in resonance, since the increase due to $\lambda_r$ is offset by 
the decrease due to $\lambda_{pl}$.
The resonant argument has a specific physical meaning which
will be described in \S \ref{ss:sr}, but from a geometrical
point of view it can be rewritten thus:
\begin{equation}
  \phi = p[\bar{\omega}_r - \lambda_{pl}(t_{peri})],
  \label{eq:phip}
\end{equation}
where $\lambda_{pl}(t_{peri})$ is the longitude of the
planet when the planetesimal is at pericenter.
In other words, $\phi/p$ defines the relative longitude
when the planetesimal is at pericenter and so the orientation
of the pattern.

\subsection{Resonant Forces}
\label{ss:sr}
The influence of a planet's gravity is to perturb
the orbits of planetesimals in the system.
These perturbations can be written down as a sum of
many terms described by the planetesimal's
disturbing function, $R$ (Murray \& Dermott 1999).
Analysis of the disturbing function shows that
these perturbations are particularly strong at the
geometrical resonance locations (eq.~[\ref{eq:apq}]),
where terms in which $\lambda_r$ and $\lambda_{pl}$ are
combined in the form $(p+q)\lambda_r - p\lambda_{pl}$ are
amplified.

The resonances I will be discussing all involve terms
including the angle $\phi$ defined in equation (\ref{eq:phi}).
The relevant terms in the disturbing function, assuming
the orbits of the planet and planetesimal are coplanar and
that the planet has a circular orbit (i.e., the
circular restricted three body problem), are:
\begin{equation}
  R = \left(\frac{GM_{pl}}{a}\right)\left[f_{s,1}e^2 +
       e^q
\left(f_d(\alpha)+\frac{f_i(\alpha)}{\alpha}\right)
\cos{\phi}\right],
\end{equation}
where $\alpha=a_{pl}/a$, and $f_{s,1}$, $f_d$, and $f_i$ are
coefficients corresponding to the secular, resonant direct, and
resonant indirect parts of the disturbing function respectively.
The effect of these perturbations on the orbital elements
of the planetesimal can be determined using
Lagrange's planetary equations (e.g., Murray \& Dermott 1999).
In particular, the variation in the planetesimal's
semimajor axis and eccentricity are:
\begin{eqnarray}
  \dot{a} & = & -2(p+q) \left(\frac{M_{pl}}{M_\star}\right)
                    \sqrt{\frac{GM_\star}{a}}e^q \nonumber \\
          &   & \times \left( f_d(\alpha)+\frac{f_i(\alpha)}{\alpha} \right)
                 \sin{\phi}, \label{eq:adotres} \\
  \dot{e} & = & -q \left(\frac{M_{pl}}{M_\star}\right)
                    \sqrt{\frac{GM_\star}{a^3}}e^{q-1} \nonumber \\
          &   & \times \left( f_d(\alpha)+\frac{f_i(\alpha)}{\alpha} \right)
                 \sin{\phi}. \label{eq:edotres}
\end{eqnarray}

A simple analysis of the consequence of resonant
forces using the circular restricted three body
problem shows that they make the resonant argument of
a planetesimal librate about a fixed value (i.e., a
sinusoidal oscillation; Murray \& Dermott 1999):
\begin{equation}
  \phi = \phi_m + \Delta \phi \sin{2\pi t/t_\phi}. \label{eq:phievol}
\end{equation}
Indeed, a planetesimal is defined to
be in resonance if its resonant argument is
librating rather than circulating (i.e., a monotonic increase
or decrease).
It is the fact that the resonant arguments of all planetesimals
in the same resonance librate about the same value that causes
their azimuthal distribution to be asymmetric.

\subsubsection{Libration Center without Migration}
\label{sss:lc}
The angle about which $\phi$ librates, $\phi_m$, can be understood by
considering the geometry of resonance (Peale 1976; Murray
\& Dermott 1999).
Most of the perturbations to a planetesimal's orbit
occur at conjunction (i.e., when the planet and the planetesimal
are at the same longitude).
Conjunctions which occur either at pericenter or
apocenter confer no angular momentum to the planetesimal.
However, conjunctions that occur before/after the planetesimal's
apocenter (and after/before the pericenter) lead to a net
decrease/increase in the planetesimal's angular momentum.
This means that resonant forces push the conjunction towards
apocenter.
Since the resonant angle can also be written thus:
\begin{equation}
  \phi = q(\lambda_c - \bar{\omega}_r),
\end{equation}
where $\lambda_c$ is the longitude at which the planet and
the planetesimal have a conjunction, apocentric libration
is one at which $\phi_m/q=180^\circ$.

The argument outlined above is not valid however for all
resonances, just those with $q=1$ and $p \ne 1$.
For a start, while conjunctions always occur at the same location
in the orbit of the planetesimal for a $q=1$ resonance,
conjunctions for $q=2$ resonances occur at two locations
$180^\circ$ apart around the planetesimal's orbit.
Apocentric conjunctions would thus have to be followed by
pericentric conjunctions.
Since the forces of the pericentric conjunction 
would be stronger than those at apocenter, such libration
is not stable.
Rather conjunctions for stable libration occur midway between
pericenter and apocenter whence the forces from alternate
conjunctions cancel, i.e., $\phi_m/2=90^\circ$.

Also, while the resonant argument for the 2:1 resonance librates
about $180^\circ$ for low eccentricity ($e<0.04$) orbits, the
libration center can take one of two values for higher
eccentricity orbits, one with $\phi_m>180^\circ$ the other with
$\phi_m<180^\circ$ (so-called asymmetric libration;
Beaug\'{e} 1994; Malhotra 1996; Chiang \& Jordan 2002).
The physical explanation for this behavior is that
perturbations to the orbits also occur when the
planetesimal is at its pericenter, as well as when it is at
conjunction.
Consider a 2:1 planetesimal which has a conjunction with
the planet just before/after its apocenter.
As already mentioned, the forces from the conjunction
act to decrease/increase the planetesimal's angular momentum.
Now consider the forces acting on the planetesimal as it continues
along the rest of its orbit.
At first these forces act to increase its angular momentum, since
the planet is in front of the planetesimal, but when
the planet's longitude is $180-360^\circ$ in front of the 
planetesimal these forces decrease the planetesimal's angular
momentum.
Since the planetesimal spends more time at longitudes relative
to the planet close to its pericenter (see Figure \ref{fig:res}),
these forces do not exactly cancel around the orbit
and there is a net increase/decrease in the planetesimal's
angular momentum (for conjunctions before/after apocenter).
These forces are important for the $p=1$ resonances
for which every pericenter passage occurs at the same longitude
relative to the planet.

Stable libration for the 2:1 resonance occurs at a value
of $\phi$ which is $<180^\circ / >180^\circ$ for which the
angular momentum loss/gain at conjunction is balanced by its
gain/loss at pericenter.
Including terms up to $O(e^2)$ in the expansion
of the disturbing function for the circular restricted three
body problem, the appropriate semimajor
axis evolution at $\alpha \approx 2^{-2/3}$ is given by:
\begin{eqnarray}
  \dot{a} & = & -2\left(\frac{M_{pl}}{M_\star}\right)
    \sqrt{\frac{GM_\star}{a}}[
    3.38e\sin{\phi} \nonumber \\
          &   & + 14.38e^2\sin{2\phi} - 2.52e\sin{\phi}
    ].
\end{eqnarray}
In this expression, the first two terms are the direct
terms of the $p=q=1$ and $p=q=2$ resonances respectively,
and the third term is the indirect term of the
$p=q=1$ resonance.
The physical interpretation of these terms is that the
first two are from perturbations caused at conjunction,
while the third term is from perturbations at pericenter.
Setting $\dot{a} = 0$ gives the result that
\begin{equation}
  \cos{\phi_m} = -0.0298/e. \label{eq:phim21an}
\end{equation}
A similar result was found by Beaug\'{e} (1994) by considering
the phase space of the 2:1 resonance with a Hamiltonian
model including harmonics up to second order.

\subsubsection{Libration Center with Migration}
\label{sss:lcwm}
The discussion of the stable libration center in \S \ref{sss:lc} was
based on the assumption that the resonant forces confer no angular
momentum to the planetesimal.
This is not the case when the planet is migrating, since the 
planetesimal must be migrating out with the planet to maintain
the resonance:
\begin{equation}
  \dot{a}_{mig} = \dot{a}_{pl} \left( \frac{p+q}{p} \right)^{2/3},
  \label{eq:adotmig}
\end{equation}
and so its angular momentum must be increasing.
Given the discussion in \S \ref{sss:lc}, stable libration should
thus be offset to slightly higher values of $\phi$ in the
presence of migration.
Setting equation (\ref{eq:adotmig}) equal to the average rate of
change of $a$ due to resonant forces in the circular restricted
three body problem (i.e., eq. [\ref{eq:adotres}]
with $\phi$ replaced with $\phi_m$) implies that
\begin{equation}
  \sin{\phi_m} \propto -(\theta/\mu)e^{-q}.
  \label{eq:sphim2}
\end{equation}

\subsubsection{Eccentricity Evolution}
\label{sss:ei}
Another consequence of the planet's perturbing forces is to
link the evolution of a planetesimal's eccentricity to
that of its semimajor axis.
This means that the same resonant forces which
cause a planetesimal's semimajor axis to increase
while the planet's orbit is migrating out,
also cause the planetesimal's eccentricity to
increase.
The eccentricity of a planetesimal which has migrated
from an orbit with an eccentricity $e_0$ at a semimajor axis
$a_0$ to one at a semimajor axis $a$ due trapping in a
$p+q:p$ resonance can be estimated from the
circular restricted three body problem.
Using equations (\ref{eq:adotres}) and (\ref{eq:edotres}),
it can be shown that the change in $a$ and $e$ due to
resonant forces are correlated by the relationship
$da/de=2(1+p/q)ae$.
Thus the planetesimal's eccentricity
can be estimated to be:
\begin{equation}
  e = \sqrt{e_0^2+\frac{q}{p+q}\ln{a/a_0}}.
  \label{eq:eres}
\end{equation}

\subsection{Numerical Results}
\label{ss:numdrp}
Based on the discussion of \S\S \ref{ss:rg} and \ref{ss:sr},
it is easy to see that the spatial distribution of a population
of planetesimals that have been trapped in resonance by a migrating
planet can be defined by the distribution of their orbital
parameters (see also \S \ref{ss:tm}).
Orbital distributions sufficient to do this are derived in this
section for the planetesimals in the migration simulations
presented in \S \ref{s:rcp}.

First the output of these simulations were used to determine the
distributions of the longitudes of the resonant planetesimals
as well as the evolution of their eccentricities
while in resonance.
Then additional runs were performed to determine the libration
parameters of the resonant planetesimals.
These used the output of the original simulations (i.e., the
instantaneous orbital parameters of the planetesimals and
planet) as their starting point and
continued the evolution for just a few libration periods
(see Appendix \ref{app:adiabatic}) sampled
at 500 timesteps.
The libration parameters $\phi_m$ and $\Delta \phi$
for each planetesimal in the run that was trapped
in resonance were then measured by a fit
to the evolution of its appropriate resonant
argument using equation (\ref{eq:phievol}).

In order to determine the mean parameters for the ensemble
of planetesimals that were trapped in resonance, a
histogram of these parameters was displayed.
The mean libration center of these planetesimals,
$\langle \phi_m \rangle$, was determined, and
a gaussian fit performed to the distribution of libration
amplitudes.
This resulted in the best fit mean libration
amplitude, $\langle \Delta \phi \rangle$ and standard deviation
of the distribution of libration amplitudes, $\sigma_{\Delta \phi}$;
a gaussian distribution always provided
a decent approximation for the distribution of libration
amplitudes.

\subsubsection{Distribution of Longitudes}
\label{sss:ndol}
It had originally been anticipated that the distribution
of the longitudes of resonant planetesimals in the runs would be
random, since they were started with random longitudes.
This was not the case;
a typical (but not unique) feature of the longitude
distributions obtained was a double-peak superimposed on a random
distribution.
This was found to be an artifact of the initial conditions,
caused by all planetesimals being started at exactly the same
semimajor axis.
When an additional population of planetesimals at a
different semimajor axis were introduced into a run, these
were also captured into resonance and their longitude distribution
was also double-peaked.
However, at any one time, the peaks for the populations which
originated at different semimajor axes occurred at different
longitudes.
Further runs with planetesimals started with
semimajor axes randomly chosen within a range, and their
longitude distribution was indeed found to be random.
Since the planetesimals that end up in resonance originate
from a range of semimajor axes, their longitudes can be assumed
to be random after resonant trapping.
This means that the instantaneous distribution of a population
of trapped planetesimals which all have the same resonant
argument and eccentricity would look like that shown in
Figure \ref{fig:res}, since this figure shows paths of such
planetesimals at regular increments in their longitudes.

\subsubsection{Eccentricity Increase}
\label{sss:nei}
The eccentricity increase for planetesimals that
are trapped in resonance was found to be well
approximated by equation (\ref{eq:eres}) for all
of the runs.

\subsubsection{Libration Centers}
\label{sss:nlc}

\subsubsubsection{3:2, 4:3 and 5:3 Resonances}
\label{ssss:324353}
As expected on physical grounds (\S \ref{sss:lc}),
for the three resonances 3:2, 4:3, and 5:3, the
libration centers were found to tend towards
$180^\circ$ for low planet migration rates.
The way $\langle \phi_m \rangle$ responded to increasing the
planet's migration rate (\S \ref{sss:lcwm})
and varying the other parameters was tested using the sets
of runs for the 3:2 resonance which varied the planet mass,
planetesimal semimajor axis and stellar mass.
These runs showed that $\langle \phi_m \rangle$ only
depends on $\theta/\mu$.

\begin{figure}
  \begin{center}
    \begin{tabular}{rlc}
      \textbf{(a)} & \hspace{-0.25in}
        \epsscale{0.88} 
\plotone{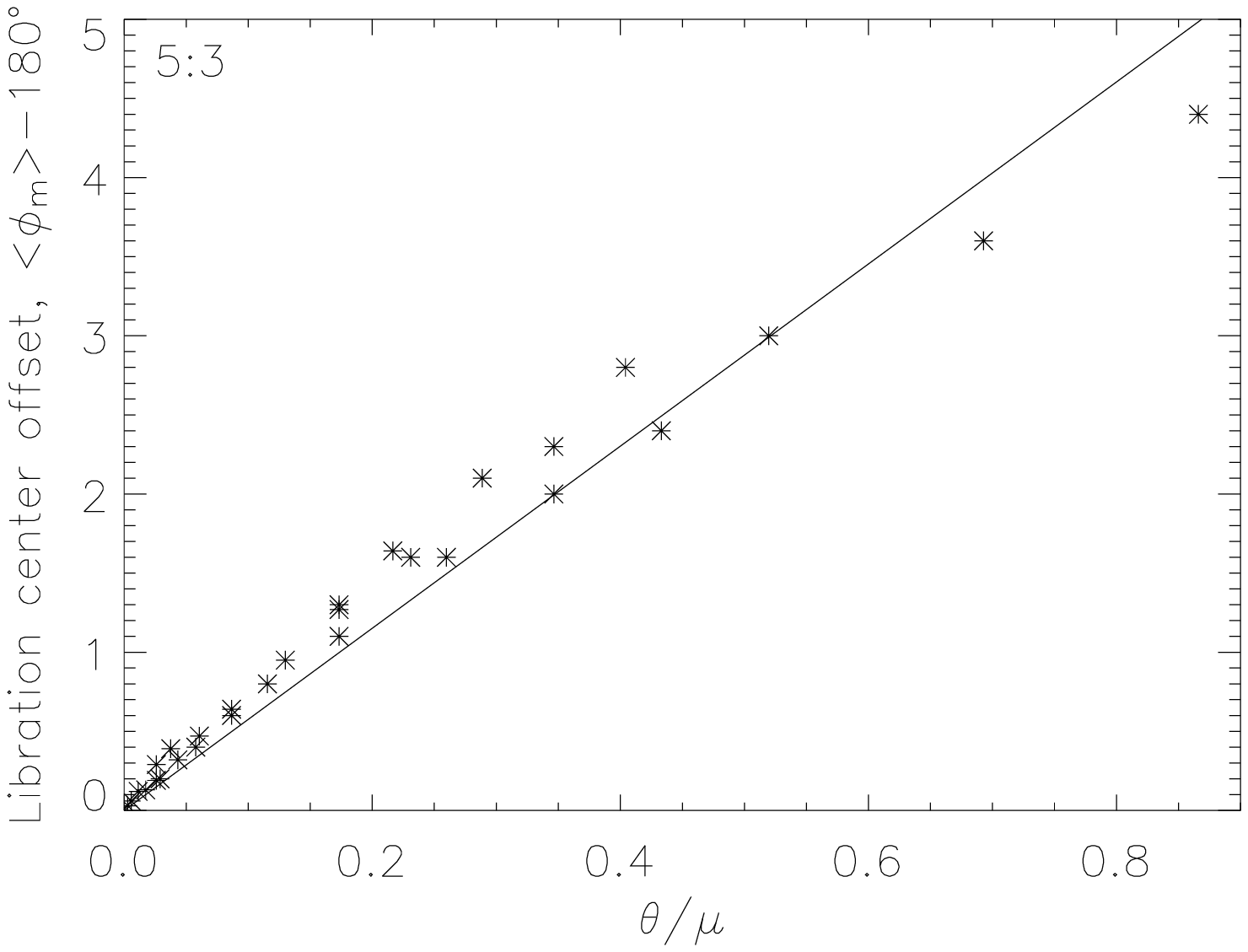} & \\[0.2in]
      \textbf{(b)} & \hspace{-0.3in} 
        \epsscale{0.9} 
\plotone{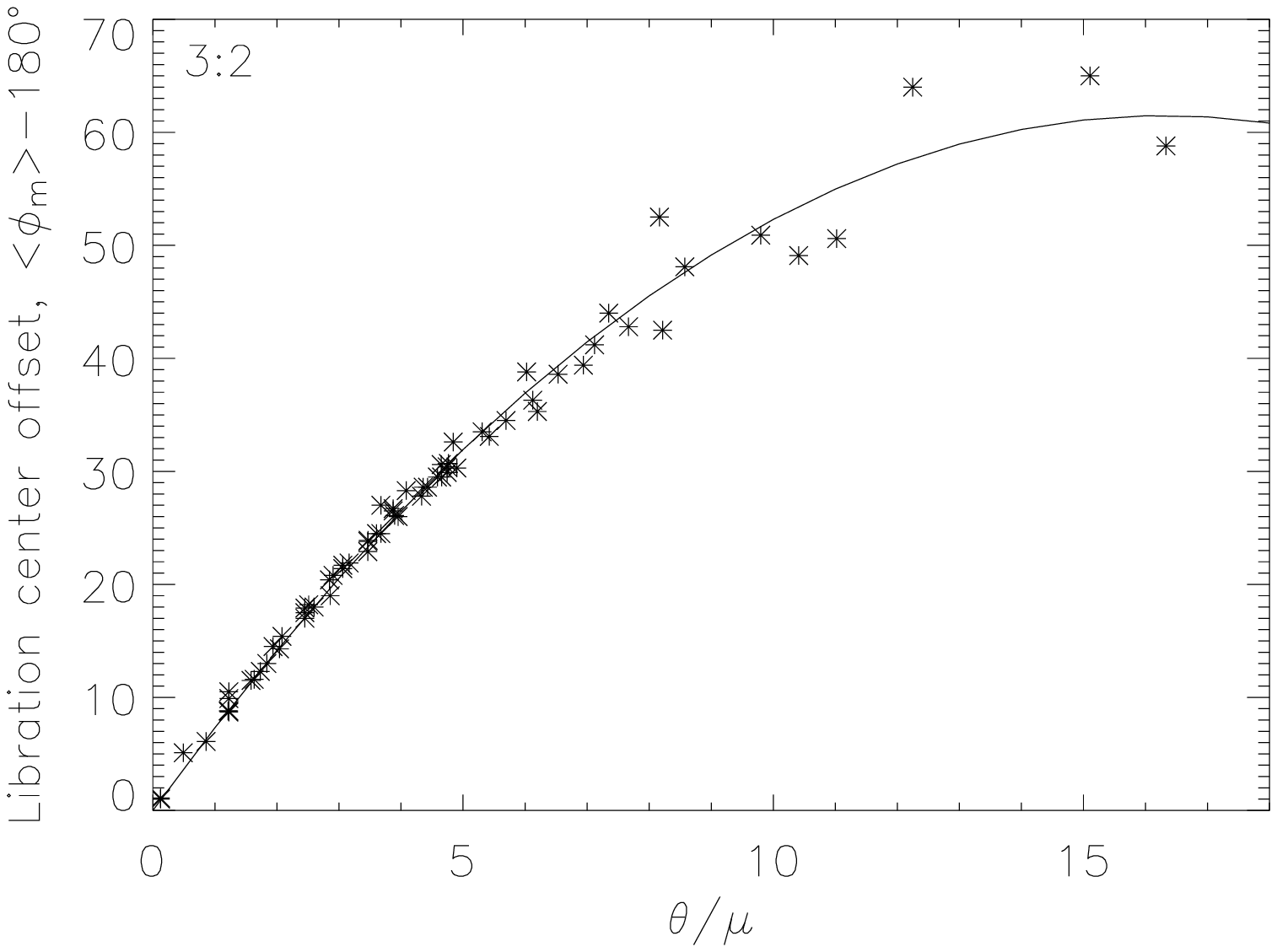} & \\[0.2in]
      \textbf{(c)} & \hspace{-0.3in}
        \epsscale{0.9} 
\plotone{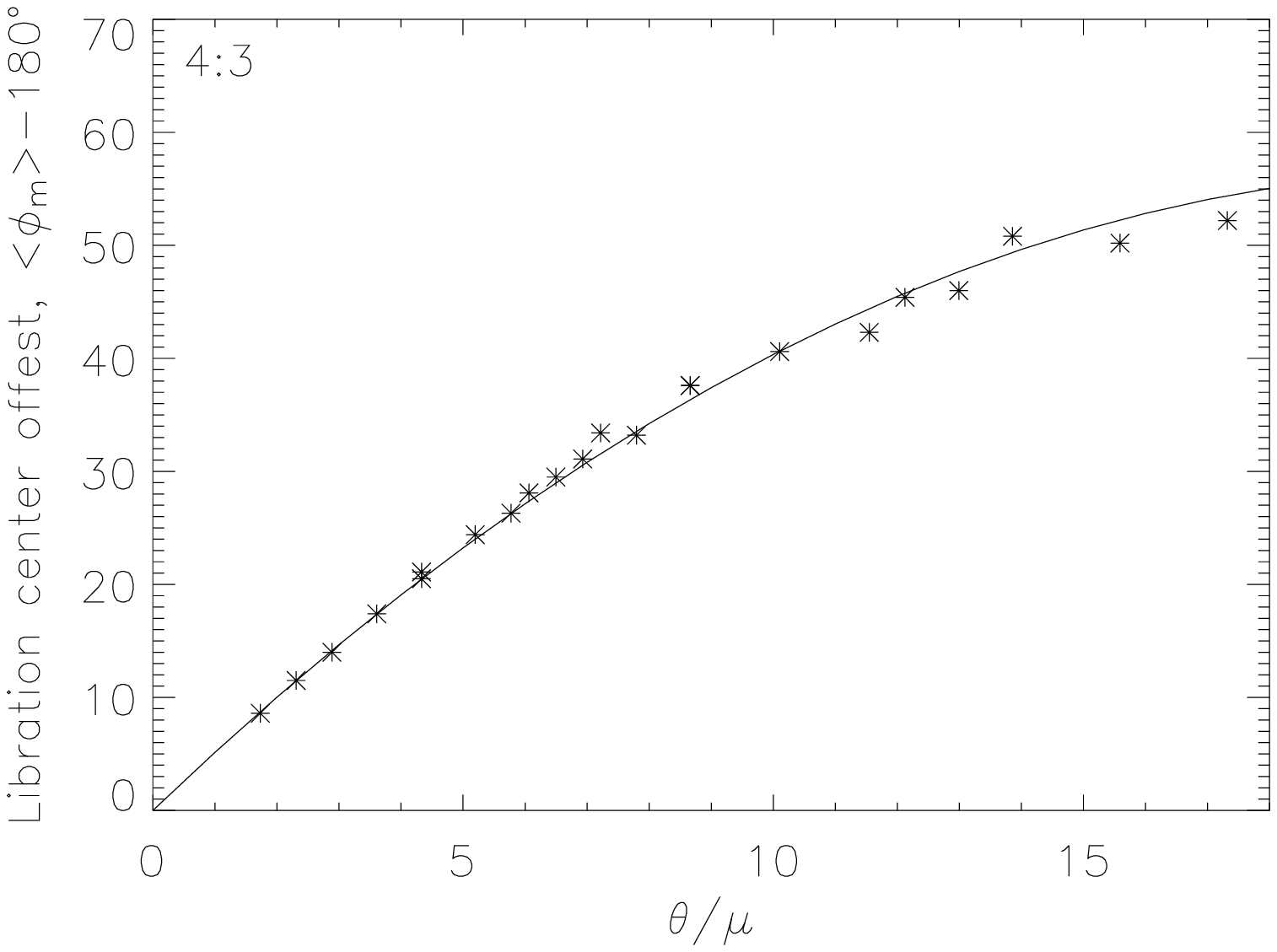} &
    \end{tabular}
  \end{center}  
  \caption{Offset of the mean libration centers, $\langle \phi_m \rangle$,
  from $180^\circ$ for planetesimals trapped in the
  \textbf{(a)} 5:3, \textbf{(b)} 3:2, and \textbf{(c)}
  4:3 resonances as a result of planet migration defined
  by the parameters $\mu$ and $\theta$ (eqs.~[\ref{eq:mu}] and
  [\ref{eq:theta}]).
  The solid lines show the fits to these offsets for each
  resonance given in equations (\ref{eq:phim53})-(\ref{eq:phim43}).
  \label{fig:phim533243}}
\end{figure}

Figure \ref{fig:phim533243} shows the deviation of
$\langle \phi_m \rangle$ from $180^\circ$ for each of the
resonances as a function of $\theta/\mu$.
These show an approximately linear correlation, as expected
from the circular restricted three body problem
(eq.~[\ref{eq:sphim2}]), but with a turnover for high
$\theta/\mu$.
However, only a very small variation of $\langle \phi_m \rangle$
was found in the course of migration as $e$ increases,
serving as a caution against using simple analytical
solutions of the circular restricted three body problem.
Fits for each of the resonances were performed
having the form
$\langle \phi_m \rangle - 180^\circ = A(\theta/\mu)+B(\theta/\mu)^2$.
The results are:
\begin{eqnarray}
  \langle {\phi_m}_{5:3} \rangle - 180^\circ & = &
    5.8(\theta/\mu), \label{eq:phim53} \\
  \langle {\phi_m}_{3:2} \rangle - 180^\circ & = &
    7.5(\theta/\mu) - 0.23(\theta/\mu)^2, \label{eq:phim32} \\
  \langle {\phi_m}_{4:3} \rangle - 180^\circ & = &
    5.3(\theta/\mu) - 0.12(\theta/\mu)^2, \label{eq:phim43}
\end{eqnarray}
and these fits are also shown on Figure \ref{fig:phim533243}.

\subsubsubsection{2:1(u) and 2:1(l) Resonances}
\label{ssss:21ul}
Again as expected on physical grounds (\S \ref{sss:lc}),
the libration center of a planetesimal in the 2:1 resonance
was found to change during the course of the migration as its
eccentricity increased.
Also, the evolution of its libration center was found to take
one of two courses:
one in which $\phi_m$ increased during the migration such
that it was always $>180^\circ$, and the other in which $\phi_m$ 
decreased and so was always $<180^\circ$.
From now on I will refer to these as two separate resonances,
the 2:1(u) and 2:1(l) resonances, respectively.

\begin{figure}
  \begin{center}
    \begin{tabular}{c}
        \epsscale{0.9} 
\plotone{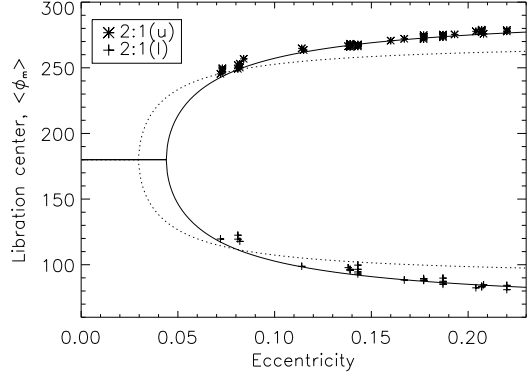}
    \end{tabular}
  \end{center}
  \caption{Variation of the mean libration center, $\langle \phi_m \rangle$,
  as a result of planet migration for planetesimals
  trapped in the 2:1(u) and 2:1(l) resonances.
  The solid line shows the fit to this variation
  given in equation (\ref{eq:phim21}).
  The analytical solution given in equation (\ref{eq:phim21an})
  is shown with the dotted line.
  \label{fig:phim21}}
\end{figure}

The mean libration centers of the resonant planetesimals in
the runs shown in Figure \ref{fig:123435mpl}a are
plotted in Figure \ref{fig:phim21};
planetesimals were separated into the 2:1(u) and 2:1(l)
resonances and their libration centers measured at
four epochs throughout the migration.
The libration centers in these runs were found to depend almost
entirely on the planetesimal's eccentricity,
with any variation due to $\mu$ or $\theta$ barely
discernible.
Neither was any significant difference found between the
2:1(u) and 2:1(l) resonances, and a simple parameterised fit
was performed of the form $\cos{\langle \phi_m \rangle} = A + B/e$.
The result is:
\begin{eqnarray}
  \cos{\langle \phi_{m_{2:1}}\rangle } & = & 0.39 - 0.061/e, \label{eq:phim21}
\end{eqnarray}
and these lines are shown on Figure \ref{fig:phim21} with the
additional constraint that $\cos{\langle \phi_{m_{2:1}} \rangle} \geq -1$.
The line derived from the circular restricted three body problem
(eq.~[\ref{eq:phim21an}]) is also shown on this figure.

\subsubsection{Trapping Probability for 2:1(l) Resonance}
\label{sss:tp21l}
\begin{figure}
  \begin{center}
    \begin{tabular}{c}
        \epsscale{0.9} 
\plotone{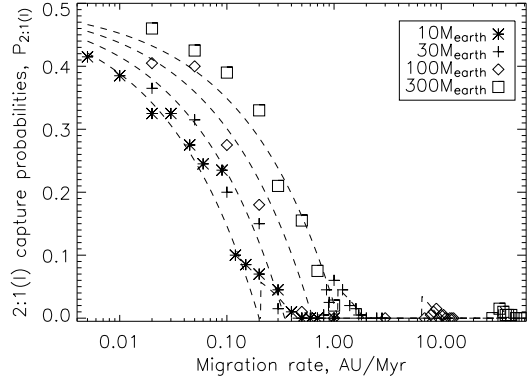}
    \end{tabular}
  \end{center}
  \caption{Capture probabilities for the 2:1(l) resonance,
  $P_{2:1(l)}$, for planetesimals initially 30 AU from a
  $2.5M_\odot$ star.
  The parameterised fits to these probabilities
  given in equations (\ref{eq:p21l}) and (\ref{eq:dp21l})
  for the four planet masses shown in this plot
  (i.e., $\mu=4$, 12, 40, and 120) are shown with dashed
  lines.
  \label{fig:21lprob}}
\end{figure}

The trapping probabilities derived in \S \ref{ss:12} do not
take into account whether the planetesimal ends up in the
2:1(u) or 2:1(l) resonance, since the semimajor axis and eccentricity
evolution is the same for both resonances.
However, it was recently reported that trapping probabilities
are not the same for the two resonances (Chiang \& Jordan 2002).
Thus I reanalysed the trapping probability runs for the 2:1 resonance
to determine which of the resonances was being populated;
the results are shown in Figure \ref{fig:21lprob}, where I
have plotted $P_{2:1(l)}$, the trapping probability for the 2:1(l)
resonance.
It was found that there are always less planetesimals trapped in
the 2:1(l) resonance than in the 2:1(u), except in the limit
where the planet's migration rate tends to zero at which point
the two resonances are equally populated.
\footnote{Since the original runs were designed to have trapping
probabilities close to 50\%, additional runs had to be carried out
to ascertain how $P_{2:1(l)}$ varies as planet migration rate is
reduced when trapping is 100\% for the 2:1 resonance as a whole.}
As the migration rate is increased, while trapping into the 2:1
resonance is still 100\%, $P_{2:1(l)}$ decreases
$\propto \theta^{0.5}\mu^{-0.25}$ until it reaches zero.
This behavior can be approximated by:
\begin{equation}
  P_{2:1(l)} = 0.5 - 0.85\theta^{0.5}\mu^{-0.25}. \label{eq:p21l}
\end{equation}

Unusually, as the migration rate is increased further to the point
where trapping into the 2:1 resonance begins to decrease,
trapping probabilities for the 2:1(l) resonance increase
again such that a few per cent can become trapped.
For modeling purposes this behavior was approximated using
the following functions:
If $\theta\mu^{-1.5} > 0.09$ then the capture probabilities given
in equation (\ref{eq:p21l}) should be increased by an amount
\begin{equation}
  dP_{2:1(l)} = 0.11 - 0.48\theta\mu^{-1.35}, \label{eq:dp21l}
\end{equation}
assuming that $dP_{2:1(l)}>0$.
The complete fits are shown in Figure \ref{fig:21lprob},
and the line delineating migrations for which $P_{2:1(l)} \approx 1/3$
(i.e., for which the 2:1(u) resonance is twice as populated as the 2:1(l)
resonance) is shown in Figure \ref{fig:tpsumm} (i.e.,
$\theta \approx 0.038\sqrt{\mu}$ from eq.~[\ref{eq:p21l}]).

The reason for this asymmetry remains unclear, however
a clue to its origin comes from the
physical interpretation of the origin of the resonance.
As the planetesimal's eccentricity increases, its resonant
argument initially librates about $180^\circ$.
Once its eccentricity reaches $\sim 0.03$ the libration center
must either increase or decrease to balance the angular
momentum imparted to it at conjunction and pericenter
(\S \ref{sss:lc}).
Nothing in the argument presented so far has hinted at either
of the resonances being stronger than the other.
However, because of the migration, the stable libration
is not exactly at $\phi_m = 180^\circ$, but at a slightly higher
value (\S \ref{sss:lcwm}), even if this offset is too small to
detect in these runs (\S \ref{sss:nlc}).
The fact that $\phi$ is more often (if not always) $>180^\circ$, 
may make the 2:1(u) resonance the more likely path.
This would tie in with a higher asymmetry expected for higher
migration rates for which the offset of the libration center
from $180^\circ$ is higher (\S \ref{sss:lcwm}).

\subsubsection{Libration Amplitude Distributions}
\label{sss:dol}
The distribution of libration amplitudes was always found to be
fairly broad with $\sigma_{\Delta \phi}$ in the range $5-20^\circ$.
No significant correlation could be found of $\sigma_{\Delta \phi}$
with either $\mu$ or $\theta$ for any of the resonances.
Thus for modelling purposes these distributions were assumed
to have values of
\begin{eqnarray}
  \sigma_{\Delta \phi_{2:1}} & = & 10^\circ, \label{eq:sdphi21} \\
  \sigma_{\Delta \phi_{5:3}} & = & 11^\circ, \label{eq:sdphi53} \\
  \sigma_{\Delta \phi_{3:2}} & = & 15^\circ, \label{eq:sdphi32} \\
  \sigma_{\Delta \phi_{4:3}} & = & 15^\circ, \label{eq:sdphi43}
\end{eqnarray}
which are the means of all the runs performed for each resonance.

\begin{figure}
  \begin{center}
    \begin{tabular}{rlc}
      \textbf{(a)} & \hspace{-0.3in}
        \epsscale{0.7} 
\plotone{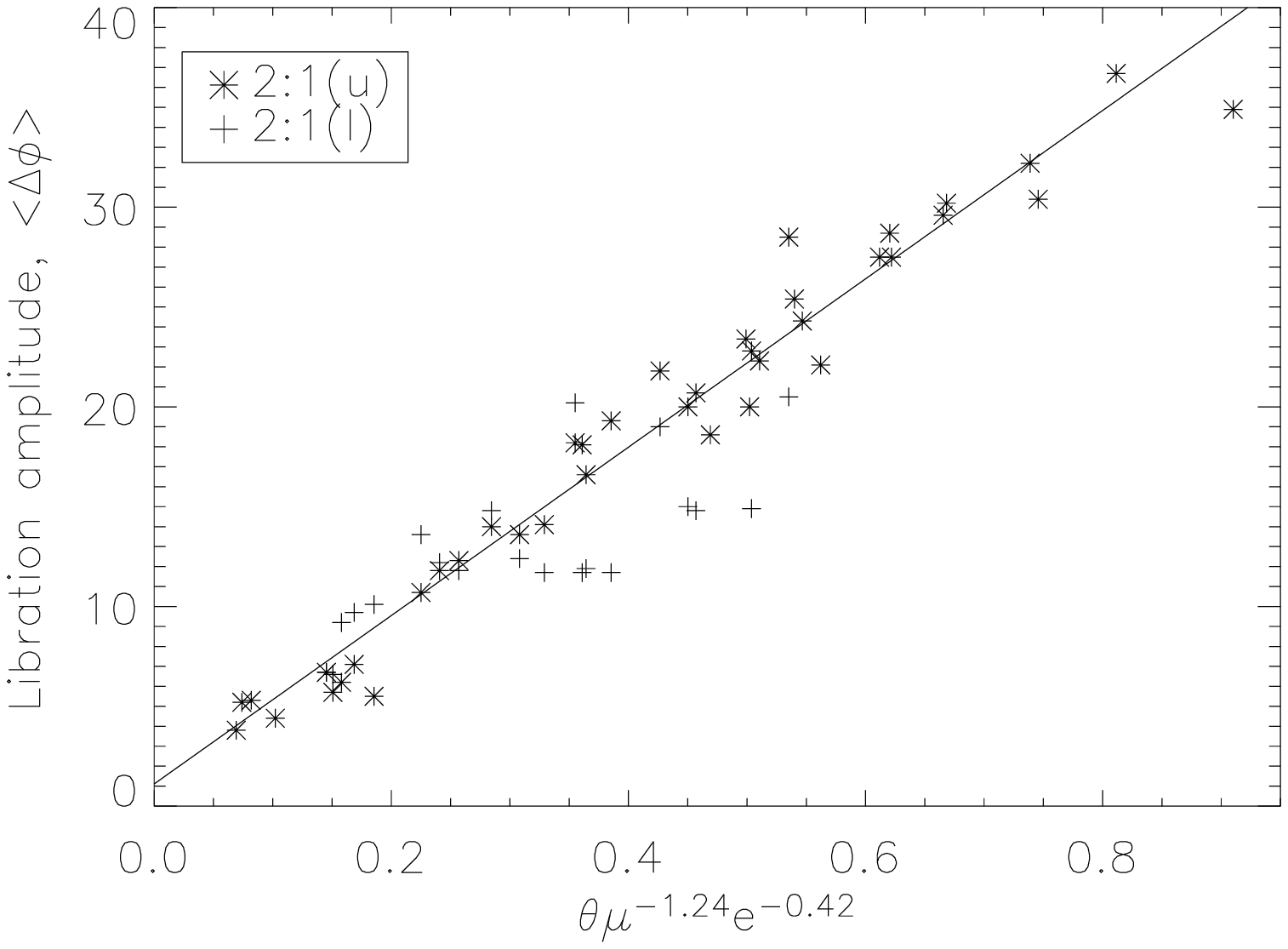} & \\[0.2in]
      \textbf{(b)} & \hspace{-0.3in} 
        \epsscale{0.7} 
\plotone{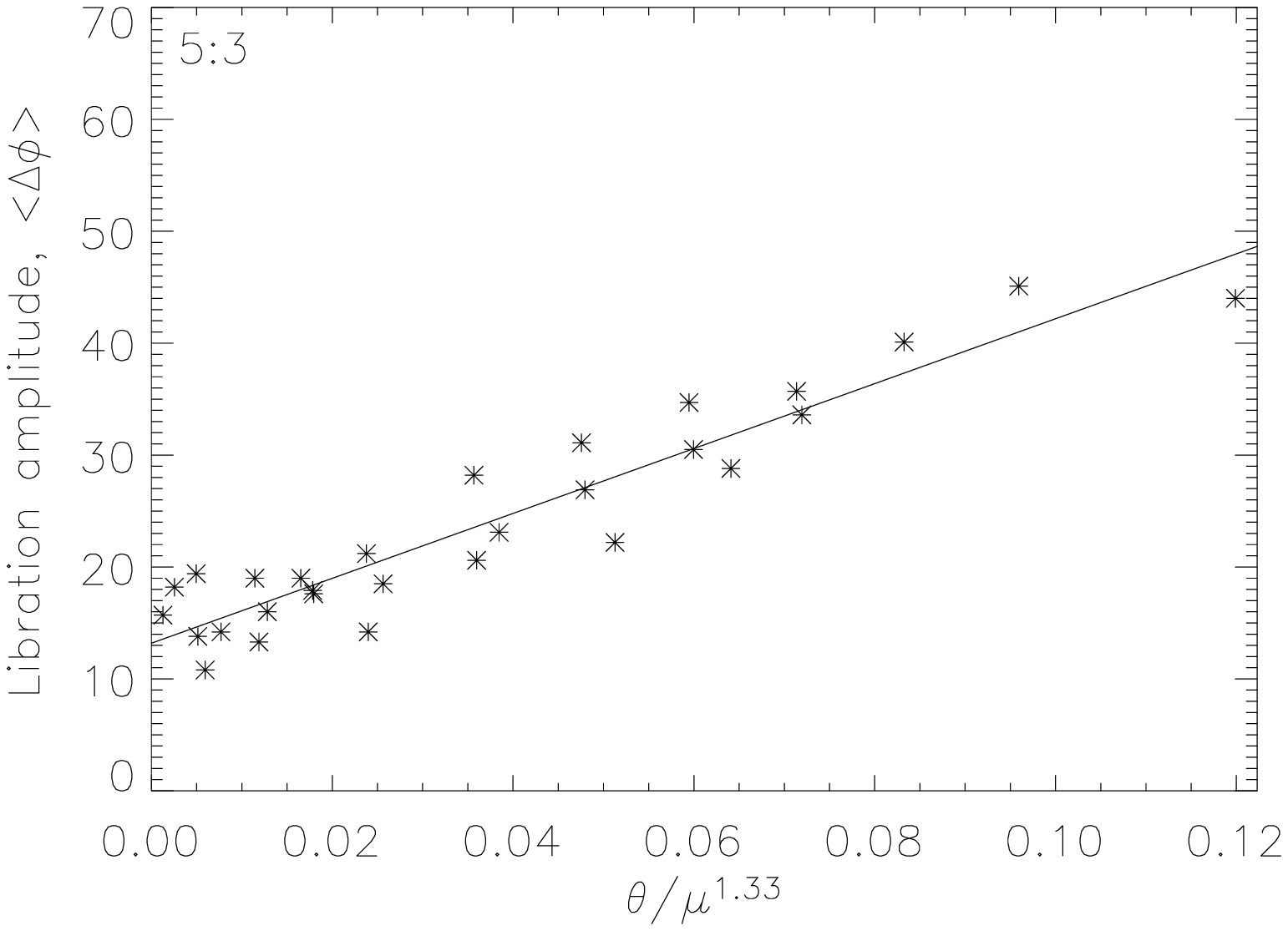} & \\[0.2in]
      \textbf{(c)} & \hspace{-0.3in}
        \epsscale{0.7} 
\plotone{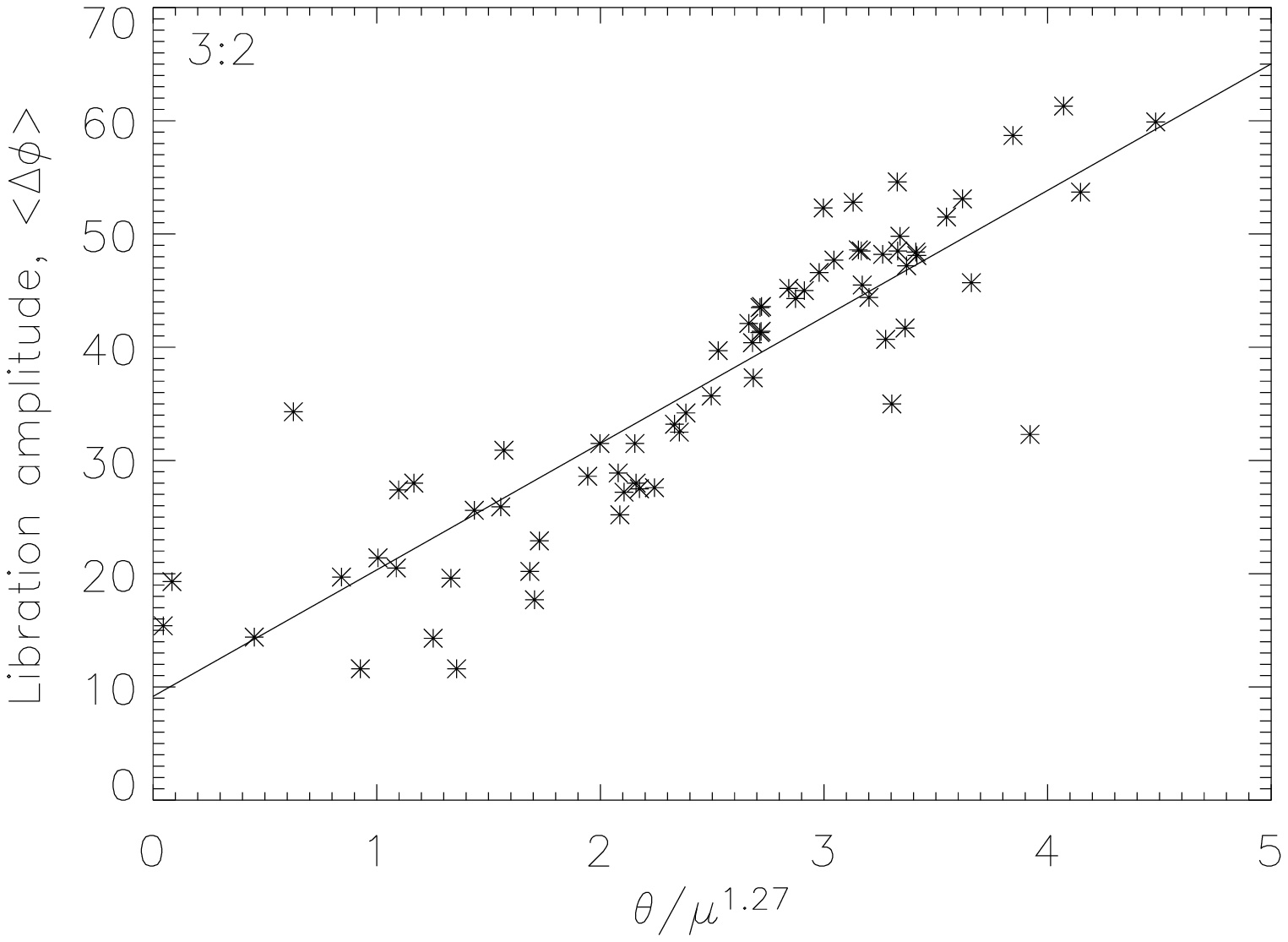} & \\[0.2in]
      \textbf{(d)} & \hspace{-0.3in}
        \epsscale{0.7} 
\plotone{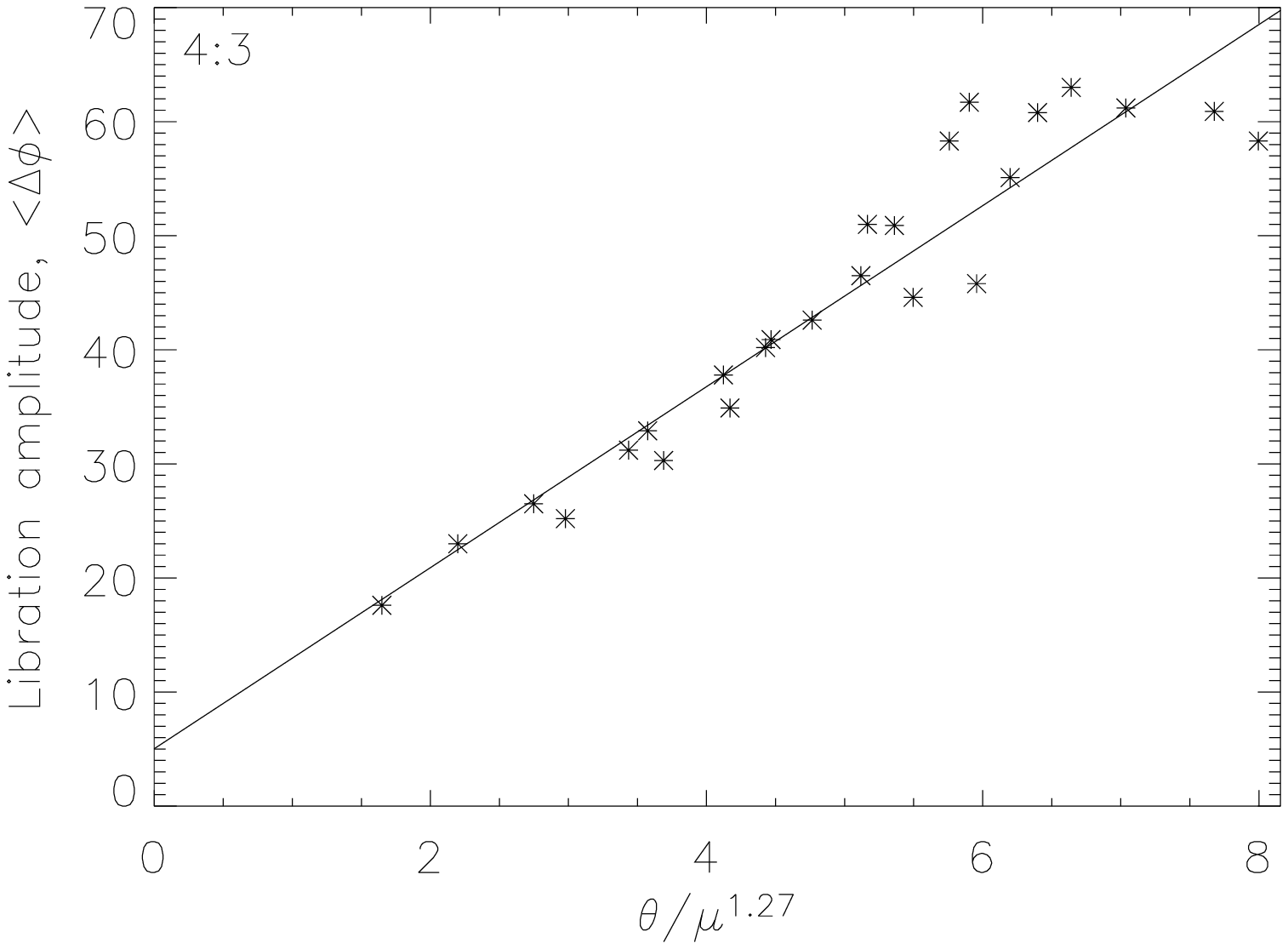} &
    \end{tabular}
  \end{center}
  \caption{Mean libration amplitudes, $\langle \Delta \phi \rangle$,
  of planetesimals captured in the \textbf{(a)} 2:1,
  \textbf{(b)} 5:3, \textbf{(c)} 3:2, and \textbf{(d)}
  4:3 resonances for migrations defined by the parameters
  $\mu$ and $\theta$ (eqs.~[\ref{eq:mu}] and [\ref{eq:theta}]).
  The solid lines show the fits to these libration amplitudes
  given in equations (\ref{eq:dphi21})-(\ref{eq:dphi43}).
  \label{fig:dphi}}
\end{figure}

However, the mean libration amplitudes, $\langle \Delta \phi \rangle$,
which are plotted in Figure \ref{fig:dphi},
were found to vary in a systematic manner determined by
the dimensionless parameters $\mu$ and $\theta$.
In particular the mean libration amplitudes depend only on the
combination $\theta/\mu^A$, so that libration amplitudes are
higher for faster migrations and more massive planets.
For the 2:1 resonance the libration amplitude was also found
to decrease during the migration as the planetesimals' eccentricities
are increased.
No significant variation of $\langle \Delta \phi \rangle$
during the migration was found for the 5:3, 3:2, and 4:3 resonances,
and only $\langle \Delta \phi \rangle$ at the end of their runs
were considered.
Also, no significant difference was found between the 2:1(u) and
2:1(l) resonances, and so their results were considered together
as showing the variation for the 2:1 resonance.

Least squares fits were performed to these results
of the form $\langle \Delta \phi \rangle = A + B(\theta/\mu^C)/e^D$,
where $D=0$ except for the 2:1 resonance.
The results of these fits are:
\begin{eqnarray}
  \langle \Delta \phi_{2:1} \rangle & = &
    1.1  + 42.2(\theta/\mu^{1.24})/e^{0.42}, \label{eq:dphi21} \\
  \langle \Delta \phi_{5:3} \rangle & = &
    13.2 + 290.0(\theta/\mu^{1.33}), \label{eq:dphi53} \\
  \langle \Delta \phi_{3:2} \rangle & = &
    9.2  + 11.2(\theta/\mu^{1.27}), \label{eq:dphi32} \\
  \langle \Delta \phi_{4:3} \rangle & = &
    5.0  + 7.9(\theta/\mu^{1.27}), \label{eq:dphi43}
\end{eqnarray}
and these fits are also shown in Figure \ref{fig:dphi}.

\subsubsection{Other Parameters}
\label{sss:op}
Other parameters were also derived from these runs which
are interesting from a celestial mechanics point of view,
but which are not directly relevant for the structure of
a planetesimal disk.
These are described in Appendix \ref{app:adiabatic}.

\section{Debris Disk Structure}
\label{s:dds}
The results given in \S\S \ref{s:rcp} and \ref{s:drp} can be used to
predict the spatial distribution of planetesimals at the end of
any given migration defined by $\mu$ and $\theta$
(eqs.~[\ref{eq:mu}] and [\ref{eq:theta}]), and in \S \ref{ss:tm}
a numerical scheme is described which does just that.
Perhaps more importantly, the converse is also true:
an observed spatial distribution of planetesimals can be used
to constrain the migration which caused that distribution.
To help with such an interpretation, some generalizations about
the kind of structures that result from different migrations are
given in \S \ref{ss:gs}.
In \S \ref{ss:vega} this model is applied to observations of
Vega's debris disk which shows that it is possible to set
quite tight constraints on the migration history of this
system.
\S \ref{ss:caveats} discusses the limitations of the model
and future developments.

\subsection{Numerical Model}
\label{ss:tm}
The model is completely defined by the following input
parameters:
\begin{itemize}
  \item \textbf{Planetesimals:} their initial distribution
        is defined by $a_{min}$, $a_{max}$, and $\delta$, where
        the number of planetesimals in the semimajor axis range
        $a$ to $a+da$ is $\propto a^\delta da$;
        in the minimum mass solar nebula model $\delta = -0.5$.
  \item \textbf{Planet:} has a mass $M_{pl}$ and a migration
        defined by $a_{{pl}_i}$, $a_{{pl}_f}$, and $\dot{a}_{pl}$,
        which are its initial and final semimajor axes and its
        migration rate, respectively.
  \item \textbf{Star:} has a mass $M_\star$.
\end{itemize}

A number of planetesimals, $N_{pp}$, are then distributed
in semimajor axis randomly according to the prescription
above, with eccentricities also randomly chosen
between 0 and 0.01;
$N_{pp}$ has to be sufficient to define the distribution of
eccentricities of resonant planetesimals in the final
distribution, and is normally set at 400.
For each planetesimal, the passing of each of the planet's
resonances is considered in the order they encounter the
planetesimal.
If a random number in the range 0-1 is less than the trapping
probability defined by equation (\ref{eq:p2}) and
Table \ref{tab:xyuv}, then the planetesimal is assumed to
become trapped in that resonance;
for planetesimals that are trapped in the 2:1 resonance,
the probability that this is the 2:1(l) resonance is also
determined from equations (\ref{eq:p21l}) and (\ref{eq:dp21l}).
Naturally no more of the resonances are then considered,
and this planetesimal's eccentricity and semimajor axis
are assumed to increase according to equations
(\ref{eq:apq}) and (\ref{eq:eres}).

Each of the resonant planetesimals is assumed to be
representative of $9600(p+q)$ more.
For each planetesimal, 400 values of $\phi$ were chosen by
first using equations (\ref{eq:phim53}) - (\ref{eq:phim21})
to determine the appropriate libration center.
Then 20 libration amplitudes were chosen at random from the appropriate
gaussian distribution defined by equations (\ref{eq:sdphi21})
- (\ref{eq:dphi43}), and for each libration amplitude 20 values of
$\phi$ were chosen from equation (\ref{eq:phievol}) with values
of $t/t_\phi$ evenly distributed between 0 and 1.
For each of these values of $\phi$, $24(p+q)$ planetesimals
were assigned evenly spaced longitudes so that their spatial
distribution matched the patterns shown in Figure \ref{fig:res}
(with an appropriate orientation defined by $\phi$).
The resulting number density distributions for each of the resonant
planetesimals was then normalised to unity (by dividing by
$9600(p+q)$).

The semimajor axes and eccentricities of non-resonant
planetesimals remain unchanged during the migration
unless they reach the planet's resonance overlap region:
the region
\begin{equation}
  |a/a_{pl} - 1| < 1.3(3\times 10^{-6}\mu)^{2/7} \label{eq:aresol}
\end{equation}
is chaotic and planetesimals entering this region
would be scattered out of the system on short timescales
(Wisdom 1980);
such planetesimals are removed from the model.
Each of the remaining non-resonant planetesimals was assumed to
be representative of 400 more with even distributions of
longitude and longitude of pericenter, and 
their resulting number density distributions normalised to unity
(by dividing by 400).

\subsection{Generic Structures}
\label{ss:gs}
First consider the distribution of planetesimals ignoring both
the change in libration center due to migration and the
amplitude of the libration, i.e. such that
$\phi(t) = \phi_m(\theta=0)$.
The spatial distribution of such planetesimals would look like that
shown in Figure \ref{fig:res}.
That is, planetesimals in the 3:2 resonance would be strongly
concentrated at $\pm 90^\circ$ longitude relative to the planet, while
those in the 4:3 and 5:3 resonances would be concentrated $\pm 60^\circ$
and $180^\circ$ from the planet, and those in the 2:1(u) resonance would be
concentrated $103-79^\circ$ behind the planet (for eccentricities 0.1-0.3)
and most of those in the 2:1(l) would be found $103-79^\circ$ in front of
the planet.
The strength and physical size of the concentrations of a population
of planetesimals that are in a given resonance depend on the
distribution of their eccentricities.

The fraction of planetesimals that end up in a particular resonance
is determined not only by the $\mu$ and $\theta$ of the migration,
but also by the initial distribution of its planetesimal population,
as well as the extent of the migration.
To help visualize what the outcome of any given migration
would be, Figure \ref{fig:migzones} and Table \ref{tab:lr} summarize
the dynamical structures resulting from migrations that are located
in the seven different zones in the $\mu-\theta$ plot of Figure 
\ref{fig:tpsumm}.
The boundaries between the zones are taken as the lines of
50\% trapping probability for the different resonances, 
as well as the line for which twice as many planetesimals are in
the 2:1(u) resonances compared with the 2:1(l) resonance.
Clearly these boundaries are not so distinct, although
the areas covered by 10-90\% trapping probabilities are
relatively small on this plot.
The application of these zones will become clearer in
\S \ref{ss:vega}.

Consider now the offset in libration center due to migration.
This does not affect the spatial distribution of 2:1 planetesimals,
but results in a rotation of the pattern for the other
resonances shown in Figure \ref{fig:res} by an amount
$(\phi_m-180^\circ)/p$.
This is plotted in Figure \ref{fig:dphimp} for migrations which
result in trapping probabilities of 10, 50, and 90\%.
Since the turnover is not well characterized for high
values of $\mu$ the line is not extrapolated above the
highest value of $\mu$ in the runs.
The rotation is always negligible for the 5:3 resonance, and is
usually small, $<30^\circ$, for the 3:2 and 4:3 resonances.
In other words, this rotation would only be noticeable for high
mass planets ($\mu \approx 100$) migrating close to the limit
where trapping probabilities are $<50$\%.
Further, this rotation is only valid while the planet is
migrating.
Simulations similar to those already described were performed
with the planet migration turned off.
In this instance the libration was about $180^\circ$ (apart from
the libration centers of the 2:1 resonances which were unchanged).
Since the rotation is small it is included in the model
for consistency, but any model which relies on this rotation
must consider the probability of our witnessing a system
mid-migration.

Consider now the effect of the libration of $\phi$ due to
a non-zero $\Delta \phi$.
Since the oscillation is sinusoidal, the distribution of
$\phi$ is not peaked at $\phi_m$, but actually has
two peaks at $\phi_m \pm \Delta \phi$.
In principle, if $\Delta \phi$ is big enough, the
planetesimal distribution will peak at $2p$ rather
than $p$ longitudes relative to the planet (Chiang \& Jordan 2003).
The maximum rotation of the pattern from the orientation defined by 
$\phi_m$ (i.e. that shown in Figure \ref{fig:res}) is
$\pm \Delta \phi / p$, and Figure \ref{fig:dphip} shows
$\Delta \phi /p$ for all resonances as
a function of $\mu$ for the values of $\theta$ for which trapping
probabilities are 10, 50, and 90\%.
For the 2:1 resonance, the libration amplitude decreases throughout
the migration, and so is plotted at two reference eccentricities,
$e=0.03$ and 0.3.
Given the physical size of the clumps, these libration amplitudes
are relatively small, even when the migration is fast enough
to result in low trapping probabilities.
Thus libration in this model results only in a slight azimuthal
smearing of each clump.

\subsection{Application to Vega}
\label{ss:vega}
Vega is a nearby (7.8 pc) main sequence A0 star
($M_\star \approx 2.5M_\odot$) with an age of
$\sim 350$ Myr (Song et al. 2001).
Its emission spectrum exhibits a strong excess above
the level of the photosphere at wavelengths longward of
12 $\mu$m (Aumann et al. 1984).
This excess originates in dust grains orbiting the
star that are continually replenished by the destruction
of larger planetesimals in its debris disk.
Imaging at submillimeter wavelengths shows the morphology
of the excess emission (Holland et al. 1998;
see Fig.~\ref{fig:vegaim}a) down to a resolution of
$14\arcsec$ (FWHM):
the lowest contours are nearly circularly symmetric
extending to $\sim 24\arcsec$, indicating that the disk
is being observed pole-on, but the disk's structure is
dominated by an emission peak $\sim 9\arcsec$ to the
northeast of the star;
the highest contours are also elongated in the southwest
direction.

The structure of the disk was recently mapped with even
higher spatial resolution ($\sim 3\arcsec$) using millimeter
interferometry (Koerner, Sargent \& Ostroff 2001; Wilner et al.
2002).
While these observations were not sensitive to the larger
scale structure, they did show that a significant fraction
of the millimeter emission could be resolved into two
clumps, one $9.5\arcsec$ northeast of the star, the other
$8\arcsec$ southwest of the star, and that the northeast
clump is brighter than that southwest.
Thus it appears that the submillimeter images are best
interpreted as emission from an extended disk which is dominated
by two roughly equidistant clumps of unequal brightness on
opposite sides of the star.

\begin{figure}
  \begin{center}
    \begin{tabular}{c}
        \epsscale{0.9} 
\plotone{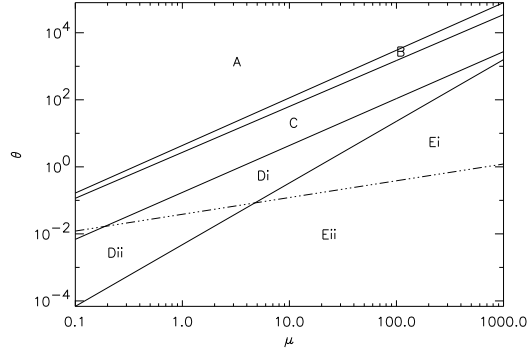}
    \end{tabular}
  \end{center}
  \caption{Definition of the migration zones A-Eii discussed
  in the text and Table \ref{tab:lr}.
  The lines are the same as those in plotted in Figure
  \ref{fig:tpsumm}.
  \label{fig:migzones}}
\end{figure}

\begin{figure}
  \begin{center}
    \begin{tabular}{c}
        \epsscale{0.9} 
\plotone{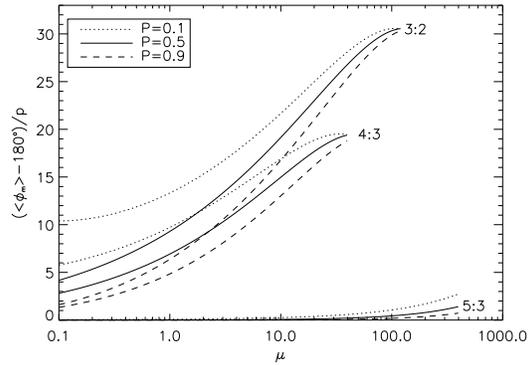}
    \end{tabular}
  \end{center}
  \caption{Rotation (clockwise) of the resonant pattern shown in Figure
  \ref{fig:res} due to the migration of the planet characterized
  in Figure \ref{fig:phim533243}.
  This is shown for migrations resulting in trapping probabilities
  of 10, 50, and 90\%.
  \label{fig:dphimp}}
\end{figure}

\begin{figure}
  \begin{center}
    \begin{tabular}{rlc}
      \textbf{(a)} & \hspace{-0.3in}
        \epsscale{0.9} 
\plotone{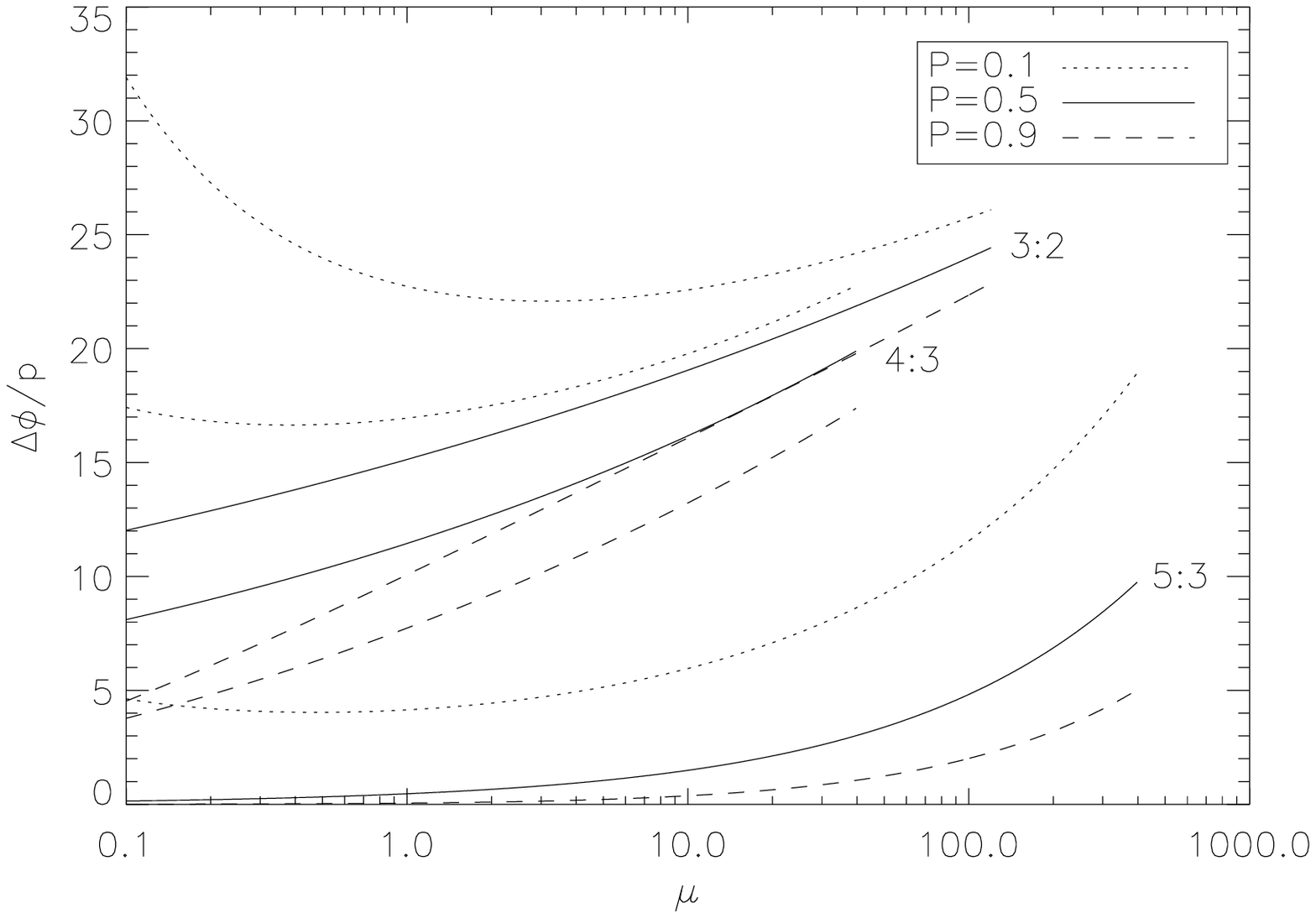} & \\[0.2in]
      \textbf{(b)} & \hspace{-0.3in} 
        \epsscale{0.9} 
\plotone{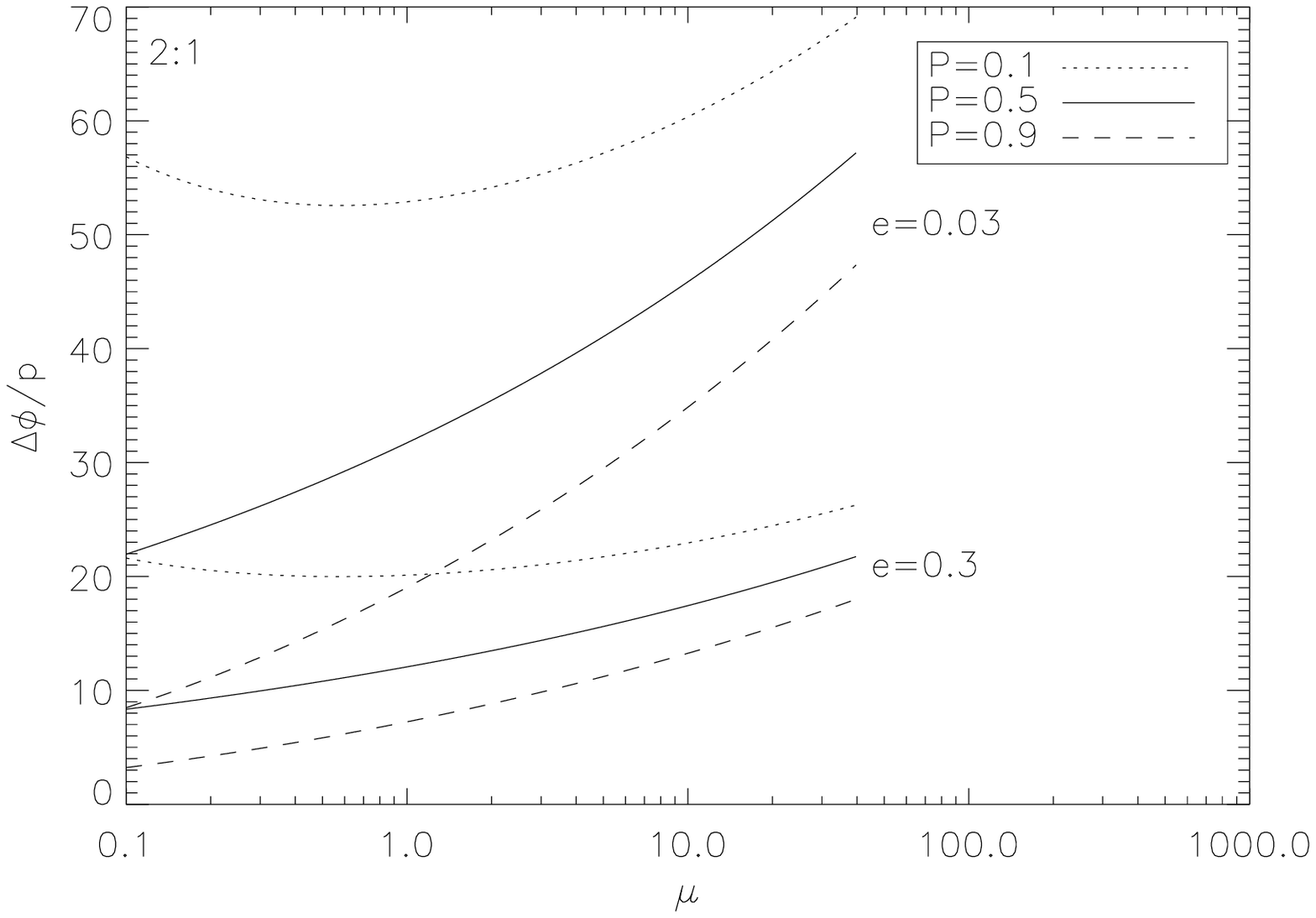} &
    \end{tabular}
  \end{center}
  \caption{Libration amplitudes for migrations resulting in trapping probabilities
  of 10, 50, and 90\% for: \textbf{(a)} the 5:3, 3:2, and 4:3 resonances;
  \textbf{(b)} the 2:1 resonance when $e=0.03$ and 0.3.
  \label{fig:dphip}}
\end{figure}

\begin{figure}
  \begin{center}
    \begin{tabular}{rlc}
      \textbf{(a)} & \hspace{-0.3in}
        \epsscale{0.9} 
\plotone{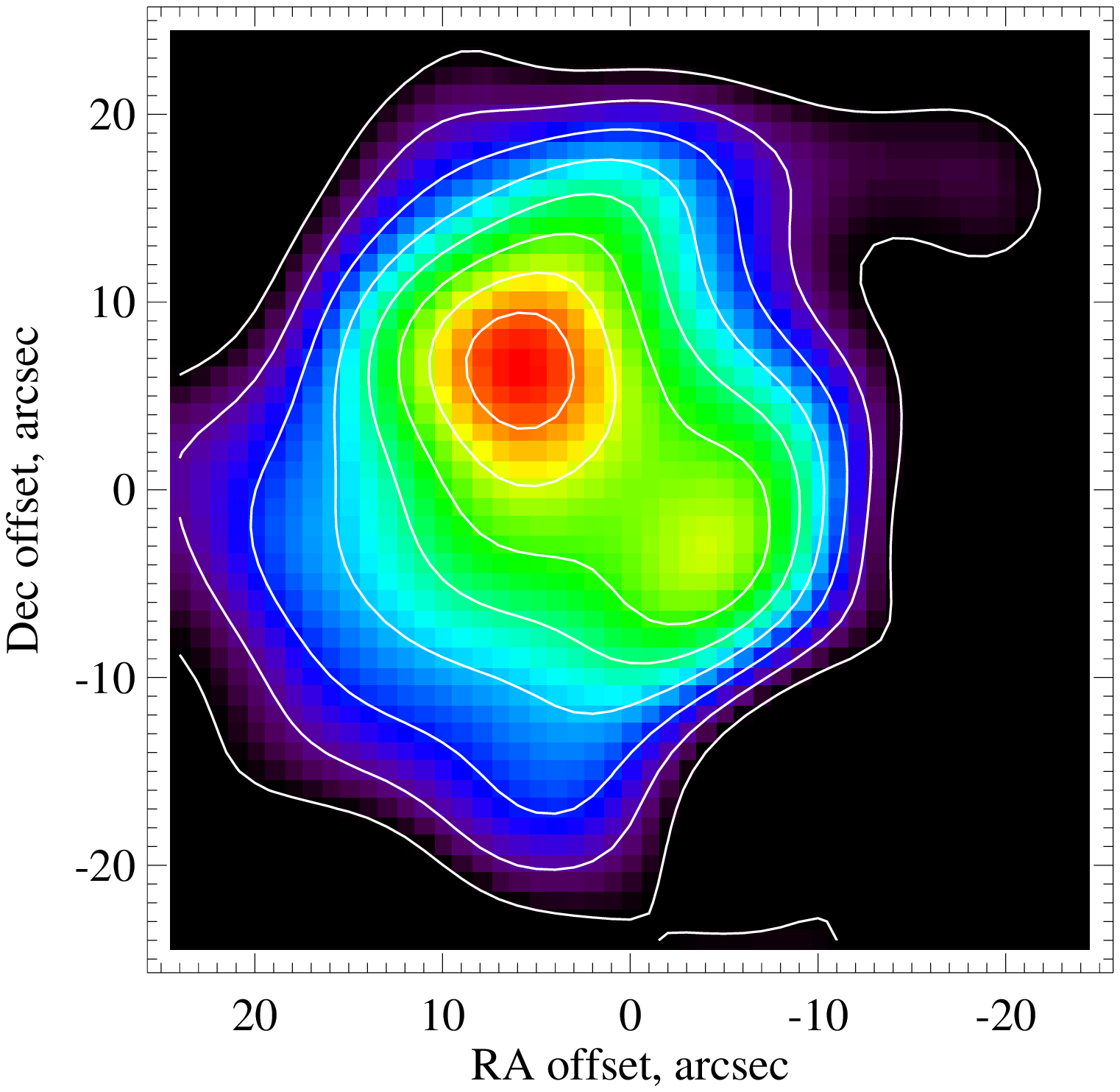} & \\[0.2in]
      \textbf{(b)} & \hspace{-0.3in}
        \epsscale{0.9} 
\plotone{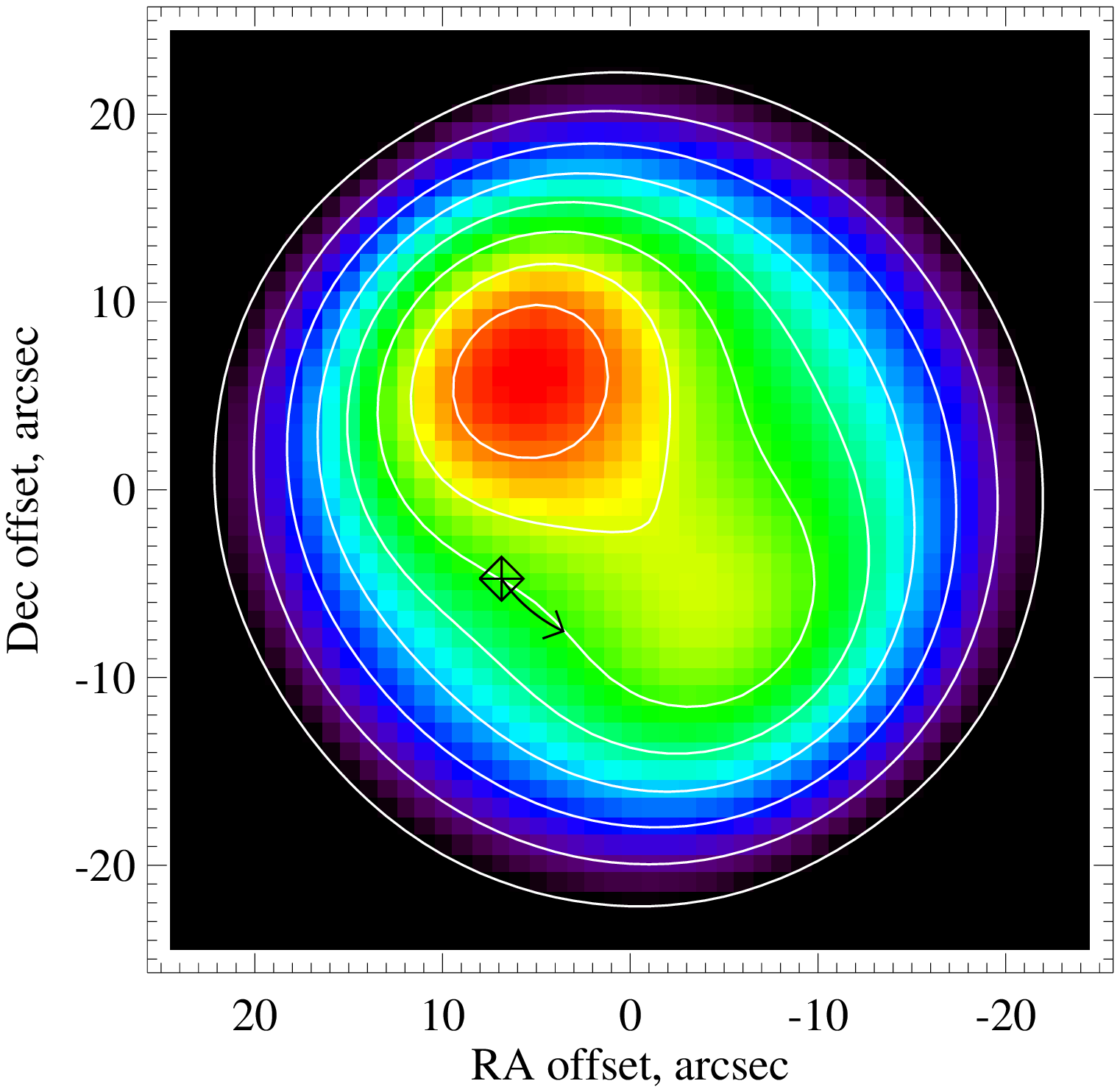} &
    \end{tabular}
  \end{center}
  \caption{850 $\mu$m images of Vega.
  \textbf{(a)} Observation taken using SCUBA at the JCMT
  (Holland et al. 1998).
  The contours start at 3.8 mJy beam$^{-1}$ and increase
  at 1.9 mJy beam$^{-1}$ intervals.
  The beam size has a $14\arcsec$ FWHM, and additional gaussian
  smoothing of $7\arcsec$ FWHM has been applied.
  Emission from the stellar photosphere of $\sim 5.9$ mJy
  has not been subtracted from the image.
  \textbf{(b)} Simulated model image of dust grains created in
  the destruction of planetesimals shown in Figure \ref{fig:vegamod}.
  The planet is shown at the location of the diamond-plus
  and its direction of motion is also shown.
  Appropriate color table, contours, smoothing and stellar
  photosphere have been included to allow a direct comparison
  with \textbf{(a)}.
  \label{fig:vegaim}}
\end{figure}

Based on the discussion of \S \ref{ss:gs},
the migration zone (Fig.~\ref{fig:migzones}) causing this
structure can be narrowed down to zone Di:
Two clumps of equal brightness could have been explained by
planetesimals trapped in the 3:2 resonance (zone C),
or in equal numbers into the 2:1(u) and 2:1(l) resonances
(zones Dii or Eii), but the asymmetry indicates that one of the
clumps is overpopulated, pointing to migration
zone Di or Ei.
Further constraints are set by the lack of evidence for an
additional 3 clump pattern rotated relative to the two clump
pattern.
While 3 clumps from the 4:3 resonance are inevitable,
zone Ei is ruled out as trapping of planetesimals
into the 5:3 resonance in this zone means there are less
available to be trapped into the 2:1(u) resonance.
In theory the outcome of migrations anywhere within
zone Di will be similar and are not constrained by this model,
however the fuzzy edges of the zones mean that further
constraints are possible.
Here I set the mass of the planet to be the same as that
of Neptune, $M_{pl} = 17.2M_\oplus$ ($\mu = 6.9$),
which means that the migration rate must be close to
0.5 AU Myr$^{-1}$.

The remaining parameters were then constrained by a best
fit to the submillimeter disk image presented in Holland
et al. (1998) and reproduced in Figure \ref{fig:vegaim}a:
the only variables were $a_{max}$, $a_{{pl}_i}$, $a_{{pl}_f}$,
and $\dot{a}_{pl}$, since $\delta$ was fixed at -0.5 and
$a_{min}$ was set at $a_{{pl}_i}$.
The observing geometry was assumed to be face-on and the
spatial distribution of planetesimals was converted into an
image of the dust emission resulting from the destruction of
those planetesimals using the following assumptions:
(i) that the spatial distribution of the dust exactly follows
the distribution of the parent bodies (i.e., so that the cross-sectional
area of emitting dust is proportional to the number of planetesimals);
(ii) that the grains' submillimeter emission is
$\propto 1/\sqrt{r}$, which is a good approximation for
the large grains which contribute to a disk's submillimeter
emission.

\begin{figure}
  \begin{center}
    \begin{tabular}{rlc}
      \textbf{(a)} & \hspace{-0.3in}
        \epsscale{0.9} 
\plotone{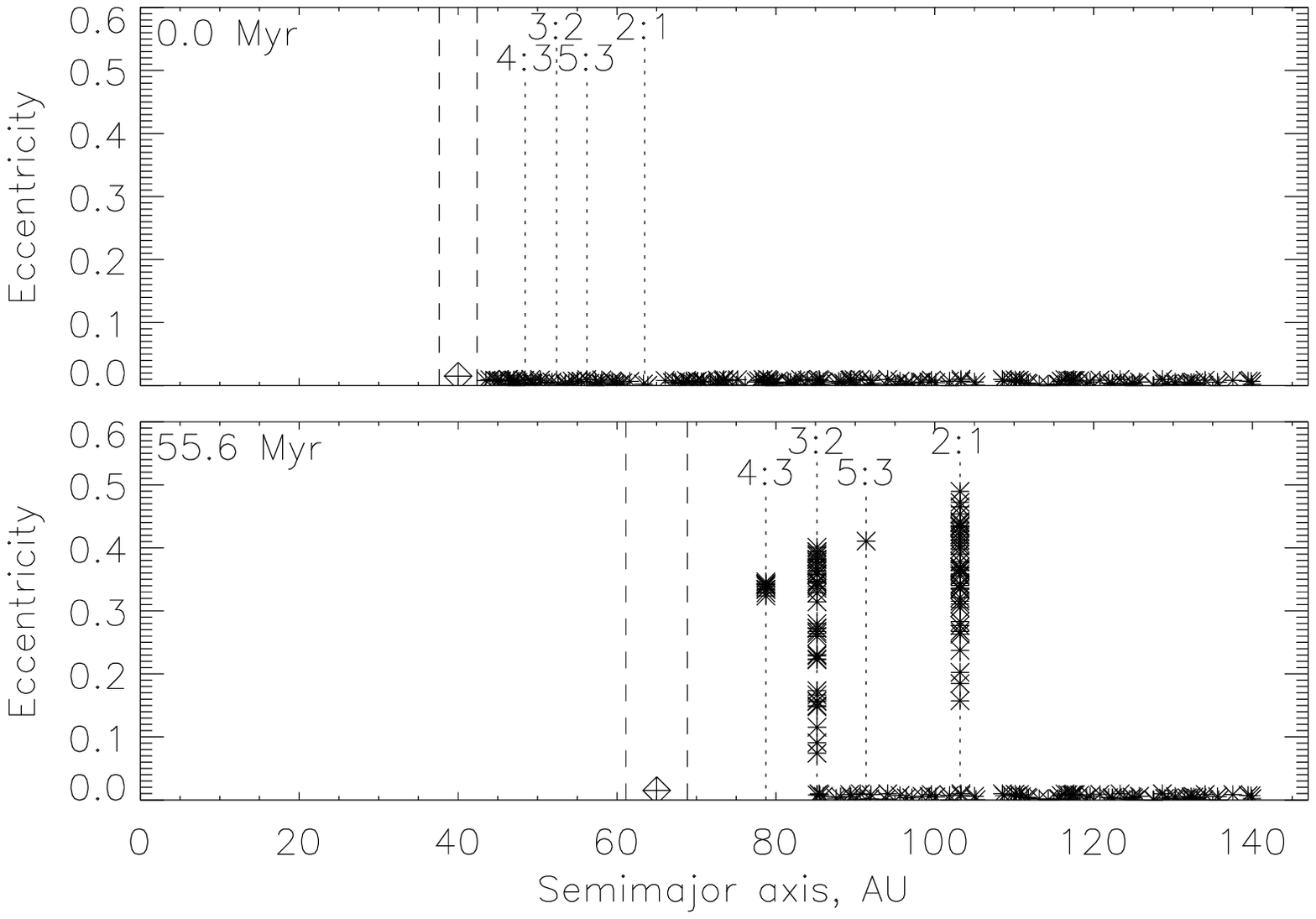} & \\[0.2in]
      \textbf{(b)} & \hspace{-0.3in}
        \epsscale{0.9} 
\plotone{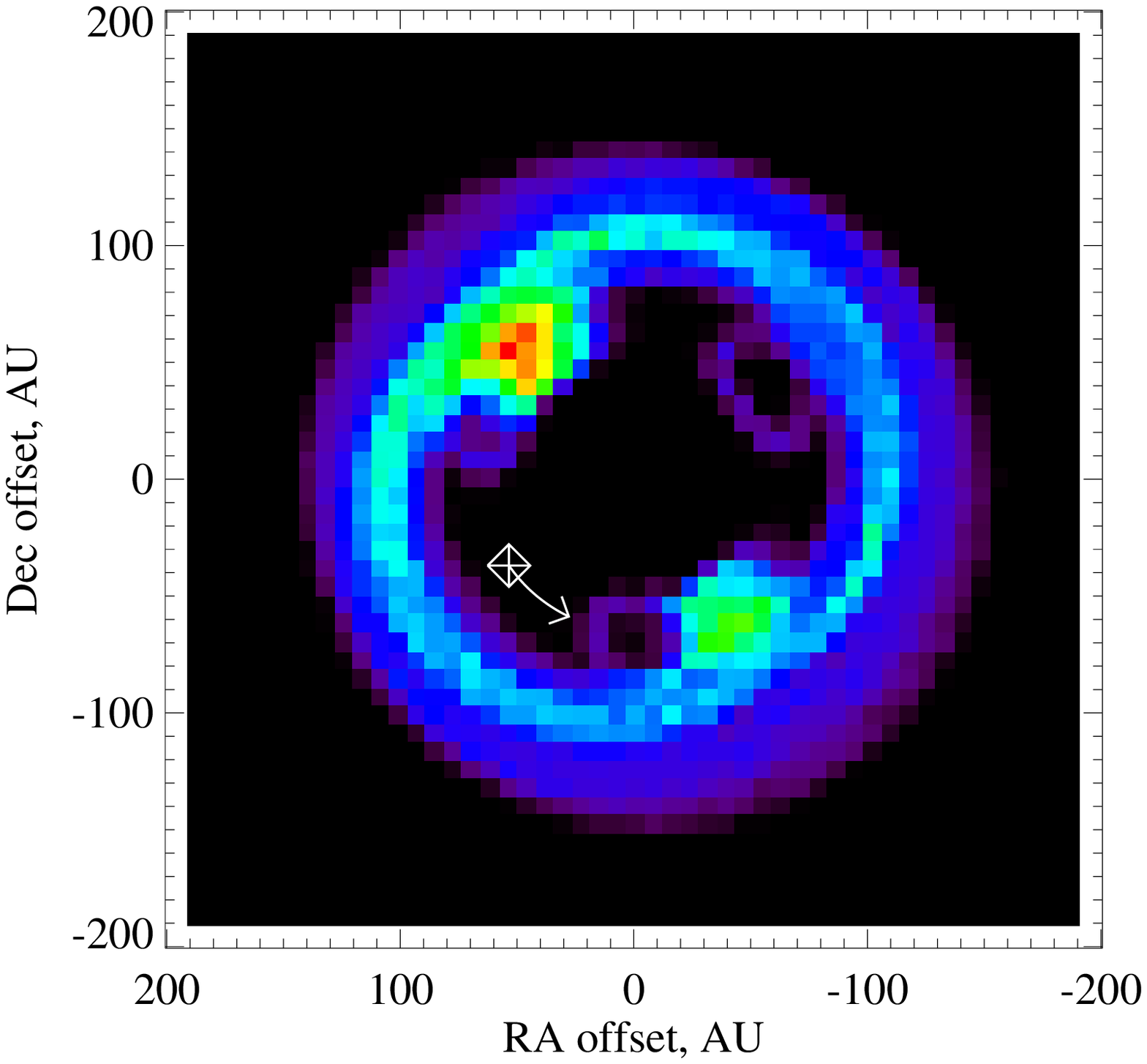} &
    \end{tabular}
  \end{center}
  \caption{Model of the origin of Vega's structure as a result
  of the migration of a Neptune mass planet from 40-65 AU over
  56 Myr.
  \textbf{(a)} The initial and final orbital distribution
  (eccentricity vs semimajor axis) of planetesimals in the
  disk.
  For clarity only the parameters of 200 planetesimals (the asterisks)
  are shown in this plot.
  The planet is located at the diamond-plus, the dotted lines
  indicate the location of the planet's resonances,
  and the dashed lines indicate the chaotic resonance overlap region.
  \textbf{(b)} Image of the number density distribution of planetesimals
  in the disk at the end of the migration.
  The planet is located at the diamond-plus.
  \label{fig:vegamod}}
\end{figure}

The resulting best fit is shown in Figure \ref{fig:vegaim}b,
and was performed by comparison of the contours of the two images;
the orbital and spatial distributions of the underlying
planetesimal population are shown in Figure \ref{fig:vegamod}.
The best fit parameters were found to be $a_{max} = 140 \pm 15$ AU,
$a_{{pl}_i} = 40 \pm 10$ AU, $a_{{pl}_f} = 65 \pm 5$ AU, and
$\dot{a}_{pl} = 0.45 \pm 0.1$ AU Myr$^{-1}$ (corresponding to
$\theta = 1.8-3.4$ at 40-140 AU and a total migration time
of 56 Myr\footnote{Given the age of the system, this implies that the migration
is now finished, thus it is important to point out that the rotation of
the resonant structure in the model due to migration (i.e.,
$\phi_m-180^\circ$; \SS \ref{sss:lcwm} and \ref{ss:gs}) is very small,
$<2^\circ$, and so its effect on the derived structure is negligible.}).
In the final distribution 5.1\% of the planetesimals
are trapped in the 4:3 resonance,
22.5\% in the 3:2 resonance, 0.4\% in the 5:3 resonance, 18.6\%
in the 2:1(u) resonance, 0.7\% in the 2:1(l) resonance,
41.0\% remain in nonresonant orbits, while 11.6\% are ejected by
resonance overlap.
The outer edge of the disk, $a_{max}$, was constrained
to within $\sim 15$ AU by a fit to the lowest contours in the
images.
The final planet location, $a_{{pl}_f}$, was constrained to within
$\sim 5$ AU by the radial offset of the northeast clump.
The contrast of the clumps as well as their morphology
were then used to constrain the initial location of the planet,
$a_{{pl}_i}$, and the migration rate, $\dot{a}_{pl}$.
Since the best fit has a migration rate for which trapping into the 2:1
resonance is $< 100$\%, and into the 5:3 resonance is $> 0$\%,
small changes in $\dot{a}_{pl}$
of the order $\pm 0.02$ AU Myr$^{-1}$ resulted in large changes in
the model structure.
However, $\dot{a}_{pl}$ could not be so well constrained, because
the extent of the migration, and so $a_{{pl}_i}$, also has a large
effect on the model structure, since this determines both the
fraction of planetesimals in each resonance as well as their
eccentricity distributions.
The errors given above are not formal errors, but approximate
limits for which acceptable fits are possible to the structure of Figure 
\ref{fig:vegaim}a.

\begin{figure}
  \begin{center}
    \begin{tabular}{c}
        \epsscale{0.9} 
\plotone{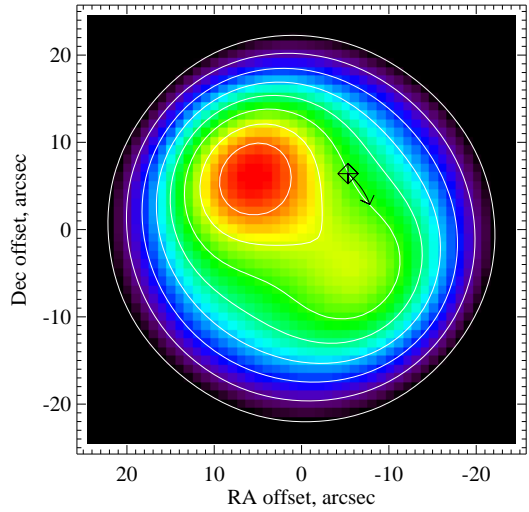}
    \end{tabular}
  \end{center}
  \caption{Alternative model explaining the 850 $\mu$m image
  of Vega's disk in which the planet causing the structure
  orbits the star clockwise (see Fig. \ref{fig:vegaim}).
  \label{fig:vegaim2}}
\end{figure}

This demonstrates that this model can fit the observations very
well and in doing so constrains important parameters regarding
the evolution of this system.
The implications of this model are as follows.
First of all the whole pattern is expected to orbit the
star with the same period as the planet.
As this is predicted to be at 65 AU, the orbital period is
330 yr.
Since the clumps are $\sim 9\arcsec$ from the star, their
motion would be $0\farcs17$ yr$^{-1}$.
With an absolute pointing uncertainty of $\pm 2\arcsec$
from the James Clerk Maxwell Telescope, such absolute motion
would not be detectable for several decades in submillimeter
images with current technology.
However, it would be easier to detect the relative
motion of the two clumps, since the position angle of
this structure should vary at $1.1^\circ$ yr$^{-1}$.
The direction of the orbital motion presented in Figure
\ref{fig:vegaim} is anticlockwise.
However this solution is not unique, since the model is
nearly symmetrical about the line joining the two clumps,
and I could also have modeled the system with
clockwise orbital motion, in which case the planet
would be currently north of the star.
Such a model is shown in Figure \ref{fig:vegaim2};
there are only slight differences in the disk
structure compared with Figure \ref{fig:vegaim}b.
The important point is that the brighter clump,
corresponding to the 2:1(u) resonant planetesimals, trails
some $75^\circ$ degrees in longitude behind the planet.

Another implication of the model is that the emission distribution
is much more intricate than that detectable with a $14\arcsec$ beam.
It is therefore interesting that in interferometric images of
Vega (Koerner et al.~2001), the northeast clump splits into three,
with two lower level clumps straddling a brighter clump;
the brighter clump in the model presented in this paper
is expected to be extended in longitude (see
Figure \ref{fig:vegamod}b).
This model also makes the prediction that a lower level
three clump pattern exists from planetesimals in the 4:3
resonance.
While two of the clumps almost blend with the two brighter
clumps from the 3:2 and 2:1 planetesimals, a faint clump
is expected on the opposite side of the star from the planet.
If the presence of such a clump could be inferred from
higher resolution and sensitivity observations of this disk,
the location of the planet and its direction of motion
could be unambiguously determined.

The mass and migration rate of the planet
have been constrained to lie at the lower edge of zone Di of
Figure \ref{fig:migzones}, although
an additional constraint is that the migration rate must be
$>0.07$ AU Myr$^{-1}$ for the 25 AU migration to have been
completed over the age of the system.
Thus the planet mass and migration rate I chose is
not unique and must be determined by a study of the
origin of the migration rate.
However, it is noteworthy that in Hahn \& Malhotra
(1999), Neptune was found to migrate from 23-30 AU over
50 Myr in their model with an initial planetesimal disk
mass of $50 M_\oplus$, a migration rate not too dissimilar
to that chosen for this model.
In other words, the mass and migration rate of the planet
causing the structure of Vega's disk could be
similar to the mass and migration rate of
Neptune in our own system.
This similarity to the solar system is contrary to
previous models for Vega's disk structure, which was
interpreted as dust grains migrating in toward the star
due to Poynting-Robertson drag that get trapped into the
2:1 and 3:1 resonances of a 3 Jupiter mass planet on an
orbit with an eccentricity of 0.6 (Wilner et al.~2002).

As for the possibility of directly detecting this planet,
observational constraints have recently been set on the
presence of planets around Vega, with an upper limit
of $\sim 3300M_\oplus$ at $10\arcsec$ from the star
(Metchev, Hillenbrand \& White 2003).
Given its proximity to the bright star Vega, it would be
very difficult to detect this planet with current technology
if it is indeed as small as $17M_\oplus$.
A larger planet, with a necessarily much faster migration
rate, may be detectable.

\subsection{Caveats}
\label{ss:caveats}
Despite the complexity of the model, it
is still just a first step toward a complete
explanation of the structures which could be caused
in extrasolar systems by planet migration.

For a start, the trapping probabilities were derived for
migration through a dynamically cold disk ($e<0.01$).
The planetesimals' eccentricities have a significant
influence on trapping probabilities, which decrease as
the eccentricities are increased, and also do not reach a
maximum at 100\% trapping probability for low migration
rates if the eccentricity is above a certain threshold
(Borderies \& Goldreich 1984; Melita \& Brunini 2000).
Most importantly, the trapping probabilities of different
resonances are affected in different ways, with higher
order ($q>1$) resonances becoming more populated relative
to lower order resonances for higher eccentricities.
Indeed Chiang et al.~(2003) propose that the
Kuiper belt must have been dynamically hot ($e\approx 0.2$)
when the migration of Neptune occurred to explain
the presence of three KBOs in the 5:2
resonance given that only seven objects have been discovered
in Neptune's 2:1 resonance.
It is difficult to predict an appropriate eccentricity
distribution for the residual planetesimal disk.
While planetesimals are thought to form in a dynamically
cold disk, the sweeping of a planet's resonances and its
secular perturbations, as well as scattering of planetesimals
from closer to the star, can all contribute to increasing the
average eccentricity of the planetesimal disk.
Certainly future models will have to consider the possibility
of high eccentricities, but in doing so will become
more complicated, since trapping into higher order resonances
will also have to be considered.

Another omission of the model is stochastic effects.
Migration in the models of Hahn \& Malhotra (1999)
is not smooth, but shows large jumps.
This is because the residual planetesimals used in their
simulation had to be large to give reasonable integration
times.
If the most massive residual planetesimals causing migration
are smaller than a certain limit, then stochastic effects
can be ignored (Chiang et al.~2003), otherwise they
must be characterized (Zhou et al.~2002).
One reason for anticipating that the residual planetesimals
causing the migration are much smaller than the planet would be
if the planet formed much closer to the star than the
planetesimals.
This could be the case if the planet was flung out in a 
gravitational interaction with other giant planets
(Thommes et al.~2002) or migrated out to a more distant
location.

Possibly the most worrying aspect of the model is
the translation of the distribution of planetesimals to
the distribution of dust.
These were assumed to be identical, but this may not be exactly true.
For a start the radial location of the resonances are
different for small dust, since they orbit the star slower than
larger grains due to radiation pressure (Wyatt et al. 1999).
Also, these particles have different orbits from 
their parents, not only because of radiation pressure, but
also due to the velocity given to them during the collision.
The effect of P-R drag and subsequent collisions should also
be taken into account.
These are important issues, but ones which merit a paper in themselves.
However, in defense of this assumption I will point out two things.
First, the dust which contributes to the submillimeter
images is expected to be large, since small grains
emit very inefficiently at such long wavelengths
(Wyatt \& Dent 2002).
Second, most resonant planetesimals are on planet-crossing
orbits (i.e., $a_{res}(1-e_{res}) < a_{pl}$), and
the only reason such orbits are stable is because resonant
forces prevent a close approach (\S \ref{s:drp}).
Small dust originating from such planetesimals, but
which is no longer in resonance, would therefore be
expected to be short-lived, since without the
protection of the resonance a close approach to the planet
is possible resulting in scattering out of the system.
Such scattering is confirmed in more recent runs following the
evolution of the dust particles' orbits (Wyatt, in prep.).

Another concern is that the dust distribution may be affected by random
events in which single massive planetesimals are disrupted.
Such events may have been witnessed in the structure of the zodiacal cloud
(Dermott et al. 2002; Nesvorn\'{y} et al. 2003).
However, Wyatt \& Dent (2002) concluded that inidividual collisions are
unlikely to be a significant source of structure in the Fomalhaut disk
(which also contains a clump, Holland et al. 2003), since this disk is so
massive that only a collision between two planetesimals
at least 1400 km in diameter could affect its structure at the level observed,
and too few such planetesimals can coexist in the disk.
Since Vega's disk, and its clumps, have a similar mass to those of Fomalhaut
(Holland et al. 1998), I conclude that random collisions are also unlikely
to be the cause of Vega's clumpy disk structure.

While I prefer to leave the discussion of the origin
of the planet's migration to a later paper, it is worth
pointing out that only certain values of $\theta$ will be possible
for a given planet mass $\mu$ and initial planetesimal disk
mass (Hahn \& Malhotra 1999).
In other words, the planet's mass and migration rate, as well
as the initial mass of Vega's planetesimal disk, could be
much better constrained with a model incorporating both
migration and resonant trapping.
Also, in the models of the migration of the solar system's
planets, Neptune's migration is outward only because of the
existence of the massive planets interior to its orbit.
This is because the planetesimals scattered inwards by Neptune
(causing its outward migration) may not return for a subsequent
scattering event (causing inward motion) due to their interaction
with the closer in planets, whereas the converse is not true.
Thus an outward migration may be said to be indicative of
at least two planets.
However, Ida et al.~(2000) presented an analytical model
for planet migration claiming that once outward migration has
started, it is self-sustaining, since there are automatically
less planetesimals interior to the planet's orbit.

\section{Conclusion}
\label{s:concl}
I have presented a model describing the consequence of the
outward migration of a planet on the dynamical and
spatial structure of a planetesimal disk residing
outside the planet's orbit.
In \S \ref{s:rcp} numerical simulations were used to show how
trapping probabilities into the 4:3, 3:2, 5:3, and 2:1
resonances can be estimated to within 5\% given just two
parameters $\mu=M_{pl}/M_\star$ and
$\theta = \dot{a}_{pl}\sqrt{a/M_\star}$.
Resonant forces cause a planetesimal's resonant argument, $\phi$,
to librate and in \S \ref{s:drp} physical arguments were used
to explain what azimuthal structure is expected in the
distribution of planetesimals as a result of this libration:
planetesimals trapped in the 4:3 and 5:3 resonances are
concentrated at 60, 180, and 300$^\circ$ longitude relative to the planet,
those in the 3:2 resonance are concentrated at $\pm 90^\circ$
longitude, and those in the 2:1 resonance are concentrated in
two clumps associated with the 2:1(u) resonance with a concentration
at relative longitude of $\sim 285^\circ$, and the 2:1(l) resonance
with a concentration at $\sim 75^\circ$.
Also in \S \ref{s:drp}, numerical simulations were used to show
how the angle about which $\phi$ librates, as well as the
amplitude of that libration, are affected by the migration parameters.
These simulations also characterized the overpopulation of the
2:1(u) resonance relative to the 2:1(l) for different
migrations.
In \S \ref{s:dds} the numerical results were used to derive a
numerical scheme to predict the spatial distribution of
planetesimals resulting from a migration defined by
$\mu$ and $\theta$.
It was shown that the dynamical structure of a post-migration
disk can have one of seven states depending on the $\mu$
and $\theta$ of the migration.

Application of the model to the structure of Vega's debris
disk presented in Holland et al.~(1998) shows that its two
clumps of unequal brightness can be explained by the migration
of a Neptune mass planet from 40-65 AU over $\sim 56$ Myr through
a planetesimal disk initially extending from 40-140 AU.
The two clumps are caused by planetesimals trapped in the
3:2 and 2:1 resonances, and the brightness asymmetry
is caused by an overabundance of planetesimals in the 2:1(u)
resonance relative to the 2:1(l) resonance.
While the extent of the planet's migration is well
constrained by the brightness of the clumps, its
mass and migration rate are not unique, although they are
constrained to lie within certain ranges defined by zone Di
in Figure \ref{fig:migzones}.
Further constraints on the planet's mass and migration rate,
as well as on the mass of the planetesimal disk, would be
possible by modeling the origin of the planet's migration
in angular momentum exchange with the planetesimal disk.
Predictions of the model which may prove its validity
are the orbital motion of Vega's clumpy pattern
($1.1^\circ$ yr$^{-1}$), the location of the planet ($8\farcs3$
from the star, $75^\circ$ in longitude in front of the
orbital motion of the NE clump), and the high resolution
structure of the clumps including the presence of a fainter
third clump on the opposite side of the star from the planet
from planetesimals in the 4:3 and 5:3 resonances.

While the mass and migration rate of Vega's perturbing planet
are not yet fully constrained, the migration parameters
derived for Neptune in the solar system (Hahn \& Malhotra 1999)
are close to the small region of parameter space which
could have caused Vega's structure.
There is also an intriguing suggestion that a planet's
outward migration requires the presence of massive planets
interior to its orbit.
Thus it seems possible that Vega's planetary system is
very similar to our own, not only in the presence and
mass of the planets in its system, but in that system's
early evolution.
It is also possible that application of this model to
the other clumpy debris disks may show these to harbor
solar-like planetary systems.
The weight of such conclusions is damped only by
the limitations of the model.
These will be addressed in future studies, and include
questions about the extent the dust distribution follows
that of the parent planetesimals, and how the conclusions
are affected if the planetesimal disk is initially
dynamically hot ($e>0.01$).

\acknowledgments
I am very grateful to Wayne Holland for providing
the published SCUBA observations Vega.

\appendix
\section{Adiabaticity}
\label{app:adiabatic}
\begin{figure}
  \begin{center}
    \begin{tabular}{rlc}
      \textbf{(a)} & \hspace{-0.3in}
        \epsscale{0.7} 
\plotone{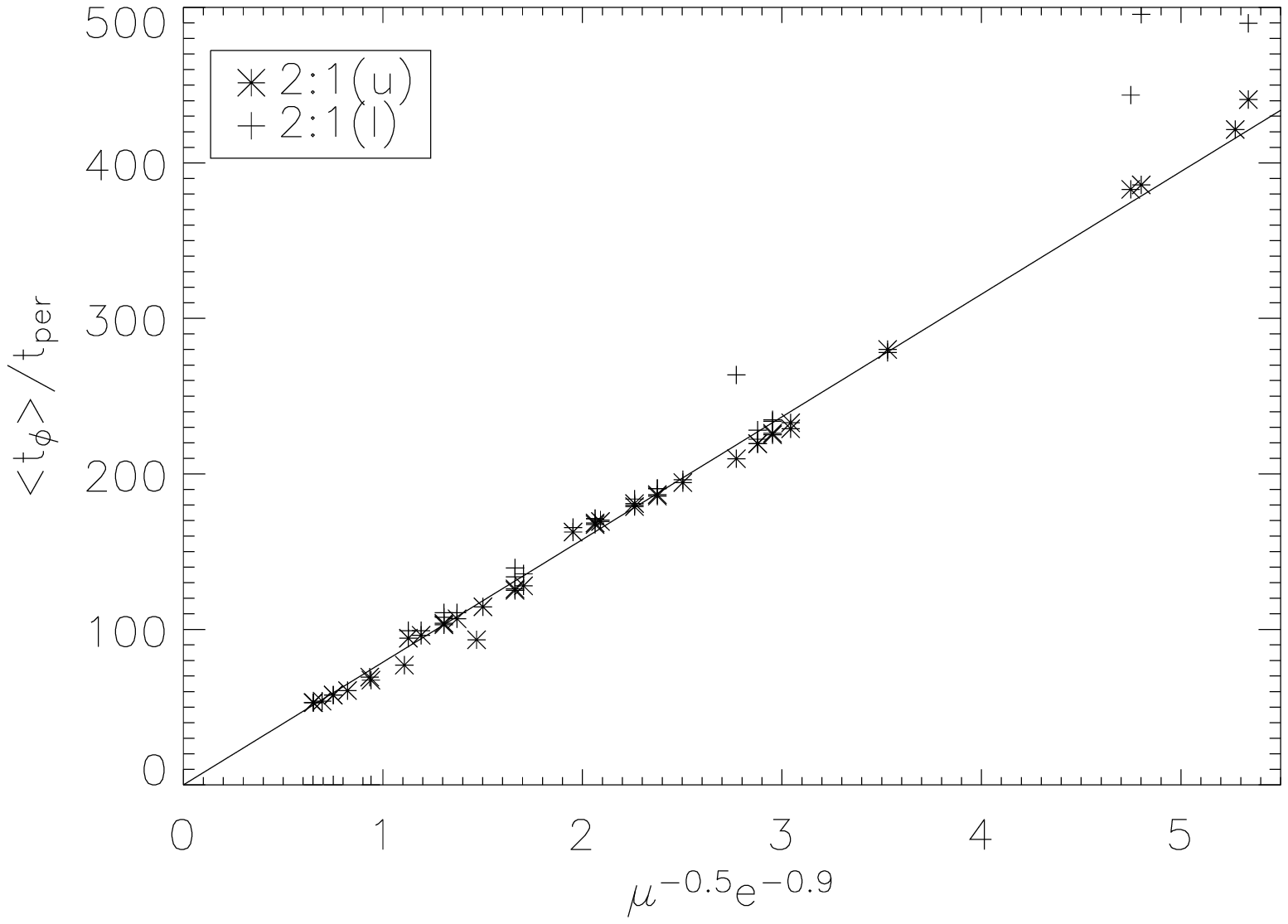} & \\[0.1in]
      \textbf{(b)} & \hspace{-0.3in} 
        \epsscale{0.7} 
\plotone{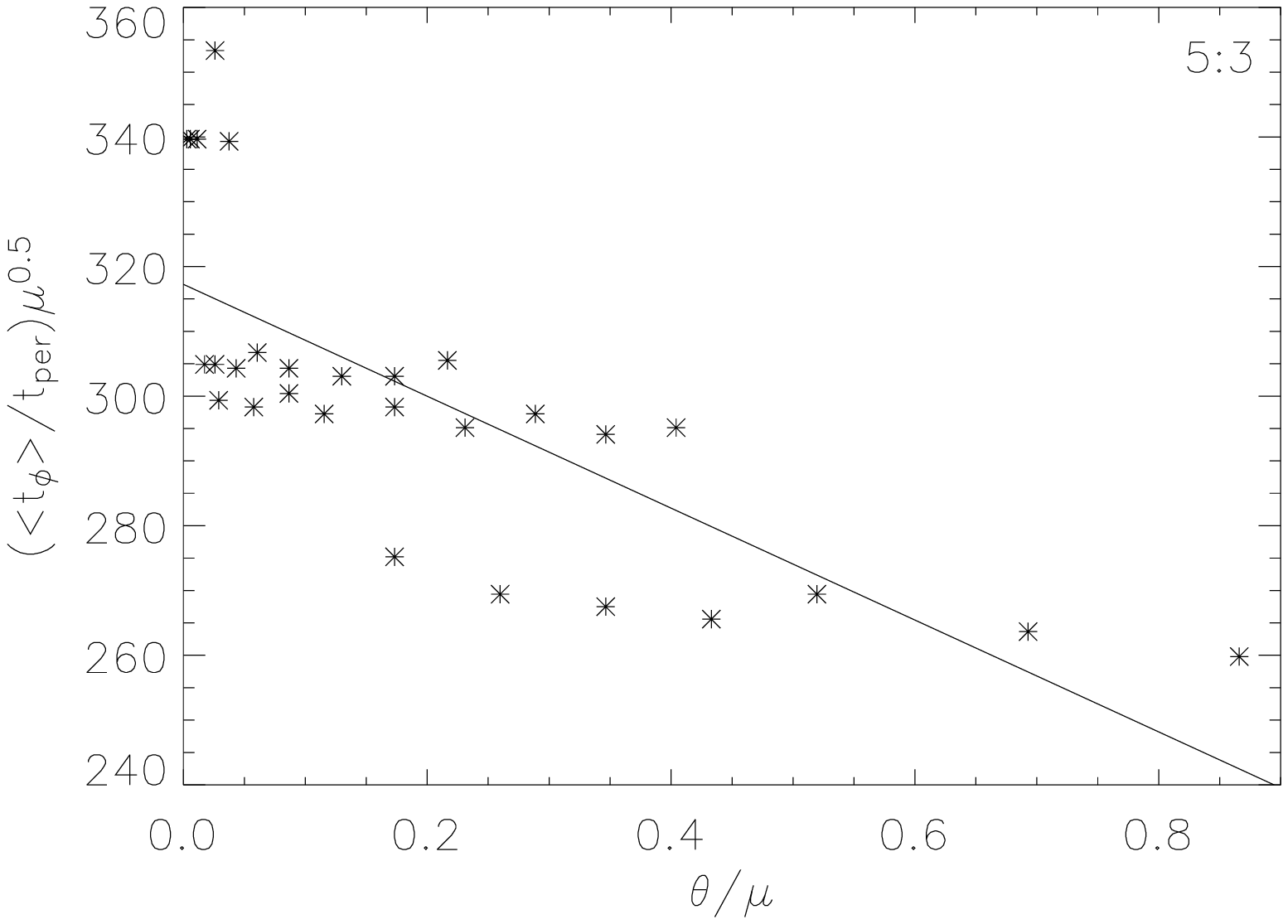} & \\[0.1in]
      \textbf{(c)} & \hspace{-0.3in}
        \epsscale{0.7} 
\plotone{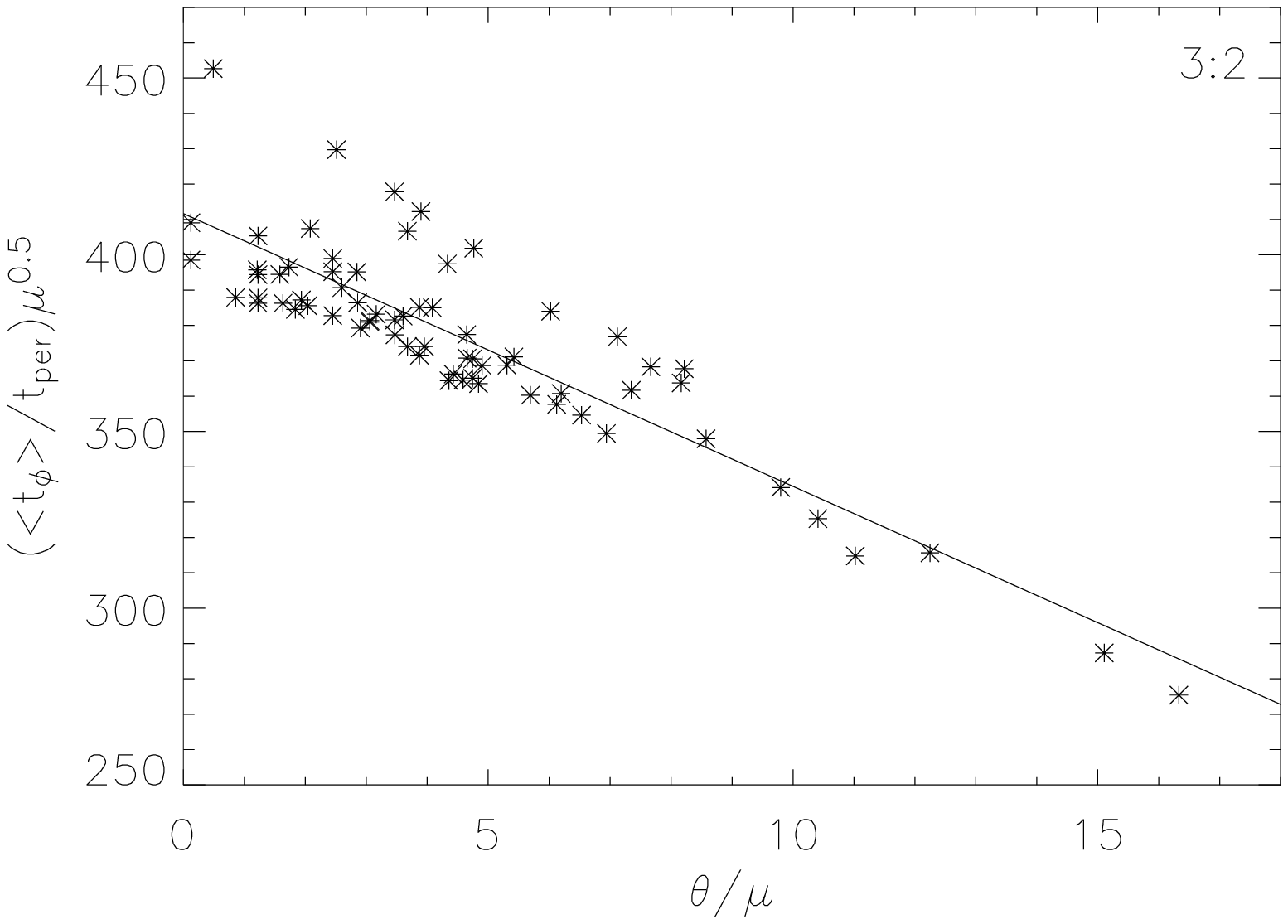} & \\[0.1in]
      \textbf{(d)} & \hspace{-0.3in}
        \epsscale{0.7} 
\plotone{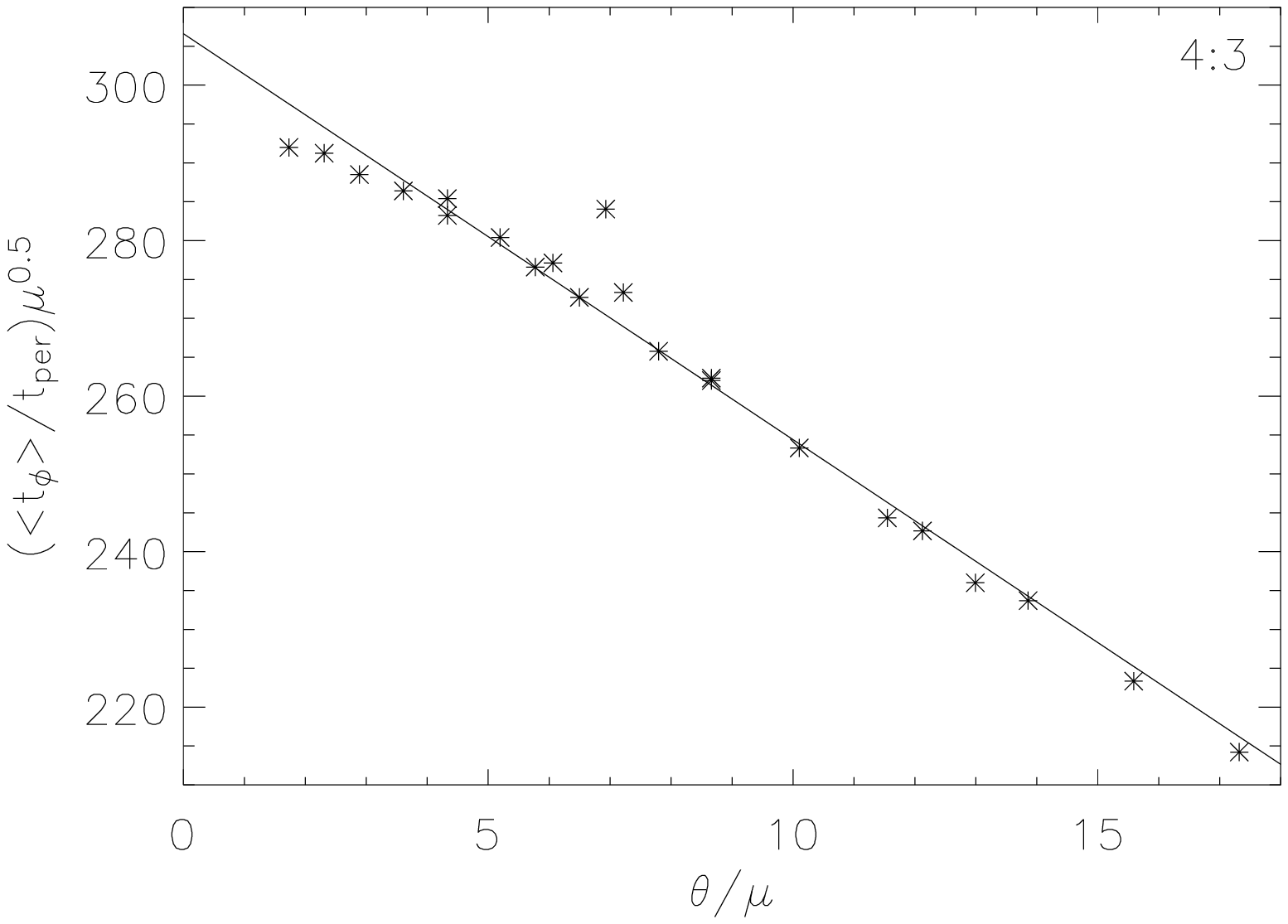} &
    \end{tabular}
  \end{center}
  \caption{Mean libration periods, $\langle t_\phi \rangle / t_{per}$,
  of planetesimals captured in the \textbf{(a)} 2:1,
  \textbf{(b)} 5:3, \textbf{(c)} 3:2, and \textbf{(d)}
  4:3 resonances for migrations defined by the parameters
  $\mu$ and $\theta$ (eqs.~[\ref{eq:mu}] and [\ref{eq:theta}]).
  The solid lines show the fits to these libration periods
  given in equations (\ref{eq:tphi21})-(\ref{eq:tphi43}).
  \label{fig:tphi}}
\end{figure}

As a result of the fit to the libration of $\phi$ performed in
\S \ref{ss:numdrp}, 
the libration period for each planetesimal, $t_\phi$ was also
determined for all resonant planetesimals in each run, as was
the mean libration period of the ensemble, $\langle t_\phi \rangle$.
The results are plotted in Figure \ref{fig:tphi}.
Fits to these were performed of the form
$(\langle t_\phi \rangle/t_{per})\sqrt{\mu} = A/e^B + C\theta/\mu$,
where $t_{per}$ is the orbital period of the planetesimal
and $B=0$ except for the 2:1 resonance.
These showed that
\begin{eqnarray}
  \langle t_{{\phi}_{2:1}} \rangle /t_{per} & = & 78.9\mu^{-0.5}e^{-0.9}, \label{eq:tphi21} \\
  \langle t_{{\phi}_{5:3}} \rangle /t_{per} & = & 317\mu^{-0.5} - 86.4\theta\mu^{-1.5}, \label{eq:tphi53} \\
  \langle t_{{\phi}_{3:2}} \rangle /t_{per} & = & 412\mu^{-0.5} - 7.7\theta\mu^{-1.5}, \label{eq:tphi32} \\
  \langle t_{{\phi}_{4:3}} \rangle /t_{per} & = & 307\mu^{-0.5} - 5.2\theta\mu^{-1.5}, \label{eq:tphi43}
\end{eqnarray}
and these fits are also shown on Figure \ref{fig:tphi}.
So, the libration period for the 2:1 resonance is
not affected by the planet's migration rate, but it does
decrease during the migration as the planetesimal's
eccentricity increases.
However, for the 5:3, 3:2, and 4:3 resonances, the libration period is constant
throughout the migration, but does depend on the $\mu$ and $\theta$ of 
the migration in that faster migrations (and larger planet masses) result 
in smaller libration periods.

\begin{figure}
  \begin{center}
    \begin{tabular}{c}
      \epsscale{0.7} 
\plotone{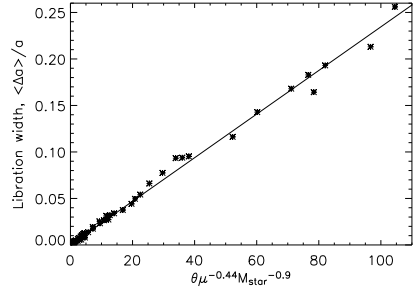}
    \end{tabular}
  \end{center}
  \caption{Mean libration width, $\langle \Delta a \rangle / a$,
  of planetesimals captured in the 3:2
  resonance for migrations defined by the parameters
  $\mu$ and $\theta$ (eqs.~[\ref{eq:mu}] and [\ref{eq:theta}])
  for stellar masses $M_\star$.
  The solid line shows the fit to these libration widths
  given in equation (\ref{eq:da32}).
  \label{fig:da}}
\end{figure}

The libration width for each planetesimal, $\Delta a = a_{max} - a_{min}$,
was also measured, as was the mean libration width
$\langle \Delta a \rangle$.
Since the runs for the 3:2 resonance showed that $\Delta a$ is not just
dependent on $\mu$ and $\theta$, but also on $M_\star$ which was not varied
for the other resonances, I only attempted to characterize the libration
width for the 3:2 resonance (see also Figure \ref{fig:da}):
\begin{equation}
  \langle \Delta a_{3:2} \rangle /a = 0.0023 \theta\mu^{-0.44}M_\star^{-0.9}.
  \label{eq:da32}
\end{equation}
The mean libration widths for all resonances were used, however, to question the
adiabaticity of this libration. 
Adiabaticity is defined as when the libration period, $t_\phi$,
is much shorter than the time it takes for the resonance to
cross the libration width, i.e., $t_\phi \ll \Delta a / \dot{a}_{pl}$.
I found that in the runs shown here, 
$N_{lib} = (\Delta a / \dot{a}_{pl})/t_\phi$ was in the ranges:
$1.5-15$ for the 3:2 resonance;
$2.5-3.9$ for the 4:3 resonance;
$10-400$ for the 5:3 resonance;
and $3.6-25$ for the 2:1 resonance.
In other words, all the migrations considered in this paper satisfy
adiabatic invariance (Henrard 1982), though many are close to this
limit.


\clearpage

\clearpage

\begin{deluxetable}{ccccccc}
  \tabletypesize{\scriptsize}
  \tablecaption{Coefficients that determine capture probabilities from
    equation (\ref{eq:p2}) for different resonances (\S \ref{s:rcp}).
    Also given are the semimajor axes of the resonances, $a_r$, with respect
    to that of the perturbing planet, $a_{pl}$, and the average errors in the
    capture probabilities using these models, $P_{err}$.
    \label{tab:xyuv}}
  \tablewidth{0pt}
  \tablehead{\colhead{Resonance} & \colhead{$a_r/a_{pl}$} &
    \colhead{$X$} & \colhead{$Y$} & \colhead{$u$} & \colhead{$v$} &
    \colhead{$P_{err}$}}
  \startdata
    4:3 & 1.21 & $0.23 \pm 0.02$ & $5.6 \pm 2.0$ & $1.42 \pm 0.01$ & $0.29 \pm 0.1$  & 0.05  \\
    3:2 & 1.31 & $0.37 \pm 0.02$ & $5.4 \pm 2.0$ & $1.37 \pm 0.01$ & $0.38 \pm 0.1$  & 0.04  \\
    5:3 & 1.41 & $210  \pm 20$   & $1.0 \pm 0.2$ & $1.84 \pm 0.02$ & $0.20 \pm 0.08$ & 0.04  \\
    2:1 & 1.59 & $5.8  \pm 0.2$  & $4.3 \pm 2.0$ & $1.40 \pm 0.02$ & $0.27 \pm 0.2$  & 0.025 \\
  \enddata
\end{deluxetable}

\clearpage

\begin{deluxetable}{cccccc}
  \tabletypesize{\scriptsize}
  \tablecaption{Which resonances planetesimals initially at semimajor axes of $a_i$
    end up in given the initial semimajor axis of the planet $a_{{pl}_i}$ for the
    different migration scenarios A-Eii shown in Figure \ref{fig:migzones}.
    Also given are the longitudes of the concentrations of planetesimals in
    the different resonances relative to the planet, $\lambda - \lambda_{pl}$.
    \label{tab:lr}}
  \tablewidth{0pt}
  \tablehead{ \colhead{Resonance} & \colhead{2:1(u)} &
    \colhead{2:1(l)} & \colhead{5:3} & \colhead{3:2} & \colhead{4:3}}
  \startdata
    $\lambda-\lambda_{pl}$ & $-(107-79^\circ)$ & $(107-79^\circ)$ & $\pm 60^\circ, 180^\circ$ & $\pm 90^\circ$ & $\pm 60^\circ, 180^\circ$ \\
    A   & - & - & - & - & - \\
    B   & - & - & - & - & $a_i > 1.21a_{{pl}_i}$ \\
    C   & - & - & - & $a_i > 1.31a_{{pl}_i}$ & $a_i = (1.21-1.31)a_{{pl}_i}$ \\
    Di  & $a_i > 1.59a_{{pl}_i}$ & - & - & $a_i=(1.31-1.59)a_{{pl}_i}$ & $a_i = (1.21-1.31)a_{{pl}_i}$ \\
    Dii & 50\% of $a_i > 1.59a_{{pl}_i}$ & 50\% of $a_i > 1.59a_{{pl}_i}$ & - & $a_i=(1.31-1.59)a_{{pl}_i}$ & $a_i = (1.21-1.31)a_{{pl}_i}$ \\
    Ei  & $a_i > 1.59a_{{pl}_i}$ & - & $a_i=(1.41-1.59)a_{{pl}_i}$ & $a_i=(1.31-1.41)a_{{pl}_i}$ & $a_i = (1.21-1.31)a_{{pl}_i}$ \\
    Eii & 50\% of $a_i > 1.59a_{{pl}_i}$ & 50\% of $a_i > 1.59a_{{pl}_i}$ & $a_i=(1.41-1.59)a_{{pl}_i}$ & $a_i=(1.31-1.41)a_{{pl}_i}$ & $a_i = (1.21-1.31)a_{{pl}_i}$ \\
  \enddata
\end{deluxetable}

\end{document}